\pdfoutput=1 %
\documentclass[11pt]{article}

\usepackage[T1]{fontenc}
\usepackage{lmodern}
\usepackage[margin=1in]{geometry}
\usepackage{microtype}

\usepackage{amsmath}   %
\usepackage{amssymb}   %
\usepackage{amsthm}    %
\usepackage{bm}        %

\usepackage{graphicx}  %
\usepackage{booktabs}  %
\usepackage{multirow}  %
\usepackage[round]{natbib} %
\usepackage{xcolor}
\usepackage[colorlinks=true, linkcolor=blue!50!black,
            citecolor=blue!50!black, urlcolor=blue!50!black]{hyperref}

\theoremstyle{plain}
\newtheorem{theorem}{Theorem}
\newtheorem{lemma}[theorem]{Lemma}

\newtheorem{assumption}{Assumption}
\theoremstyle{definition}

\theoremstyle{remark}
\newtheorem{remark}[theorem]{Remark}

\newcommand{\ve}[1]{{\mbox{\boldmath ${#1}$}}}

\DeclareMathOperator*{\argmin}{argmin}

\newenvironment{keywords}
  {\par\medskip\noindent\textbf{Keywords:}\ }
  {\par\medskip}
\newcommand{\acks}[1]{\section*{Acknowledgments and Disclosure of Funding}#1}
\newcommand{\name}{\normalfont\bfseries}
\newcommand{\email}{\quad\normalfont\normalsize\ttfamily}
\newcommand{\addr}{\normalfont}
\newcommand{\AND}{\end{tabular}\par\vskip 1.5ex\begin{tabular}[t]{c}}
\newcommand{\jmlrheading}[7]{}
\newcommand{\ShortHeadings}[2]{}
\newcommand{\firstpageno}[1]{}
\newcommand{\editor}[1]{}

\begin{document}

\title{Direct estimation and inference of differential Granger causality between two high-dimensional time series}

\author{\name Yue Wang \email yue.2.wang@cuanschutz.edu \\
       \addr Department of Biostatistics and Informatics\\
       University of Colorado Anschutz Medical Campus\\
       Aurora, CO 80045, USA
       \AND
       \name Jing Ma \email jingma@fredhutch.org \\
       \addr Public Health Sciences Division\\
       Fred Hutchinson Cancer Center\\
       Seattle, WA 98109, USA
       \AND
        \name Ali Shojaie \email ashojaie@uw.edu \\
       \addr Department of Biostatistics\\
       University of Washington\\
       Seattle, WA 98195, USA
       }

\date{}

\maketitle

\begin{abstract}
Comparing dependence structures between two related multivariate time series is of fundamental interest in many scientific applications, where changes may occur in both directed temporal interactions and contemporaneous connectivity.
Modeling each time series by a vector autoregressive (VAR) model, we propose a new framework for estimation and inference of differential Granger causality (DiffGC) and differential network (DiffNet) structures in high dimensions. The proposed method is based on a novel estimating equation derived from the Yule--Walker equations that directly links the difference between VAR transition matrices to the corresponding difference between precision matrices. Unlike separate estimation strategies that estimate the two VAR models individually and then take their difference, the proposed method directly targets the differential structures and requires sparsity only of the differences, thereby accommodating potentially dense individual networks, including hub structures.
Building on the direct estimators, we develop selective inference procedures for both DiffNet and DiffGC parameters, providing valid post-selection inference while accounting for the data-driven screening process. Theoretically, we establish convergence rates and support recovery guarantees for the proposed estimators, derive asymptotic distributions for the selective inference targets, and obtain new consistency results for DiffNet estimation under weaker assumptions than existing methods. Simulation studies confirm the theoretical convergence rates and demonstrate accurate support recovery and robust inferential performance. An application to resting-state electroencephalography (EEG) data identifies substantial changes in both contemporaneous and Granger-causal connectivity across recording sessions.
\end{abstract}


\section{Introduction}

Vector autoregressive (VAR) models are a fundamental tool for analyzing multivariate time series and have been widely used in economics, finance, and neuroscience \citep{de2008forecasting, fan2011sparse, moller2001instantaneous, sims1980macroeconomics}. In a VAR($p$) model, the current observation is represented as a linear function of its previous $p$ lags, with the corresponding transition matrices encoding directed temporal dependencies among the component time series. Nonzero entries of these transition matrices characterize Granger causal relationships and are often of primary scientific interest \citep{shojaie2021}. When the dimension is large relative to the sample size, estimation of the transition matrices requires additional structural assumptions, most commonly sparsity. A large literature has therefore developed regularized estimation procedures and corresponding theoretical guarantees for high-dimensional VAR models \citep{hamilton1994time,
fujita2007modeling, hsu2008subset, lozano2009grouped, haufe2010sparse, pereira2010learning,
shojaie2010discovering, song2011large, basu2015regularized, han2015}.

While substantial progress has been made in estimating a single high-dimensional VAR model, many scientific questions concern differences between two related VAR models. Such questions arise in a variety of settings. For example, researchers may be interested in comparing connectivity patterns before and after a major event within a single time series, such as the onset of an epileptic seizure or a financial market disruption. In longitudinal studies, one may seek to understand how Granger causal relationships evolve across repeated measurements from the same subject. Similar comparisons arise between distinct populations, such as treatment and control groups in clinical studies. In each of these settings, the primary objective is to identify changes in Granger causal relationships between two multivariate time series.

Suppose two stationary $d_0$-dimensional time series satisfy
\begin{equation}
\label{var1}
{\bf x}^{(l)}_t
=
\sum_{k=1}^{p}
A_k^{(l)\top}
{\bf x}^{(l)}_{t-k}
+
{\ve \epsilon}^{(l)}_t,
\qquad
l=1,2, ~~ t=1,\ldots, n_l
\end{equation}
where $A_k^{(l)}$ denotes the $k$th-order transition matrix.
We assume ${\bf x}^{(l)}_t \sim N(0, \Sigma_l)$ for $l=1,2$.

Our goal is to estimate
\[
\Delta_{A,k}
=
A_k^{(1)}-A_k^{(2)},
\qquad
k=1,\ldots,p,
\]
which characterize changes in Granger causal relationships between the two systems.
Using the standard companion representation \citep{lutkepohl2005new}, define
\[
\widetilde{\bf x}^{(l)}_t
=
\left(
{\bf x}^{(l)\top}_{t+p-1},
{\bf x}^{(l)\top}_{t+p-2},
\ldots,
{\bf x}^{(l)\top}_{t}
\right)^\top,
\qquad l=1,2,
\]
where $\widetilde{\bf x}^{(l)}_t \in \mathbb{R}^d$ with $d = d_0p$.
Then the VAR($p$) model in \eqref{var1} can be rewritten as a VAR(1) model:
\begin{align}\label{varp-var1}
\widetilde{\bf x}^{(l)}_t
=
\widetilde A^{(l)\top}
\widetilde{\bf x}^{(l)}_{t-1}
+
\widetilde{\ve\epsilon}^{(l)}_t,
\qquad l=1,2,
\end{align}
where $\widetilde A^{(l)}$ is the corresponding companion transition matrix. Let
$\Omega_l=\Sigma_l^{-1}$.
The two key quantities considered in this paper are
\[
\Delta_A
=
\widetilde A^{(1)}-\widetilde A^{(2)}
\qquad\mbox{and}\qquad
\Delta_\Omega
=
\Omega_1-\Omega_2
=
\Sigma_1^{-1}-\Sigma_2^{-1}.
\]
The quantity $\Delta_\Omega$ coincides with the differential precision matrix studied in non-time-series differential network analysis, which we refer to as DiffNet \citep{Zhao2014, yuan2017differential}. The first block column of $\Delta_A$ contains $\Delta_{A,1},\ldots,\Delta_{A,p}$ and therefore characterizes changes in Granger causal relationships between the two systems, which we refer to as differential Granger causality (DiffGC).

To the best of our knowledge, DiffGC has not been directly estimated in previous work. A straightforward strategy is to estimate the two VAR models separately and then take the differences between the corresponding transition matrices. We refer to this strategy as \emph{separate estimation}.
In high-dimensional settings, consistency of separate estimation procedures based on $l_1$-regularization typically requires each transition matrix to be sparse. Specifically, one must assume that $A_k^{(l)}$ is sparse for every $k=1,\ldots,p$ and $l=1,2$.
However, sparsity of the individual VAR models may be substantially stronger than what is required for DiffGC analysis.
Even when the transition matrices $\{A_k^{(1)}\}_{k=1}^{p}$ and $\{A_k^{(2)}\}_{k=1}^{p}$ are individually dense, the DiffGC matrices
\(
\{\Delta_{A,k}\}_{k=1}^p
\)
may remain sparse if only a relatively small number of Granger causal relationships change across the two systems.
An important example arises in networks containing \emph{hub nodes} that interact with a large number of other variables and therefore induce dense rows or columns in the corresponding transition matrices \citep{buckner2009cortical, power2013evidence}.
As a result, the individual VAR models may violate standard sparsity assumptions even when only a small number of Granger causal relationships differ between the two systems. In these settings, separate estimation may perform poorly because it requires accurate estimation of two potentially dense VAR models.
This limitation highlights the need for new methodologies that directly estimate DiffGC under sparsity assumptions on the changes themselves.

To address the limitations of separate estimation strategy, we propose a unified framework for estimation and inference of DiffGC in high-dimensional VAR models. The main contributions of this paper are summarized as follows.

\begin{enumerate}

\item \textbf{Direct estimation of DiffGC.}
We develop a new estimating equation based on the Yule--Walker equations \citep{yule1927method, walker1931periodicity} that directly links DiffGC $\Delta_A$ to DiffNet
\(
\Delta_\Omega.
\)
Building on this equation, we propose a direct estimator of the DiffGC matrix $\Delta_A$ that only requires sparsity of the differential structure.
In contrast to separate estimation approaches, the proposed method does not require sparsity of the individual VAR models and therefore accommodates hub structures and other dense connectivity patterns.

\item \textbf{New theoretical guarantees.}
We establish convergence rates and support recovery guarantees for both $\Delta_\Omega$ and $\Delta_A$. For DiffNet estimation, we derive consistency results under substantially weaker assumptions than existing approaches.
The resulting convergence rates are also potentially faster than those established in existing work. For DiffGC estimation, we establish element-wise, Frobenius-norm, and variable-selection consistency.

\item \textbf{Selective inference for DiffNet and DiffGC.}
Beyond estimation, we develop a selective inference framework for both the DiffNet parameter $\Delta_\Omega$ and the DiffGC parameter $\Delta_A$. The proposed procedure combines sample splitting and post-selection inference to provide valid $p$-values while accounting for the data-driven selection process. A key technical contribution is the derivation of asymptotic distributions for the resulting inferential targets, which enables valid post-selection inference and uncertainty quantification for both DiffNet and DiffGC. To the best of our knowledge, this is the first inference framework for DiffGC in high-dimensional VAR models.

\end{enumerate}

Extensive simulation studies verify the theoretical convergence rates, demonstrate accurate support recovery, and confirm the effectiveness of the proposed estimation and inference procedures. An application to resting-state EEG data further illustrates the practical utility of the proposed framework for detecting changes in both contemporaneous connectivity, as characterized by DiffNet, and directed Granger-causal connectivity, as characterized by DiffGC.

\subsection*{Notation}

Throughout the paper, scalars, vectors, and matrices are denoted by lowercase letters, bold lowercase letters, and uppercase letters, respectively. For a vector ${\bf v}$, let $\|{\bf v}\|_0$, $\|{\bf v}\|_q$, and $\|{\bf v}\|_\infty$ denote the $l_0$, $l_q$, and $l_\infty$ norms, respectively. For a matrix $M=(m_{jk})$, we define
\[
\|M\|_{\max}
=
\max_{j,k}|m_{jk}|,
\qquad
\|M\|_{\mathrm F}
=
\Big(
\sum_{j,k}m_{jk}^2
\Big)^{1/2},
\]
and let $\|M\|_q=\max_{\|{\bf v}\|_q=1}\|M{\bf v}\|_q$ denote the induced matrix norm. We further define
\[
|M|_0
=
\sum_{j,k}
I(m_{jk}\neq 0),
\qquad
|M|_1
=
\sum_{j,k}
|m_{jk}|.
\]

For a matrix $M$, $\mathrm{vec}(M)$ denotes the vector obtained by stacking the columns of $M$. We use $\lambda_{\min}(M)$ and $\lambda_{\max}(M)$ to denote the smallest and largest eigenvalues of $M$, respectively. For matrices $A$ and $B$ of compatible dimensions, $A\otimes B$ denotes their Kronecker product and
\[
\langle A,B\rangle
=
\mathrm{tr}(AB^\top).
\]

\section{Direct estimation}\label{sec:directest}

We develop the direct estimation procedure based on the companion VAR(1) representation in \eqref{varp-var1}. In addition to $\Sigma_l=\operatorname{cov}\left(\widetilde{\bf x}^{(l)}_t\right)$, define the lag-one autocovariance matrix
$\Theta_l=\operatorname{cov}\left(\widetilde{\bf x}^{(l)}_t,\widetilde{\bf x}^{(l)}_{t+1}\right)$, $l=1,2$.
Rather than separately estimating $\widetilde A^{(1)}$ and $\widetilde A^{(2)}$, we derive an estimating equation that directly links $\Delta_A$ to $\Sigma_1$, $\Sigma_2$, $\Theta_1$, $\Theta_2$, and $\Delta_\Omega$. The covariance and lag-one autocovariance matrices can be estimated directly from sample moments, while $\Delta_\Omega$ can itself be estimated directly without separately estimating $\Sigma_1^{-1}$ and $\Sigma_2^{-1}$. This construction therefore allows direct estimation of $\Delta_A$ under sparsity assumptions on the differential structures rather than on the individual VAR models.

Specifically,
by the Yule-Walker equation, we have
\[
    \widetilde{A}^{(l)} = \Sigma_l^{-1} \Theta_l.
\]
Thus, we write
\begin{align*}
    \Delta_A &~ = \widetilde{A}^{(1)} - \widetilde{A}^{(2)} = \Sigma_1^{-1}\Theta_1 - \Sigma_2^{-1}\Theta_2  \nonumber\\
    &~ = \Delta_\Omega \Theta_1 + \Sigma_2^{-1} (\Theta_1 - \Theta_2)
    \nonumber \\
    &~ = \Delta_\Omega \Theta_2 + \Sigma_1^{-1}(\Theta_1 - \Theta_2).
\end{align*}
Therefore, some algebra yields
\begin{align}\label{050420:1}
 (\Sigma_1 + \Sigma_2)\Delta_A = \Sigma_1\Delta_\Omega\Theta_2 + \Sigma_2\Delta_\Omega\Theta_1 + 2(\Theta_1 - \Theta_2).
\end{align}

Natural estimators of $\Sigma_l$ and $\Theta_l$ are given by
\begin{align}
\label{est:1}
\widehat{\Sigma}_l
=
\frac{1}{n_l-p+1}
\sum_{t=1}^{n_l-p+1}
\widetilde{\bf x}_t^{(l)}
\widetilde{\bf x}_t^{(l)\top},
\qquad
\widehat{\Theta}_l
=
\frac{1}{n_l-p}
\sum_{t=1}^{n_l-p}
\widetilde{\bf x}_t^{(l)}
\widetilde{\bf x}_{t+1}^{(l)\top},
\end{align}
for $l=1,2$. Thus, the primary remaining challenge is estimation of the DiffNet matrix
\(
\Delta_\Omega.
\)
A natural estimator is $\widehat{\Sigma}_1^{-1}-\widehat{\Sigma}_2^{-1}$ whenever the sample covariance matrices are nonsingular. In high-dimensional settings, however, this estimator is ill-posed. One possible remedy is to separately estimate $\Sigma_1^{-1}$ and $\Sigma_2^{-1}$ using sparse precision-matrix estimation techniques such as the graphical lasso \citep{friedman2008sparse}, neighborhood selection \citep{meinshausen2006high}, or constrained $l_1$ minimization \citep{cai2011constrained, han2015}. Similar to the discussion in the introduction, these approaches require sparsity of the individual precision matrices and may perform poorly when individual sparsity is violated.

To avoid separate estimation, we exploit the equation
\begin{align}
\label{diffest:add:1}
\Sigma_2\Delta_\Omega\Sigma_1
=
\Sigma_2-\Sigma_1,
\end{align}
or equivalently,
\(
\Sigma_1\Delta_\Omega\Sigma_2
=
\Sigma_2-\Sigma_1.
\)
Based on \eqref{diffest:add:1}, we directly estimate $\Delta_\Omega$ using the D-trace approach of \citet{yuan2017differential}, which only requires sparsity of $\Delta_\Omega$. Specifically, we define the D-trace loss function
\begin{align}
\label{dtrace:1}
L_{\mathrm D}
\left(
\Delta_\Omega,
\Sigma_1,
\Sigma_2
\right)
=
\frac{1}{4}
\Big(
\langle
\Sigma_1\Delta_\Omega,
\Delta_\Omega\Sigma_2
\rangle
+
\langle
\Sigma_2\Delta_\Omega,
\Delta_\Omega\Sigma_1
\rangle
\Big)
-
\langle
\Delta_\Omega,
\Sigma_2-\Sigma_1
\rangle.
\end{align}

The D-trace loss is convex and uniquely minimized at $\Delta_\Omega$ because
\begin{align}
\frac{\partial L_{\mathrm D}
\left(
\Delta_\Omega,
\Sigma_1,
\Sigma_2
\right)}
{\partial \Delta_\Omega}
&=
\frac{1}{2}
\left(
\Sigma_1\Delta_\Omega\Sigma_2
+
\Sigma_2\Delta_\Omega\Sigma_1
\right)
-
(\Sigma_2-\Sigma_1),
\label{dtrace:gradient}
\\
\frac{\partial^2 L_{\mathrm D}
\left(
\Delta_\Omega,
\Sigma_1,
\Sigma_2
\right)}
{\partial \Delta_\Omega
\partial \Delta_\Omega^\top}
&=
\frac{1}{2}
\left(
\Sigma_1\otimes\Sigma_2
+
\Sigma_2\otimes\Sigma_1
\right)
\succ 0.
\label{hessian}
\end{align}

Replacing $\Sigma_l$ by $\widehat{\Sigma}_l$ for $l=1,2$, we estimate $\Delta_\Omega$ by
\begin{align}
\label{dtrace:2}
\widehat{\Delta}_\Omega(\nu)
=
\argmin_{\Delta_\Omega}
\left\{
L_{\mathrm D}
\left(
\Delta_\Omega,
\widehat{\Sigma}_1,
\widehat{\Sigma}_2
\right)
+
\nu |\Delta_\Omega|_1
\right\},
\end{align}
where $\nu>0$ is a tuning parameter. An efficient algorithm for solving \eqref{dtrace:2} was proposed in \citet{yuan2017differential}.

\begin{remark}
\citet{Zhao2014} proposed an alternative approach for estimating $\Delta_{\Omega}$ via the $l_1$-constrained minimization \citep{cai2011constrained}, when data are \textit{i.i.d.} within each group.
However, their approach is computationally less efficient than the approach based on the D-trace loss, especially for large $d$. As discussed in \citet{yuan2017differential},
the computational complexity and the memory requirement of the $l_1$-constrained minimization approach both scale as $O(d^4)$, whereas
the D-trace loss approach requires $O(d^3)$ computational complexity and $O(d^2)$ memory.
\end{remark}

Given $\widehat{\Sigma}_l$ and $\widehat{\Theta}_l$ in (\ref{est:1}) and $\widehat{\Delta}_\Omega(\nu)$ in (\ref{dtrace:2}), we next estimate $\Delta_A$ based on the estimating equation (\ref{050420:1}).
Let $\ve\beta_j$ denote the $j$th column of $\Delta_A$ and let ${\bf w}_j$ denote the $j$th column of
\[
\Sigma_1\Delta_\Omega\Theta_2
+
\Sigma_2\Delta_\Omega\Theta_1
+
2(\Theta_1-\Theta_2).
\]
Then (\ref{050420:1}) can be written as
\begin{align}
\label{050420:2}
(\Sigma_1+\Sigma_2)\ve\beta_j
-
{\bf w}_j
=
0,
\qquad
j=1,\ldots,d.
\end{align}

Equation (\ref{050420:2}) motivates the quadratic loss function
\[
L_A
\left(
\ve\beta_j;
\Sigma_1,
\Sigma_2,
\Delta_\Omega,
\Theta_1,
\Theta_2
\right)
=
\frac{1}{2}
\ve\beta_j^\top
(\Sigma_1+\Sigma_2)
\ve\beta_j
-
\ve\beta_j^\top
{\bf w}_j,
\]
which has a unique minimizer at $\ve\beta_j$. We therefore estimate $\ve\beta_j$ by
\begin{align}
\label{add:0514:01}
\widehat{\ve\beta}_j(\lambda_j)
=
\argmin_{\ve\beta_j}
\left\{
L_A
\left(
\ve\beta_j;
\widehat{\Sigma}_1,
\widehat{\Sigma}_2,
\widehat{\Delta}_\Omega,
\widehat{\Theta}_1,
\widehat{\Theta}_2
\right)
+
\lambda_j
\|\ve\beta_j\|_1
\right\},
\end{align}
where $\lambda_j>0$ is a tuning parameter. For $j=1,\ldots,d$, we solve (\ref{add:0514:01}) using the coordinate descent algorithm of \citet{wright2015coordinate} in parallel.
The direct estimator of $\Delta_A$ is then given by
\[
\widehat{\Delta}_A(\ve\lambda)
=
\left[
\widehat{\ve\beta}_1(\lambda_1)
\ \cdots \
\widehat{\ve\beta}_d(\lambda_d)
\right].
\]

An attractive feature of \eqref{add:0514:01} is that estimation of $\Delta_A$ decomposes into $d$ independent problems, one for each column of the DiffGC matrix. Consequently, the $d$ optimization problems can be solved in parallel, leading to substantial computational savings in high-dimensional settings. Moreover, the decomposition permits column-specific regularization parameters, allowing the amount of shrinkage to adapt to heterogeneous sparsity patterns across different columns of $\Delta_A$.

The numerical performance of $\widehat{\Delta}_A(\ve \lambda)$ and $\widehat{\Delta}_\Omega(\nu)$
depends on the choice of the tuning parameters. We select the tuning parameters  using an approximate Bayesian information criterion (aBIC), similar to the criterion in \citet{Zhao2014}. Specifically, letting $a_{n_1, n_2, p} = n_1 + n_2 - 2(p-1)$,
the optimal $\nu_{\text{opt}}$ is chosen to minimize
\begin{align*}%
   & a_{n_1, n_2, p} \left \| \frac{1}{2} \left( \widehat{\Sigma}_1 \widehat{\Delta}_\Omega(\nu) \widehat{\Sigma}_2  + \widehat{\Sigma}_2 \widehat{\Delta}_\Omega(\nu) \widehat{\Sigma}_1 \right)- \left(\widehat{\Sigma}_2 - \widehat{\Sigma}_1\right) \right \|_{\max}
   + \log\left(a_{n_1, n_2, p}\right)\left|\widehat{\Delta}_\Omega(\nu)\right|_0. %
\end{align*}
With $\nu_{\text{opt}}$, each $\lambda_j$ is determined by minimizing
\begin{align*}%
   a_{n_1, n_2, p} \left \| \left(\widehat{\Sigma}_1 + \widehat{\Sigma}_2 \right) \widehat{\ve \beta}_j(\lambda_j) - \widehat{\bf w}_j \right \|_ { \max } + \log\left(a_{n_1, n_2, p}\right)\left \|\widehat{\ve \beta}_j(\lambda_j)\right\|_0,
\end{align*}
where $\widehat{\bf w}_j$ is the $j$-th column of  $\widehat{\Sigma}_1\widehat{\Delta}_\Omega(\nu_{\text{opt}})\widehat{\Theta}_2 + \widehat{\Sigma}_2\widehat{\Delta}_\Omega(\nu_{\text{opt}})\widehat{\Theta}_1 + 2\left(\widehat{\Theta}_1 - \widehat{\Theta}_2\right)$.
\section{Theoretical properties of the direct estimator}\label{sec:thm:est}

\subsection{Convergence rates of $\widehat{\Delta}_\Omega(\nu)$}
In this subsection, we first establish nonasymptotic convergence rates for $\widehat{\Delta}_\Omega(\nu)$, which will serve as a basis for the analysis of $\widehat{\Delta}_A(\bm{\lambda})$.
While similar results for non-time-series data ({\it i.i.d.} data) exist in the literature, we shall establish a new convergence rate that is potentially faster than the existing ones with less stringent conditions.

We first introduce some additional notations. Let $S_\Omega$ denote the indices of the nonzero entries in $\Delta_\Omega$; that is,
$S_\Omega = \{(j,k): (\Delta_\Omega)_{jk} \neq 0\}$. Also, denote
by $s_\Omega$ the number of nonzero entries in $\Delta_\Omega$, i.e.,
$s_\Omega = |\Delta_\Omega|_0$.
For an arbitrary stationary stochastic process $\{{\bf z}_t\}_{t \in \mathbb{N}}$,
we define $\Gamma(h) = \mbox{cov}({\bf z}_t, {\bf z}_{t+h})$ for all $t, h \in \mathbb{N}$ as the lag-$h$ auto-covariance matrix of ${\bf z}_t$, with $\Gamma(-h) = \Gamma(h)^\top$.
The following condition
controls the stability of
$\{{\bf z}_t\}_{t \in \mathbb{N}}$ in terms of its spectral density.
\begin{assumption}
    The spectral density function
    \begin{align*}%
        f(\theta) = \frac{1}{2\pi}\sum_{h = -\infty}^{\infty} \Gamma(h)e^{-ih\theta}, ~\theta \in [-\pi, \pi]
    \end{align*}
    exists, and its maximum eigenvalue is bounded almost everywhere on $[-\pi, \pi]$; that is,
    \begin{align*}%
        \mathcal{M}(f) = \mbox{ess sup}_{\theta \in [-\pi, \pi]} \lambda_{\max}(f(\theta)) < \infty.
    \end{align*}
\end{assumption}
The spectral density is the analogue of the probability density function. Processes with larger $\mathcal{M}(f)$ are considered less stable. As discussed in \citet{douc2014nonlinear},
a large class of linear processes, including stable and invertible ARMA processes, satisfy Assumption 1.
An alternative assumption for controlling the stability is that singular values of the transition matrix of the VAR model have modulus less than 1 \citep{loh2012high, han2015}, referred to as the modulus assumption hereafter.
We adopt the spectral density assumption rather than the modulus assumption for two reasons.
First, the modulus assumption constrains each transition matrix. However, since we aim to directly estimate the difference of two transition matrices, constraints on individual transition matrices are not desirable.
Second, the spectral density assumption is less restrictive than the modulus assumption. This can be seen for a VAR(1) model with transition matrix $A$, where the modulus assumption reduces to $\|A\|_2 = c$ for some constant $c < 1$. On the one hand, $\|\Gamma(h)\|_2 \leq c^{|h|}\|\Gamma(0)\|_2$. This indicates that the spectral density function $f(\theta)$ exists, and $\mathcal{M}(f) \leq (1 - c)^{-1}\pi^{-1}\|\Gamma(0)\|_2 < \infty$. On the other hand, one can construct examples where Assumption 1 is satisfied but $\|A\|_2 > 1$; see Examples 1 and 2 in \citet{basu2015regularized}.

The following theorem establishes convergence rates for the lasso penalized D-trace estimator $\widehat{\Delta}_\Omega(\nu)$ in terms of the Frobenius and element-wise norms.
\begin{theorem}\label{thm1}
Suppose that ${\bf x}_t^{(l)}$ satisfies
Assumption 1 with the spectral density $f_l(\cdot)$ for $l = 1,2$. Denote $C_M = \max\left(\mathcal{M}(f_1), \mathcal{M}(f_2)\right)$ and $C_\Sigma = \max\left( \|\Sigma_1\|_{\max}, \|\Sigma_2\|_{\max} \right)$.
   Consider
\[
\nu =
192\sqrt{6}\pi C_M(1 + C_\Sigma |\Delta_\Omega|_1)\left(\frac{\log d}{\min(n_1, n_2) - (p-1)}\right)^{1/2}.
\]
If $d > 1$,
  $ \min(n_1, n_2) \geq
   C_G s_\Omega^2 \log d + p - 1$ with $C_G$
given in equation (\ref{CG}) in the Appendix, then with probability at least $1-16d^{-1}$, we have
\begin{align}
     \left\|\widehat{\Delta}_\Omega(\nu) - \Delta_\Omega\right\|_F ~&\leq
    12s_\Omega^{1/2}\lambda_{\min}^{-1}(\Sigma_1)\lambda_{\min}^{-1}(\Sigma_2)\nu,
    \label{thm1:eq1} \\
      \left\|\widehat{\Delta}_\Omega(\nu) - \Delta_\Omega\right\|_{\max} ~&\leq 12s_\Omega^{1/2}\lambda_{\min}^{-1}(\Sigma_1)\lambda_{\min}^{-1}(\Sigma_2)\nu
      \label{thm1:eq2}.
\end{align}
\end{theorem}

Theorem~\ref{thm1} establishes estimation consistency of $\widehat{\Delta}_\Omega(\nu)$, but does not guarantee variable selection consistency.
In the following, we provide a refinement of Theorem \ref{thm1} that can lead to consistent variable selection.
More specifically, for $\tau > 0$, we define the hard thresholding function $\mbox{HT}_\tau(t) = tI(|t| > \tau)$, and let $\mbox{sgn}(t)$ denote the sign function. For any matrix $M$, denote $\mbox{HT}_\tau(M) = \left( \mbox{HT}_\tau(M_{jk}) \right)_{jk}$ and $\mbox{sgn}(M) = \left( \mbox{sgn}(M_{jk}) \right)_{jk}$.
The following result characterizes the sign consistency of $\mbox{HT}_{\tau_\Omega}\left(\widehat{\Delta}_\Omega(\nu)\right)$ for an appropriately chosen $\tau_\Omega$, when the non-zero entries in $\Delta_\Omega$ are sufficiently large. %
\begin{theorem}\label{thm2}
Suppose that ${\bf x}_t^{(l)}$ satisfies
Assumption 1 with the spectral density $f_l(\cdot)$ for $l = 1,2$.
Let $n_1, n_2, d$ and $\nu$ satisfy the same conditions as in Theorem \ref{thm1}.
If
    \[
    \tau_{\Omega} \geq 12s_\Omega^{1/2}\lambda_{\min}^{-1}(\Sigma_1)\lambda_{\min}^{-1}(\Sigma_2)\nu,
    \]
    and $\min_{(j,k): (\Delta_\Omega)_{jk} \neq 0} \left|(\Delta_\Omega)_{jk}\right| > 2\tau_\Omega$, then with probability at least $1 - 16d^{-1}$,
    \[
    \mbox{sgn}\left(\mbox{HT}_{\tau_\Omega}\left(\widehat{\Delta}_\Omega(\nu)\right)\right) = \mbox{sgn}(\Delta_\Omega).
    \]
\end{theorem}

We note that, when the data within each group are {\it i.i.d.}, Theorem~\ref{thm1} requires less stringent assumptions than those in existing theoretical results for DiffNet estimation,
such as the mutual incoherence (MI) condition in \citet{Zhao2014}, and the irrepresentability (IR) condition in \citet{yuan2017differential}. Both conditions are related to the sparsity $s_\Omega$, dimension $d$, and maximum correlation in $\Sigma_l$ for $l = 1,2$, and can be restrictive in practice, especially for large $d$ and highly correlated variables.

We first show that the MI condition seldom holds if some variables are strongly correlated. Let $\sigma_{l,jk}$ be the $(j,k)$-th entry of $\Sigma_l$, and denote $\sigma_{l,\max} = \max_{j}\sigma_{l,jj}$, $\sigma_{l,\min} = \min_{j}\sigma_{l,jj}$ and
$\mu_l = \max_{j \neq k}|\sigma_{l,jk}|$ for $l = 1,2$.
Then, the MI condition is
\begin{align}\label{MI}
\mu = 4\max(\mu_1\sigma_{2,\max}, \mu_2\sigma_{1,\max}) \leq \sigma^S_{\min} (2s_\Omega')^{-1},
\end{align}
where $\sigma^S_{\min} = \min _ { j , k } ( \sigma _ {2,j j } \sigma _ {1,j j }, \sigma _ {2, k k } \sigma _ {1, j j }  + 2 \sigma _ {2, k j } \sigma _ {1, j k } + \sigma _ {2, j j } \sigma _ {1, k k } )$, and $s_\Omega'$ is the number of nonzero entries in the upper triangular part of the true $\Delta_\Omega$. To gain more intuition about this condition, we consider a special example where all the variables have been standardized; that is, $\sigma_{l,jj} = 1$ for $l = 1, 2$ and $j = 1, \ldots, d$. In this case, $\sigma_{\min}^S \leq 1$, and (\ref{MI}) reduces to $\max(\mu_1 , \mu_2) \leq (8s_\Omega')^{-1}$. This implies that the maximum correlation among the variables cannot exceed $1/8$ because $s_\Omega' \geq 1$; the case of $s_\Omega' = 0$ is trivial.
Clearly, this constraint on between-variable correlations is unrealistic in many applications.

The IR condition in \citet{yuan2017differential} requires that
\begin{align}\label{IR}
   \max_{e \in S_\Omega^c}\left\|\Gamma_{e,S_\Omega}(\Gamma_{S_\Omega, S_\Omega})^{-1} \right\|_1 < 1,
\end{align}
where $\Gamma$ denotes the Hessian matrix in (\ref{hessian}), and $S_\Omega$ is the support of $\Delta_\Omega$.
We examine how often (\ref{IR}) is satisfied using a simulation study. Specifically, we generated $\Sigma_1^{-1} \in \mathbb{R}^{d \times d}$
from Erd\H{o}s-R\"{e}nyi graphs with $d$ nodes and $d(d-1)/5$ nonzero off-diagonal entries for $d = 10, 20, 30, 40,$ and 50.
The values in the off-diagonal nonzero entries were generated from a uniform distribution with support $[-1, -0.5] \bigcup [0.5, 1]$, and the diagonal entries were set to be $10$.
Then, $\Sigma_2^{-1}$ was generated by changing the sign of $s_\Omega$ nonzero entries of $\Sigma_1^{-1}$. Let $S_\Omega$ denote the indices of the $s_\Omega$ entries, that is,
$S_\Omega = \{(j,k): (\Sigma_2^{-1})_{jk} \neq (\Sigma_1^{-1})_{jk}\}$.
The matrices $\Sigma_1$ and $\Sigma_2$ were obtained by inverting $\Sigma_1^{-1}$ and $\Sigma_2^{-1}$, respectively. The maximum correlation in the covariance matrix $\Sigma_l$ is around $0.1$ for $l = 1,2$, which represents a setting with weakly correlated variables.
We then calculated $\Gamma  = 0.5(\Sigma_1 \otimes \Sigma_2 + \Sigma_2 \otimes \Sigma_1)$ and examined (\ref{IR}). %
Table~\ref{irres} shows how often (\ref{IR}) holds with various $d$ and $s_\Omega$. For small $d$, e.g., $d = 10$, (\ref{IR}) holds with high probability. However, as $d$ grows, the percentage of cases where (\ref{IR}) holds drops quickly. Particularly, when $d \geq 30$, even for the very sparse case $s_\Omega = 6$, (\ref{IR}) never holds. These results indicate that the IR condition in (\ref{IR}) is not realistic even with moderate $d$, small $s_\Omega$, and weakly correlated variables.
\begin{table}[!ht]
\centering
\caption{Percentages of simulated $\Gamma$ in $1000$ simulations that meet the IR condition (\%)}
\label{irres}
\begin{tabular}{llllll}
  $d$     & 10 & 20 & 30 & 40 & 50  \\
$s_\Omega = 6$ &   100     &  12.9      &   0     &    0     & 0     \\
$s_\Omega = 12$ &   100     &   4.9     &   0    &   0     &   0   \\
$s_\Omega = 18$ &   100     &   3.1     &   0     &   0      &   0
\end{tabular}
\end{table}

Theorem \ref{thm1} relaxes these stringent conditions because when data in each group are {\it i.i.d.}, our Assumption~1 becomes trivial.
Despite requiring less restrictive conditions, our convergence rate in Theorem~\ref{thm1} may be faster than that in \citet{yuan2017differential}. Take the bound of $\left\|\widehat{\Delta}_\Omega(\nu) - \Delta_\Omega\right\|_{\max}$ as an example.
\citet{yuan2017differential} show that for a suitable $\nu$, $\|\widehat{\Delta}_{\Omega}(\nu) - \Delta_\Omega\|_{\max} \leq M_Y (\eta\log d + \log 4)^{1/2} \{\min(n_1, n_2) \}^{-1/2}$ for some constants $M_Y > 0$ and $\eta > 2$. By carefully examining their proof, we find that $M_Y = O(s_\Omega^3)$.
In comparison, Theorem~\ref{thm1} shows that $\left\|\widehat{\Delta}_\Omega(\nu) - \Delta_\Omega\right\|_{\max} \leq
M_W (\log d)^{1/2}  \{\min(n_1, n_2)\}^{-1/2}$, where $M_W = O\left( |\Delta_\Omega|_1 s_\Omega^{1/2} \lambda_{\min}^{-1}(\Sigma_1) \lambda_{\min}^{-1}(\Sigma_2) \right)$ and $|\Delta_\Omega|_1 \leq s_\Omega\|\Delta_\Omega\|_{\max}$. Hence, if $\|\Delta_\Omega\|_{\max} = O(1)$ and $\lambda_{\min}^{-1}(\Sigma_1)\lambda_{\min}^{-1}(\Sigma_2) = o(s_\Omega^{3/2})$,
then $M_W = o(s_\Omega^3)$, indicating that our convergence rate is potentially faster than that in \citet{yuan2017differential}.

\subsection{Convergence rates of $\widehat{\Delta}_A(\ve \lambda)$}
The theoretical analysis of $\widehat{\Delta}_A(\ve \lambda)$ follows the same general framework as that for
\(\widehat{\Delta}_\Omega(\nu)\),
but requires additional control of the estimation error arising from the plug-in estimator
\(\widehat{\Delta}_\Omega(\nu)\).
For \(j=1,\ldots,d\), let
\(\ve\beta_j\)
denote the \(j\)-th column of
\(\Delta_A\),
and let
\(S_{A,j}\)
and
\(s_{A,j}=|S_{A,j}|\)
denote its support and sparsity level, respectively.
Also let $S_A = \{(i,j): (\Delta_A)_{ij}\neq 0\}
= \bigcup_{j=1}^d\{(i,j): i\in S_{A,j}\}$
denote the global support of $\Delta_A$.

The next theorem establishes error bounds for
\(\widehat{\Delta}_A(\ve\lambda)\)
under the Frobenius and element-wise norms.
\begin{theorem}
\label{thm3}
Suppose that
\({\bf x}_t^{(l)}\)
satisfies Assumption 1 with spectral density function
\(f_l(\cdot)\)
for
\(l=1,2\).
Consider the tuning parameter
\(\nu\)
defined in Theorem \ref{thm1}, and let
\[
\lambda_j
= \bar{\lambda} :=
2C_1
\left(
\frac{\log d}
{\min(n_1,n_2)-p}
\right)^{1/2},
\qquad
j=1,\ldots,d,
\]
where \(C_1\) is a positive constant whose explicit form is provided in the Appendix.
If \(n_1\) and \(n_2\) are sufficiently large,
then, with probability at least
\(1-32d^{-1}\),
\begin{align}
\left\|
\widehat{\Delta}_A(\ve\lambda)
-
\Delta_A
\right\|_F
\le
&
\frac{
12 \bar{\lambda}
}{
\lambda_{\min}(\Sigma_1)
+
\lambda_{\min}(\Sigma_2)
}
\left(
\sum_{l=1}^{d}
s_{A,l}
\right)^{1/2},
\label{thm3:eq1}
\\
\left\|
\widehat{\Delta}_A(\ve\lambda)
-
\Delta_A
\right\|_{\max}
\le
&
\frac{
12 \bar{\lambda}
}{
\lambda_{\min}(\Sigma_1)
+
\lambda_{\min}(\Sigma_2)
}
\left(\max_{1\le l\le d}
s_{A,l}\right)^{1/2}.
\label{thm3:eq2}
\end{align}
\end{theorem}

Theorem \ref{thm3} establishes nonasymptotic error bounds for the proposed DiffGC estimator under both the Frobenius and element-wise norms. The explicit sample size requirements ensuring the validity of Theorem \ref{thm3} are provided in Section \ref{sec:appenC} of the Appendix.

An interesting feature of Theorem \ref{thm3} is that the element-wise error bound in \eqref{thm3:eq2} depends only on
\(\max_{1\le l\le d} s_{A,l} \),
whereas the Frobenius-norm bound in \eqref{thm3:eq1} depends on
\(\sum_{l=1}^{d} s_{A,l}\).
Consequently, the element-wise convergence rate can be substantially faster when the nonzero entries of
\(\Delta_A\)
are distributed across many sparse columns. This phenomenon reflects the column-wise structure of the proposed estimator, which estimates the columns of
\(\Delta_A\)
through \(d\) parallel optimization problems; see the proof of Theorem \ref{thm3} in the Appendix for additional details.

Similar to Theorem \ref{thm2}, the next result establishes sign consistency of the thresholded estimator
\(\mathrm{HT}_{\tau_A}\left(\widehat{\Delta}_A(\ve\lambda)\right)\)
under a beta-min condition on the nonzero entries of
\(\Delta_A\).

\begin{theorem}
\label{thm4}
Suppose all assumptions in Theorem \ref{thm3} hold and
let
$\bar{\lambda}$
be chosen as in Theorem \ref{thm3}.
If
\[
\tau_A
\ge
12 \bar{\lambda}
\left\{
\lambda_{\min}(\Sigma_1)
+
\lambda_{\min}(\Sigma_2)
\right\}^{-1}
\max_{1\le l\le d}
\sqrt{s_{A,l}},
\]
and
\[
\min_{(j,k):(\Delta_A)_{jk}\neq 0}
|(\Delta_A)_{jk}|
>
2\tau_A,
\]
then, with probability at least
\(1-32d^{-1}\),
\[
\operatorname{sgn}
\left(
\mathrm{HT}_{\tau_A}
\left(
\widehat{\Delta}_A(\ve\lambda)
\right)
\right)
=
\operatorname{sgn}
(\Delta_A).
\]
\end{theorem}

\section{Direct inference}\label{sec:DiffGC:inf}

In this section, we develop a selective inference framework for the DiffGC matrix \(\Delta_A\), which characterizes changes in Granger causal relationships between two high-dimensional VAR processes.
Specifically, our goal is to provide valid statistical inference for selected differential effects while accounting for the data-dependent model selection step. A particular challenge is that the differential parameters are not directly identifiable regression coefficients but are instead implicitly defined through moment equations involving the covariance and precision matrices of the two VAR models. We therefore begin by carefully defining the post-selection inferential targets before establishing their asymptotic distributions. As a byproduct, the proposed framework also yields valid asymptotic inference for the DiffNet parameter \(\Delta_\Omega\), thereby complementing the growing literature on DiffNet estimation with formal uncertainty quantification.

\subsection{Selective inference with sample splitting} \label{subsec:DiffGC:inf:selinf}

For the $l$-th time series \({{\bf x}_{t}^{(l)}}_{t = 1,\ldots,n_l}\), \(l=1,2\), we split the observations into three parts,
\[
{{\bf x}_{t}^{(l)}}_{t = 1,\ldots,n_l^{(1)}}, \quad
{{\bf x}_{t}^{(l)}}_{t = n_l^{(1)}+1,\ldots,n_l^{(2)}}, \quad
{{\bf x}_{t}^{(l)}}_{t = n_l^{(2)}+1,\ldots,n_l}.
\]
We apply the proposed direct estimation procedure to the first segment of each time series and obtain estimators \(\widehat{\Delta}_A\) and \(\widehat{\Delta}_\Omega\), together with their estimated supports \(\mathcal{M}_A\) and \(\mathcal{M}_\Omega\). The third segment is reserved exclusively for inference.

Let
$g_l=n_l^{(2)}-n_l^{(1)}$ denote the length of the middle segment. We assume that each time series is $\beta$-mixing with mixing coefficient $\beta_l(\cdot)$ and that $g_l$ lengths satisfy
$\beta_l(g_l)\to0$. Under this condition, the selection and inference samples are asymptotically independent. As formalized in the Appendix through a coupling argument, the inference sample can be replaced by an independent copy with the same marginal distribution at an error of order $\beta_l(g_l)$.

In classical selective inference for linear models, the target is typically defined as the subset of regression coefficients indexed by the selected variables. Such a construction is natural because the regression coefficients are directly identifiable from the underlying model.
In the present setting, however, the DiffGC matrix $\Delta_A$ and the DiffNet matrix $\Delta_\Omega$ are not directly identifiable from the observed data.
Rather, they are implicitly defined through a collection of moment equations involving the covariance and precision matrices of the two VAR processes.
Consequently, after selecting supports \(\mathcal{M}_A\) and \(\mathcal{M}_\Omega\), it is generally inappropriate to define the inferential targets simply as \( (\Delta_A)_{\mathcal{M}_A}\) and \( (\Delta_\Omega)_{\mathcal{M}_\Omega}\).
Instead, we define lower-dimensional parameters that are identified by restricting the underlying moment equations to the selected coordinates.

For ease of presentation, define
\begin{equation}\label{VA:VOmega}
V_A=\{(k-1)d+j:(j,k)\in\mathcal{M}_A\},
\qquad
V_\Omega=\{(k-1)d+j:(j,k)\in\mathcal{M}_\Omega\}.
\end{equation}

Recalling \eqref{diffest:add:1} and \eqref{050420:1},
\begin{align*}
\Sigma_2 \Delta_\Omega \Sigma_1
&=
\Sigma_2-\Sigma_1,\\ %
(\Sigma_1+\Sigma_2)\Delta_A
&=
\Sigma_1\Delta_\Omega\Theta_2
+
\Sigma_2\Delta_\Omega\Theta_1
+
2(\Theta_1-\Theta_2).
\end{align*}

After vectorization,
\begin{align}
(\Sigma_1\otimes\Sigma_2)\mathrm{vec}(\Delta_\Omega)
&~ =
\mathrm{vec}(\Sigma_2-\Sigma_1),
\nonumber\\
{I_d\otimes(\Sigma_1+\Sigma_2)}\mathrm{vec}(\Delta_A)
&~ =
(\Theta_1^\top\otimes\Sigma_2)\mathrm{vec}(\Delta_\Omega)
+
(\Theta_2^\top\otimes\Sigma_1)\mathrm{vec}(\Delta_\Omega)
+
2\mathrm{vec}(\Theta_1-\Theta_2).
\label{selinf:add:1}
\end{align}

Motivated by these equations, we define the selective inferential targets as the solutions to the corresponding restricted systems:
\begin{align}
\psi_{\Omega}
&~ =
[(\Sigma_1\otimes\Sigma_2)_{(V_\Omega,V_\Omega)}]^{-1}
{\mathrm{vec}(\Sigma_2-\Sigma_1)}_{V_\Omega},
\nonumber\\
\psi_A
&~ =
\left[{I_d\otimes(\Sigma_1+\Sigma_2)}_{(V_A,V_A)}\right]^{-1}
\Big[
(\Theta_1^\top\otimes\Sigma_2+\Theta_2^\top\otimes\Sigma_1)_{(V_A,V_\Omega)}
\psi_\Omega
+
2{\mathrm{vec}(\Theta_1-\Theta_2)}_{V_A}
\Big].
\label{selinf:add:2}
\end{align}

\begin{lemma}\label{lem:target:equivalence}
Suppose that the selected supports contain the true supports, i.e.,
$\mathcal{M}_A\supseteq S_A$ and $\mathcal{M}_\Omega\supseteq S_\Omega$.
Then
\[
\psi_A
=
\{\operatorname{vec}(\Delta_A)\}_{V_A},
\qquad
\psi_\Omega
=
\{\operatorname{vec}(\Delta_\Omega)\}_{V_\Omega},
\]
where $V_A$ and $V_\Omega$ are defined in \eqref{VA:VOmega}.
\end{lemma}

Lemma~\ref{lem:target:equivalence} shows that the proposed inferential
targets have a direct interpretation whenever the selected models contain
the true supports: they coincide exactly with the corresponding entries of
the underlying differential parameters. The support recovery results in
Section~\ref{sec:thm:est} imply that this event occurs with probability
tending to one under the stated regularity conditions. When the selected
supports fail to contain the true supports, $\psi_A$ and $\psi_\Omega$
remain well-defined as the population parameters identified by the
restricted moment equations, although they need not coincide with the
corresponding entries of $\Delta_A$ and $\Delta_\Omega$.

An additional advantage of defining the inferential targets through the restricted systems in \eqref{selinf:add:2} is that the resulting parameters are intrinsically low-dimensional. In particular, the matrices requiring inversion involve only the selected coordinates \(V_A\) and \(V_\Omega\), and therefore have dimensions \(|V_A|\times |V_A|\) and \(|V_\Omega|\times |V_\Omega|\), respectively. This should be contrasted with the full moment equations in \eqref{selinf:add:1}, whose coefficient matrices have dimensions \(d^2\times d^2\).
Consequently, the support reduction induced by the selection step not only determines the inferential targets but also renders their estimation feasible in finite samples.
Specifically, we define the plug-in estimators
\begin{align}
\widehat{\psi}_{\Omega}
&~ =
\left[\left(\widehat{\Sigma}_1\otimes\widehat{\Sigma}_2\right)_{(V_\Omega,V_\Omega)}\right]^{-1}
{\mathrm{vec}\left(\widehat{\Sigma}_2-\widehat{\Sigma}_1\right)}_{V_\Omega},
\nonumber\\
\widehat{\psi}_A
&~ =
\left[{I_d\otimes\left(\widehat{\Sigma}_1+\widehat{\Sigma}_2\right)}_{(V_A,V_A)}\right]^{-1}
\Big[
\left(\widehat{\Theta}_1^\top\otimes\widehat{\Sigma}_2+\widehat{\Theta}_2^\top\otimes\widehat{\Sigma}_1\right)_{(V_A,V_\Omega)}
\widehat{\psi}_\Omega
+
2{\mathrm{vec}\left(\widehat{\Theta}_1-\widehat{\Theta}_2\right)}_{V_A}
\Big],
\label{selinf:add:3}
\end{align}
where $\widehat{\Sigma}_l, \widehat{\Theta}_l$ for $l = 1, 2$ are given in \eqref{est:1}.
To preserve the asymptotic independence between the selection and inference stages, all estimates in \eqref{selinf:add:3} are computed exclusively from the inference samples \(\{{\bf x}_t^{(l)}\}_{t=n_l^{(2)}+1,\ldots,n_l}\) for \(l=1,2\).
In the following discussions, we establish the asymptotic distribution of $\widehat{\psi}_A$ and $\widehat{\psi}_\Omega$, and use these results to construct hypothesis tests for the selected differential effects.

\subsection{Asymptotic distributions of $\widehat{\psi}_{\Omega}$ and $\widehat{\psi}_A$}

Let
\(
n_l^{\rm inf}=n_l-n_l^{(2)}
\)
denote the size of the inference sample for the $l$th process for $l=1,2$.
For simplicity of presentation, we consider a sample-splitting scheme satisfying
\(
n_1^{\rm inf}=n_2^{\rm inf}\equiv n^{\rm inf},
\)
which can always be achieved by allocating the same number of observations from each process to the inference sample. We introduce three additional assumptions.

\begin{assumption}[Bounded eigenvalues]\label{assump:eigen}
 For $l = 1, 2$, there exist constant $C, c > 0$ such that $c < \lambda_{\min}(\Sigma_l) < \lambda_{\max}(\Sigma_l) < C$.
\end{assumption}

\begin{assumption}[Support size condition for $\psi_\Omega$]\label{assump:supp:omega}
Recall $m_\Omega = |\mathcal{M}_\Omega|$.
As $n^{\rm inf} \rightarrow \infty$,
    \[
    \frac{m_\Omega^5 (\log d)^4}{n^{\rm inf}} \rightarrow 0.
    \]
\end{assumption}

\begin{assumption}[Support size condition for $\psi_A$]\label{assump:supp:A}
Let $m_A = |\mathcal{M}_A|$.
As $n^{\rm inf} \rightarrow \infty$,
    \[
    \frac{m_A^5 (\log d)^4}{n^{\rm inf}} \rightarrow 0.
    \]
\end{assumption}

Assumption 2 is a standard regularity condition ensuring that the covariance matrices remain well-conditioned and that the restricted systems defining the inferential targets are stable. Assumptions 3 and 4 impose sparsity requirements on the selected supports.
 These conditions reflect the fact that inference is conducted only on the selected coordinates.
  Moreover, in time-series applications, the sample size is determined by the number of observed time points, which is often considerably larger than the dimensionality of the selected supports. Therefore, Assumptions 3 and 4 are generally reasonable in practice.

Let
\begin{align}\label{phi-mu}
\Phi=(\Sigma_1\otimes\Sigma_2)_{(V_\Omega,V_\Omega)}
\quad\text{and}\quad
\ve\mu=\{\operatorname{vec}(\Sigma_2-\Sigma_1)\}_{V_\Omega}. %
\end{align}
Our first result establishes an asymptotically linear representation of
$\widehat{\psi}_\Omega$.
Such a representation plays a central role in the subsequent analysis,
as it characterizes the first-order behavior of the estimator through
the sample covariance matrices and serves as the foundation for deriving
its asymptotic distribution.

\begin{theorem}\label{inference:thm1:101724}
Suppose Assumptions 1--3 are satisfied. As $n^{\rm inf}\to\infty$, conditioning on $\mathcal{M}_\Omega$,
\begin{align}
\Biggl\|
&\sqrt{n^{\rm inf}}\left(\widehat{\psi}_\Omega-\psi_\Omega\right)
-
\Phi^{-1}
\left\{
\sqrt{n^{\rm inf}}
\operatorname{vec}\left(\widehat{\Sigma}_2-\Sigma_2\right)
-
\sqrt{n^{\rm inf}}
\operatorname{vec}\left(\widehat{\Sigma}_1-\Sigma_1\right)
\right\}_{V_\Omega}
\nonumber\\
&+
(\ve\mu^\top\otimes I_{m_\Omega})
(\Phi^{-1}\otimes\Phi^{-1})
\operatorname{vec}
\biggl[
\biggl\{
\Sigma_1
\otimes
\sqrt{n^{\rm inf}}
\left(\widehat{\Sigma}_2-\Sigma_2\right)
\nonumber\\
&\qquad+
\sqrt{n^{\rm inf}}
\left(\widehat{\Sigma}_1-\Sigma_1\right) \otimes \Sigma_2
\biggr\}_{(V_\Omega,V_\Omega)}
\biggr]
\Biggr\|
=o_p(1).
\label{inference:thm1:eq1}
\end{align}
where $I_{m_\Omega}$ denotes the $m_\Omega\times m_\Omega$ identity matrix, and
$\Phi$ and $\bm{\mu}$ are defined in \eqref{phi-mu}.
\end{theorem}
Theorem \ref{inference:thm1:101724} establishes an asymptotically linear representation of \(\widehat{\psi}_\Omega\). In particular, the first-order behavior of \(\widehat{\psi}_\Omega\) is determined by a linear transformation of the sample covariance estimation errors from the two VAR processes. This representation serves as the key ingredient for deriving the asymptotic distribution of \(\widehat{\psi}_\Omega\).

For notational convenience, write
\[
V_\Omega=\{l_1,l_2,\ldots,l_{m_\Omega}\},
\qquad
l_1<l_2<\cdots<l_{m_\Omega},
\]
and let ${\bf e}_{j,m}$ denote the $m$-dimensional vector whose $j$-th entry is one and all remaining entries are zero. Define
\(
\Pi_{V_\Omega,m}
=
\left[
{\bf e}_{l_1,m},
\ldots,
{\bf e}_{l_{m_\Omega},m}
\right].
\)
For $l=1,2$, let
\(
D_l
=
\mathrm{avar}
\left\{
\sqrt{n^{\rm inf}}
\,\mathrm{vec}
\left(\widehat{\Sigma}_l\right)
\right\}
\)
denote the long-run covariance matrix of the sample covariance estimator. Specifically, for
$j_1,j_2,k_1,k_2=1,\ldots,d$, \cite{roy1989asymptotic} establishes that
the
$\big((j_1-1)d+k_1,(j_2-1)d+k_2\big)$-th entry of $D_l$ is
\begin{align}
\label{Di:entry}
\sum_{u=-\infty}^{\infty}
\Big\{
\sigma_{l,j_1j_2}(u)
\sigma_{l,k_1k_2}(u)
+
\sigma_{l,k_1j_2}(u)
\sigma_{l,j_1k_2}(u)
\Big\},
\end{align}
where $\sigma_{l,j_1j_2}(u)$ denotes the $(j_1,j_2)$-th entry of the lag-$u$ covariance matrix
\(
\mathrm{cov}
\left(
{\bf x}_t^{(l)},
{\bf x}_{t+u}^{(l)}
\right).
\)

Next, let
\begin{align}\label{SBi}
S_{B} = \left\{ (j-1)d^2 + d\left\lfloor \frac{j-1}{d} \right \rfloor + k \mbox{ with } j = 1, \ldots, d^2, k = 1, \ldots, d  \right\}.
\end{align}
Furthermore, let \(\mathcal F_{1j}\) and \(\mathcal F_{2j}\), respectively,
denote the submatrix of
\begin{align}\label{f_ij}
\left\{
\left(\Pi_{V_\Omega,d^2}^{\top} K_{d,d} \right)
\otimes
\left(
\Pi_{V_\Omega,d^2}^{\top} K_{d,d}
(\Sigma_2\otimes I_d)
\right)
\right\}_{(*, S_B)}
\mbox{ and }
\left\{
\Pi_{V_\Omega,d^2}^{\top}
\otimes
\left(
\Pi_{V_\Omega,d^2}^{\top}
(\Sigma_1\otimes I_d)
\right)
\right\}_{(*,S_B)}
\end{align}
with columns indexed by
\(
\mathcal{E}_j =
\left\{
(j-1)d^2+k:
k=1,\ldots,d^2
\right\},
\)
where $K_{d,d}$ is the $d^2 \times d^2$ commutation matrix.

With these new notations, we can write
\begin{align}
\label{add:101724:5:main}
\mathrm{vec}
\left[
\left\{
\Sigma_1
\otimes
\sqrt{n^{\rm inf}}
\left(\widehat{\Sigma}_2-\Sigma_2\right)
\right\}_{(V_\Omega,V_\Omega)}
\right]
= &~
\left(
\sum_{j=1}^{d}
\mathcal F_{2j}
\right)
\mathrm{vec}
\left[
\sqrt{n^{\rm inf}}
\left(\widehat{\Sigma}_2-\Sigma_2\right)
\right] \nonumber \\
\mathrm{vec}
\left[
\left\{
\sqrt{n^{\rm inf}}
\left(\widehat{\Sigma}_1-\Sigma_1\right) \otimes \Sigma_2
\right\}_{(V_\Omega,V_\Omega)}
\right] = &~
\left(
\sum_{j=1}^{d}
\mathcal F_{1j}
\right)
\mathrm{vec}
\left[
\sqrt{n^{\rm inf}}
\left(\widehat{\Sigma}_1-\Sigma_1\right)
\right].
\end{align}

The following theorem establishes the asymptotic distribution of
$\widehat{\psi}_\Omega$.

\begin{theorem}
\label{asymp:delta:omega:sel:inf}
Suppose all assumptions in Theorem
\ref{inference:thm1:101724}
hold. Then, as $n^{\rm inf} \rightarrow \infty$,  conditional on $\mathcal{M}_\Omega$,
\[
\sqrt{n^{\rm inf}}
\left(
\widehat{\psi}_\Omega
-
\psi_\Omega
\right)
\overset{d}{\longrightarrow}
N_{m_\Omega}
\left(
0,
\sum_{l=1}^{2}
G_l
D_l
G_l^{\top}
\right).
\]
Here,
\begin{align*}
G_1
= &~
(\ve\mu^\top\otimes I_{m_\Omega})
(\Phi^{-1}\otimes\Phi^{-1})
\left(
\sum_{j=1}^{d}
\mathcal F_{1j}
\right)
+
\Phi^{-1}
\Pi_{V_\Omega,d^2}^{\top} \\
G_2
= &~
(\ve\mu^\top\otimes I_{m_\Omega})
(\Phi^{-1}\otimes\Phi^{-1})
\left(
\sum_{j=1}^{d}
\mathcal F_{2j}
\right)
-
\Phi^{-1}
\Pi_{V_\Omega,d^2}^{\top},
\end{align*}
where $\bm{\mu}$ and $\Phi$ are defined in \eqref{phi-mu}.

\end{theorem}

Theorem \ref{asymp:delta:omega:sel:inf} establishes the asymptotic normality of
$\widehat{\psi}_\Omega$ conditional on the selected support
$\mathcal{M}_\Omega$.
The asymptotic covariance matrix takes the sandwich form
\(
\sum_{l=1}^{2}
G_l D_l G_l^{\top},
\)
where $D_1$ and $D_2$ characterize the long-run covariance structure of the two VAR processes and $G_1$ and $G_2$ describe how covariance estimation errors propagate through the linear representation established in Theorem \ref{inference:thm1:101724}. Consequently, Theorem \ref{asymp:delta:omega:sel:inf} provides the theoretical foundation for performing hypothesis tests for the selected DiffNet effects.

The analysis of $\widehat{\psi}_A$ parallels that of
$\widehat{\psi}_\Omega$. However, because $\psi_A$ depends on
$\psi_\Omega$ as well as $\Sigma_1$, $\Sigma_2$, $\Theta_1$, and
$\Theta_2$, its asymptotic behavior involves additional sources of
uncertainty. To simplify the presentation,
define
\(
\Upsilon
=
\{I_d\otimes(\Sigma_1+\Sigma_2)\}_{(V_A,V_A)}.
 \)
Let
$\mathcal{L}_j$ be the submatrix of  %
\(
\left(\Pi_{V_A, d^2}^\top \otimes \Pi_{V_A, d^2}^\top \right)_{(*, S_B)}
\)
with columns indexed by the set $\mathcal{E}_j$.
Furthermore, for $i=1,2$ and $j=1,\ldots,d$, define $\mathcal{H}_{ij}$ to be the submatrix of
\(
  \left\{
    \Pi_{V_\Omega,d^2}^{\top}
    \otimes
    \left(
      \Pi_{V_A,d^2}^{\top}
      \left(\Theta_i^{\top} \otimes I_d\right)
    \right)
  \right\}_{(*,\,S_B)}
\)
with columns indexed by $\mathcal{E}_j$,
so that for $i, l = 1, 2$,
\[
\operatorname{vec}\left[\left\{\Theta_i^{\top}
\otimes \sqrt{n^{\mathrm{inf}}}\left(\widehat\Sigma_l-\Sigma_l\right)\right\}_{(V_A,V_\Omega)}\right]
= \left(\sum_{j=1}^d \mathcal{H}_{ij}\right)
\operatorname{vec}\left[\sqrt{n^{\mathrm{inf}}}\left(\widehat\Sigma_l-\Sigma_l\right)\right].
\]
Similarly, define $\mathcal{G}_{ij}$ to be the submatrix of
\(
  \left\{
    \left(\Pi_{V_\Omega,d^2}^{\top} K_{d,d}\right)
    \otimes
    \left(
      \Pi_{V_A,d^2}^{\top}K_{d,d}
      \left(\Sigma_i \otimes I_d \right)
    \right)
  \right\}_{(*,\,S_B)}
\)
with columns indexed by $\mathcal{E}_j$, right-multiplied by $K_{d,d}$.
With that, for $i, l=1,2$, we can write
\[
\operatorname{vec} \left[\left\{\sqrt{n^{\mathrm{inf}}}\left(\widehat\Theta_l-\Theta_l\right)^{\top}
\otimes \Sigma_i\right\}_{(V_A,V_\Omega)}\right]
= \left(\sum_{j=1}^d \mathcal{G}_{ij}\right)
\operatorname{vec} \left[\sqrt{n^{\mathrm{inf}}}\left(\widehat\Theta_l-\Theta_l\right)\right].
\]

With all above notations, we then define
\begin{align*}
M_1
=&~
\psi_\Omega^\top
\otimes
\Upsilon^{-1},
\\
M_2
=&~
\Upsilon^{-1}
(\Theta_1^\top \otimes \Sigma_2
+
\Theta_2^\top \otimes \Sigma_1)_{(V_A,V_\Omega)}
\Phi^{-1}
\Pi_{V_\Omega,d^2}^\top,
\\
M_3
=&~
\Upsilon^{-1}
(\Theta_1^\top \otimes \Sigma_2
+
\Theta_2^\top \otimes \Sigma_1)_{(V_A,V_\Omega)}
(\ve\mu^\top \otimes I_{m_\Omega})
(\Phi^{-1}\otimes\Phi^{-1}),
\\
M_4
=&~
\left\{
\left[
(\Theta_1^\top \otimes \Sigma_2
+
\Theta_2^\top \otimes \Sigma_1)_{(V_A,V_\Omega)}
\psi_\Omega
+
2\{\mathrm{vec}(\Theta_1-\Theta_2)\}_{V_A}
\right]^\top
\otimes
I_{m_A}
\right\}
(\Upsilon^{-1}\otimes\Upsilon^{-1}),
\\
A_{\Sigma,1}
=&~
M_1
\sum_{j=1}^{d}
\mathcal H_{2j}
-
M_2
-
M_3
\sum_{j=1}^{d}
\mathcal F_{1j}
-
M_4
\sum_{j=1}^{d}
\mathcal L_j,
\\
A_{\Sigma,2}
=&~
M_1
\sum_{j=1}^{d}
\mathcal H_{1j}
+
M_2
-
M_3
\sum_{j=1}^{d}
\mathcal F_{2j}
-
M_4
\sum_{j=1}^{d}
\mathcal L_j,
\\
A_{\Theta,1}
=&~
M_1
\sum_{j=1}^{d}
\mathcal G_{2j}
+
2\Upsilon^{-1}
\Pi_{V_A,d^2}^\top,
\\
A_{\Theta,2}
=&~
M_1
\sum_{j=1}^{d}
\mathcal G_{1j}
-
2\Upsilon^{-1}
\Pi_{V_A,d^2}^\top, %
\end{align*}
where $\mathcal{F}_{ij}$ is defined in \eqref{f_ij}.

The matrices
$A_{\Sigma,1}$,
$A_{\Sigma,2}$,
$A_{\Theta,1}$,
and
$A_{\Theta,2}$
respectively characterize the first-order effects of estimating
$\Sigma_1$,
$\Sigma_2$,
$\Theta_1$,
and
$\Theta_2$.

\begin{theorem}
\label{thm:inf:delta_A:101724}
Suppose Assumptions 1--4 hold. Then, as
$n^{\text{inf}}\rightarrow\infty$, conditioning on $\mathcal{M}_A$ and $\mathcal{M}_\Omega$,
\[
\left\|
\sqrt{n^{\rm inf}}
\left(
\widehat{\psi}_A
-
\psi_A
\right)
-
A_{\Sigma,1}
\,
\mathrm{vec}
\left\{
\sqrt{n^{\rm inf}}
\left(\widehat{\Sigma}_1-\Sigma_1\right)
\right\}
-
A_{\Sigma,2}
\,
\mathrm{vec}
\left\{
\sqrt{n^{\rm inf}}
\left(\widehat{\Sigma}_2-\Sigma_2\right)
\right\}
\right.
\]
\[
\left.
-
A_{\Theta,1}
\,
\mathrm{vec}
\left\{
\sqrt{n^{\rm inf}}
\left(\widehat{\Theta}_1-\Theta_1\right)
\right\}
-
A_{\Theta,2}
\,
\mathrm{vec}
\left\{
\sqrt{n^{\rm inf}}
\left(\widehat{\Theta}_2-\Theta_2\right)
\right\}
\right\|
=
o_p(1).
\]
\end{theorem}

Theorem \ref{thm:inf:delta_A:101724} expresses
$\sqrt{n^{\rm inf}}\left(\widehat{\psi}_A-\psi_A\right)$
as a linear transformation of the estimation errors of
$\widehat{\Sigma}_1$,
$\widehat{\Sigma}_2$,
$\widehat{\Theta}_1$,
and
$\widehat{\Theta}_2$.
Combining this representation with the joint asymptotic distribution of
the covariance and log-one autocovariance estimators yields the following
result.

\begin{theorem}
\label{asymp:delta:A:sel:inf}
Suppose all assumptions in
Theorem \ref{thm:inf:delta_A:101724}
hold.
Then, conditional on
$\mathcal{M}_A$
and
$\mathcal{M}_\Omega$,
\[
\sqrt{n^{\rm inf}}
\left(
\widehat{\psi}_A-\psi_A
\right)
\overset{d}{\longrightarrow}
N_{m_A}
\left(
0,
\sum_{l=1}^{2}
\begin{bmatrix}
A_{\Sigma,l}
&
A_{\Theta,l}
\end{bmatrix}
\begin{pmatrix}
D_l & W_l \\
W_l^\top & \Lambda_l
\end{pmatrix}
\begin{bmatrix}
A_{\Sigma,l}^\top
\\
A_{\Theta,l}^\top
\end{bmatrix}
\right),
\]
where entries of
$D_l$
are defined in
\eqref{Di:entry}.
The
$\big((j_1-1)d+k_1,(j_2-1)d+k_2\big)$-th entries of
$W_l$
and
$\Lambda_l$
are given by
\begin{align}
\label{Wi:Lambdai:entry}
(W_l)_{(j_1-1)d+k_1,\,(j_2-1)d+k_2}
=
&~
\sum_{u=-\infty}^{\infty}
\Big\{
\sigma_{l,j_1j_2}(u)
\sigma_{l,k_1k_2}(u+1)
+
\sigma_{l,k_1j_2}(u)
\sigma_{l,j_1k_2}(u+1)
\Big\},
\\
(\Lambda_l)_{(j_1-1)d+k_1,\,(j_2-1)d+k_2}
=
&~
\sum_{u=-\infty}^{\infty}
\Big\{
\sigma_{l,j_1j_2}(u)
\sigma_{l,k_1k_2}(u+2)
+
\sigma_{l,k_1j_2}(u)
\sigma_{l,j_1k_2}(u+2)
\Big\}.
\end{align}
\end{theorem}

Theorem \ref{asymp:delta:A:sel:inf} establishes the asymptotic normality of
$\widehat{\psi}_A$
conditional on the selected supports
$\mathcal{M}_A$
and
$\mathcal{M}_\Omega$.
The asymptotic covariance matrix is obtained by propagating the joint
asymptotic covariance structure of
$\mathrm{vec}\left(\widehat{\Sigma}_l\right)$
and
$\mathrm{vec}\left(\widehat{\Theta}_l\right)$
through the linear representation established in
Theorem \ref{thm:inf:delta_A:101724}.
Here,
$D_l$
denotes the asymptotic covariance matrix of
$\mathrm{vec}\left(\widehat{\Sigma}_l\right)$,
$\Lambda_l$
denotes the asymptotic covariance matrix of
$\mathrm{vec}\left(\widehat{\Theta}_l\right)$,
and
$W_l$
denotes the asymptotic cross-covariance matrix between
$\mathrm{vec}\left(\widehat{\Sigma}_l\right)$
and
$\mathrm{vec}\left(\widehat{\Theta}_l\right)$.
Consequently, Theorem \ref{asymp:delta:A:sel:inf} provides the
theoretical basis for performing hypothesis
tests for DiffGC effects.

\subsection{Variance estimation and hypothesis testing}

Although Theorems
\ref{asymp:delta:omega:sel:inf}
and
\ref{asymp:delta:A:sel:inf}
establish the asymptotic distributions of
$\widehat{\psi}_\Omega$
and
$\widehat{\psi}_A$,
these results do not immediately yield practical inference procedures.
A key difficulty is that the asymptotic covariance matrices depend on the
long-run covariance quantities
$D_l$,
$W_l$,
and
$\Lambda_l$,
which involve infinite sums of lagged covariance terms.
Consequently, estimation of the asymptotic variance requires
estimation of an infinite-dimensional dependence structure.

A natural approach is to estimate these quantities using truncated
plug-in estimators. For example, one may estimate the entries of
$W_l$
using
\(
\sum_{u=-L_{n_l}}^{L_{n_l}}
\widehat{\sigma}_{l,j_1j_2}(u)
\widehat{\sigma}_{l,k_1k_2}(u+1),
\)
where $L_{n_l}$ is the truncation lag for the $l$-th time series for $l=1,2$.
However, such estimators are generally inconsistent and do not guarantee
positive definiteness of the resulting covariance matrix.

To address this issue,
\cite{melard1991consistent}
proposed a class of weighted estimators that are both consistent and
positive definite. Specifically, they consider estimators of the form
\[
\sum_{u=-L_{n_l}}^{L_{n_l}}
w(ub_{n_l})
\widehat{\sigma}_{l,j_1j_2}(u)
\,\,
w((u+1)b_{n_l})
\widehat{\sigma}_{l,k_1k_2}(u+1),
\]
where
$w(\cdot)$
is a continuous, positive, bounded, and square-integrable weight
function satisfying
$w(0)=1$,
and
$b_n$
satisfies
$b_n\to0$
and
$nb_n\to\infty$
as
$n\to\infty$.

Motivated by this result, we consider the family of Gaussian weights
\[
w_\alpha(x)
=
\exp\!\left(
-\frac{x^2}{2\alpha}
\right),
\qquad
b_n=n^{-\rho},
\]
where
$\alpha>0$
and
$\rho\in(0,1)$.
The resulting estimator inherits the consistency and positive-definiteness
properties established by
\cite{melard1991consistent}.
Moreover, the parameters
$\alpha$
and
$\rho$
provide direct control over the contribution of long-range temporal
dependence.
Smaller values of
$\alpha$
and
$\rho$
downweight large-lag autocovariances more aggressively, whereas larger
values retain more long-range dependence information.

It is also important to note that the estimation of
$D_l$,
$W_l$,
and
$\Lambda_l$
serves only as a nuisance step for variance estimation and is not part
of the selective inference procedure itself.
Consequently, unlike the estimators
$\widehat{\psi}_A$
and
$\widehat{\psi}_\Omega$,
which are constructed exclusively from the inference sample,
the sample autocovariances
are computed using all available time points.
Since consistency of the long-run covariance estimators does not require
independence from the selection stage, this choice improves estimation
accuracy while preserving the validity of the subsequent inference
procedure.

With pre-selected $\alpha$ and $\rho$, the weighted estimators described above yield
\(
\widehat D_l(\alpha,\rho),
\widehat W_l(\alpha,\rho),
\)
and
\(
\widehat\Lambda_l(\alpha,\rho).
\)
In addition, the matrices
\(G_l\)
appearing in Theorem~\ref{asymp:delta:omega:sel:inf}
and
\(A_{\Sigma,l}\),
\(A_{\Theta,l}\)
appearing in Theorem~\ref{asymp:delta:A:sel:inf}
depend only on population quantities and are estimated using their
natural plug-in counterparts.
Using these quantities, we obtain feasible estimators of the asymptotic
covariance matrices in
Theorems~\ref{asymp:delta:omega:sel:inf}
and
\ref{asymp:delta:A:sel:inf},
denoted by
\(
\widehat\Xi_\Omega(\alpha,\rho)
\)
and
\(
\widehat\Xi_A(\alpha,\rho), %
\)
respectively.

For the $j$-th entry of $\psi_A$, define the following Wald statistic for testing $H_0: \psi_{A,j} = 0$:
\[
Z_{A,j}(\alpha,\rho)
=
\frac{
\sqrt{n^{\rm inf}}
\,\widehat\psi_{A,j}
}{
\sqrt{
\widehat\Xi_{A,jj}(\alpha,\rho)
}
}
\stackrel{d}{\rightarrow} N(0,1).
\]
The associated two-sided $p$-value is
\[
p_{A,j}(\alpha,\rho)
=
2
\left\{
1-
\Phi
\left(
|Z_{A,j}(\alpha,\rho)|
\right)
\right\}.
\]
Since the inferential procedure for $\psi_\Omega$ is entirely analogous, we omit the details. Specifically, the corresponding Wald statistics and $p$-values are obtained by replacing
$\widehat{\psi}_A$
and
$\widehat{\Xi}_A(\alpha,\rho)$
with
$\widehat{\psi}_\Omega$
and
$\widehat{\Xi}_\Omega(\alpha,\rho)$,
respectively.
\section{Simulations}

\subsection{Simulation I}

The goal of this simulation is to evaluate the finite-sample performance of the proposed direct estimator in a setting where the individual VAR models are dense but the differential structures remain sparse. Such settings are particularly challenging for separate estimation approaches because they require accurate estimation of two high-dimensional VAR models before estimating their difference.

We considered VAR(1) models with $d=20,50$, and $100$. Following \citet{Zhao2014}, we first generated two precision matrices $\Sigma_1^{-1}$ and $\Sigma_2^{-1}$. The support of $\Sigma_1^{-1}$ was generated according to a network with approximately $60\%$ nonzero entries, and the nonzero values were sampled uniformly from $[-0.5,-0.2]\cup[0.2,0.5]$. To ensure positive definiteness, each row of $\Sigma_1^{-1}$ was scaled by $3$, $4$, and $5$ for $d=20$, $50$, and $100$, respectively. The diagonal entries were then set to one and the resulting matrix was symmetrized by averaging it with its transpose. The matrix $\Sigma_2^{-1}$ was generated by changing the signs of the largest $0.4d$ entries (in absolute value) of $\Sigma_1^{-1}$.

Similarly, the transition matrix $A^{(1)}$ was generated from a directed network with approximately $70\%$ nonzero entries. The nonzero coefficients were sampled uniformly from $[-0.8,-0.5]\cup[0.5,0.8]$ and subsequently scaled to satisfy $\|A^{(1)}\|_2=0.6$. The second transition matrix $A^{(2)}$ was generated by changing the signs of the largest $0.5d$ entries (in absolute value) of $A^{(1)}$, resulting in a sparse DiffGC matrix $\Delta_A=A^{(1)}-A^{(2)}$ despite the individual VAR models being dense. The resulting matrix satisfies $\|A^{(2)}\|_2\approx0.7$.

Next, we generated time series data following \citet{han2015}. Specifically, for $l=1,2$, we computed
\[
\Psi_l
=
\Sigma_l
-
{A^{(l)}}^\top
\Sigma_l
A^{(l)}.
\]
We verified numerically that the matrix $\Psi_l$ is positive definite. We then simulated
\[
{\bf x}_t^{(l)}
=
{A^{(l)}}^\top
{\bf x}_{t-1}^{(l)}
+
\ve\epsilon_t^{(l)},
\qquad
\ve\epsilon_t^{(l)}
\sim
N({\bf 0},\Psi_l),
\]
for $t=2,\ldots,100$ and $l=1,2$.

To the best of our knowledge, there are currently no existing methods for directly estimating $\Delta_A$. We therefore compared the proposed method with two separate estimation approaches that first estimate the individual transition matrices and then compute their difference.

\begin{enumerate}
\item[(S1)]
The $l_1$-penalized least squares estimator of \citet{hsu2008subset}, where for tuning parameter $\eta_{1,l}>0$,
\[
\widehat A_{s1}^{(l)}
=
\argmin_{A^{(l)}}
\left\{
\frac{1}{n_l-1}
\sum_{t=1}^{n_l-1}
\left\|
{\bf x}_{t+1}^{(l)}
-
{A^{(l)}}^\top
{\bf x}_t^{(l)}
\right\|_{\mathrm F}^2
+
\eta_{1,l}
|A^{(l)}|_1
\right\}.
\]

\item[(S2)]
The constrained $l_1$-minimization estimator of \citet{han2015}, where for tuning parameter $\eta_{2,l}>0$,
\[
\widehat A_{s2}^{(l)}
=
\argmin
|A^{(l)}|_1,
\qquad
\text{subject to}
\qquad
\left\|
\widehat\Sigma_l A^{(l)}
-
\widehat\Theta_l
\right\|_{\max}
\le
\eta_{2,l}.
\]
\end{enumerate}
The corresponding estimators of $\Delta_A$ are given by
\(
\widehat A_{s1}^{(1)}
-
\widehat A_{s1}^{(2)}
\)
and
\(
\widehat A_{s2}^{(1)}
-
\widehat A_{s2}^{(2)},
\)
respectively.

\begin{figure}[!t]
    \centering
    \includegraphics[width=\textwidth]{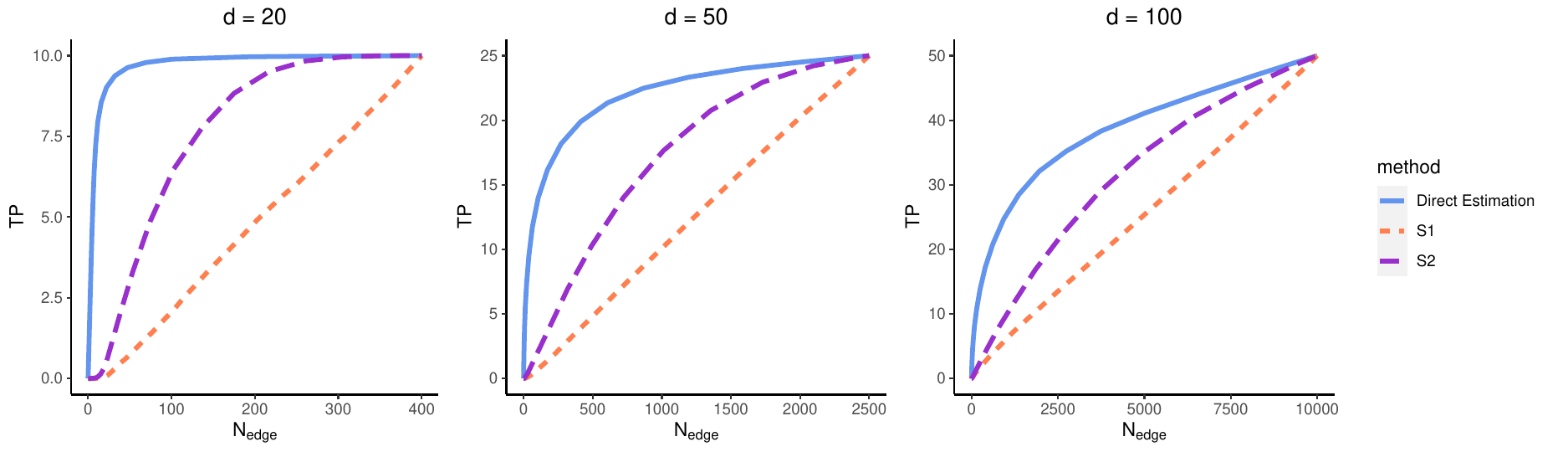}
    \caption{Support recovery performance under the VAR(1) model. The proposed direct estimator is compared with two separate estimation approaches by plotting the number of true discoveries ($\mathrm{TP}$) against the number of selected edges ($N_{\mathrm{edge}}$). The simulation is designed so that the individual VAR models are dense while the DiffGC matrix is sparse. }
    \label{roc1}
\end{figure}

For an arbitrary estimator $\widehat{\Delta}_A$, we define
\[
N_{\mathrm{edge}}
=
\sum_{i,j=1}^{d}
I\left\{\left(\widehat{\Delta}_A\right)_{ij}\neq 0\right\},
\]
and
\[
\mathrm{TP}
=
\sum_{i,j=1}^{d}
I\{
\left(\widehat{\Delta}_A\right)_{ij}\neq 0
\ \text{and}\
(\Delta_A)_{ij}\neq 0
\}.
\]

Figure~\ref{roc1} plots $\mathrm{TP}$ against $N_{\mathrm{edge}}$ for all three methods. Different points on each curve correspond to different choices of tuning parameters. For the proposed method, $\nu$ was fixed at the value selected by aBIC using an independent data set, and the curve was obtained by varying $\lambda_j=\lambda$ for $j=1,\ldots,d$. For S1 and S2, the curves were generated by varying $\eta_{1,1}=\eta_{1,2}=\eta$ and $\eta_{2,1}=\eta_{2,2}=\eta$, respectively.

Figure~\ref{roc1} shows that the proposed direct estimator consistently achieves a larger number of true discoveries for a given model size than either separate estimation approach.
This behavior is expected because the simulation is specifically designed so that the individual VAR models are dense whereas the DiffGC matrix is sparse.
Consequently, separate estimation must recover two difficult high-dimensional models before estimating their difference, while the proposed method directly targets the sparse differential structure. We also observe that S2 outperforms S1, which is consistent with the weaker sparsity assumptions required by the constrained $l_1$-minimization estimator \citep{han2015}.

\subsection{Simulation II}
The purpose of this simulation is twofold. First, we investigate the performance of the proposed method beyond the VAR(1) setting considered in Simulation I. Second, we examine its robustness when the sparsity assumption on $\Delta_\Omega$ is violated.

We generated data from the VAR(2) model
\[
{\bf x}_t^{(l)}
=
{A_1^{(l)}}^\top
{\bf x}_{t-1}^{(l)}
+
{A_2^{(l)}}^\top
{\bf x}_{t-2}^{(l)}
+
\ve\epsilon_t^{(l)},
\qquad
l=1,2.
\]
The transition matrices $A_1^{(l)}$ and $A_2^{(l)}$ were generated similarly to Simulation I. Specifically, the support of $A_1^{(1)}$ was generated according to independent Bernoulli random variables with success probability $0.5$, and the nonzero entries were sampled uniformly from $[-0.8,-0.5]\cup[0.5,0.8]$. The support of $A_2^{(1)}$ was generated according to independent Bernoulli random variables with success probability $0.3$, and the nonzero entries were sampled uniformly from $[-0.5,-0.3]\cup[0.3,0.5]$. The entries of $A_1^{(1)}$ and $A_2^{(1)}$ were subsequently scaled by factors of $5$ and $3$, respectively, ensuring that all eigenvalues of the companion matrix associated with the VAR(2) model have modulus less than one.
The matrices $A_k^{(2)}$ were generated by changing the signs of the largest $d$ entries (in absolute value) of $A_k^{(1)}$ for $k=1,2$. We then defined
\[
\Delta_{A_k}
=
A_k^{(1)}
-
A_k^{(2)},
\qquad
k=1,2.
\]
By construction, the companion matrix associated with the second VAR(2) model is also stable. The error terms $\ve\epsilon_t^{(l)}$ were independently generated from $N({\bf 0},0.1I_d)$.

Note that, under this simulation design, the resulting DiffNet $\Delta_\Omega$ is no longer sparse. This deliberate violation of the sparsity assumption on $\Delta_\Omega$ allows us to assess the robustness of the proposed method when one of its key theoretical assumptions is not satisfied.
The proposed direct estimation procedure was implemented using the VAR(1) companion representation introduced in Section~\ref{sec:directest}. The separate estimation approach S1 was implemented directly on the original VAR(2) model, whereas S2 was implemented using the companion representation, following the recommendation of \citet{han2015}.

\begin{figure}[!t]
    \centering
    \includegraphics[width = 0.75\textwidth]{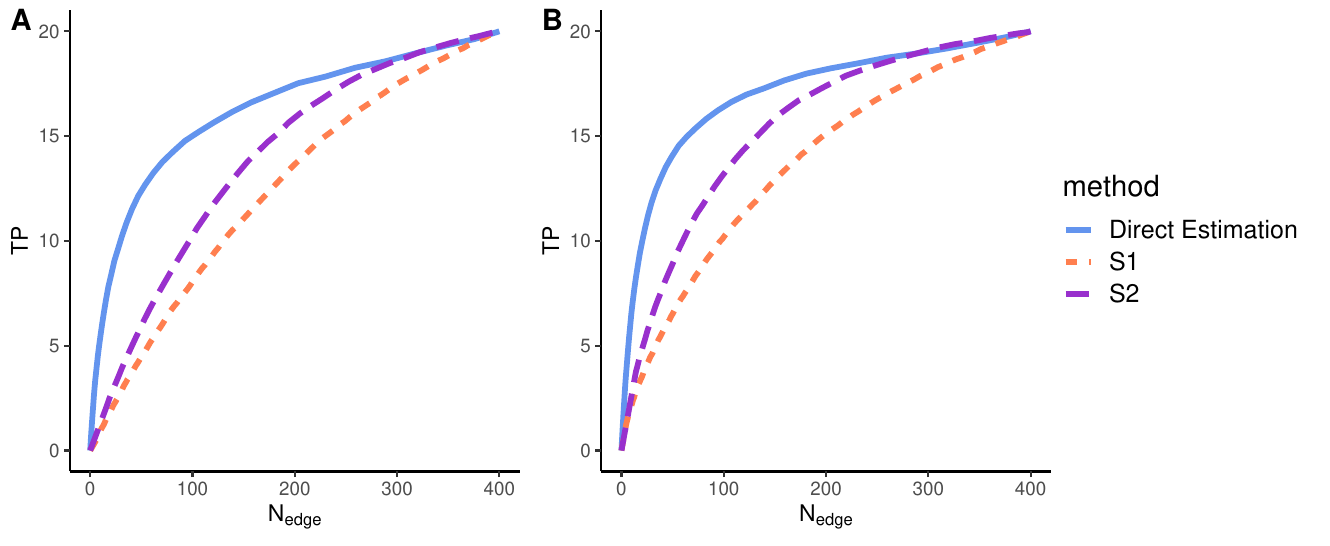}
    \caption{Support recovery performance under the VAR(2) model when the sparsity assumption on $\Delta_\Omega$ is violated. The proposed direct estimator is compared with two separate estimation approaches by plotting the number of true discoveries ($\mathrm{TP}$) against the number of selected edges ($N_{\mathrm{edge}}$). Panel (A) corresponds to recovery of $\Delta_{A_1}$ and panel (B) corresponds to recovery of $\Delta_{A_2}$.}
    \label{roc2}
\end{figure}

We considered $d=20$, so that the companion matrices $\widetilde A^{(1)}$, $\widetilde A^{(2)}$, and their difference are all $40\times40$ matrices.
Figure~\ref{roc2} presents the support recovery performance of all three methods. Similar to Simulation I, the proposed direct estimator consistently achieves a larger number of true discoveries for a given model size than the two separate estimation approaches when recovering both $\Delta_{A_1}$ and $\Delta_{A_2}$. These results demonstrate that the proposed method remains effective beyond the VAR(1) setting and exhibits a considerable degree of robustness to violations of the sparsity assumption on $\Delta_\Omega$.

\subsection{Simulation III}

The purpose of this simulation is to empirically verify the convergence rate established in Theorem~\ref{thm1} for the D-trace estimator of the DiffNet matrix $\Delta_\Omega$.
We considered
\(
\Sigma_1^{-1}
=
\left(
0.4^{|i-j|}
\right)_{i,j=1,\ldots,d},
\)
and defined
\(
S
=
\left\{
(i,j):
|i-j|=1
\ \text{and}\
\max(i,j)\le 11
\right\}.
\)
The matrix $\Sigma_2^{-1}$ was generated by changing the signs of the $(i,j)$th entries of $\Sigma_1^{-1}$ for all $(i,j)\in S$. The resulting DiffNet matrix
\(
\Delta_\Omega
=
\Sigma_1^{-1}
-
\Sigma_2^{-1}
\)
satisfies $|\Delta_\Omega|_0=20$ and $|\Delta_\Omega|_1=16$.

We considered various combinations of $d$ and $n$, with $n_1=n_2=n$. For each pair $(d,n)$, we generated two VAR(1) processes according to
\[
{\bf x}_t^{(l)}
=
A^{(l)\top}
{\bf x}_{t-1}^{(l)}
+
\ve\epsilon_t^{(l)},
\qquad
l=1,2,
\]
where $A^{(l)}$ was generated as in Simulation~I and
\[
\ve\epsilon_t^{(l)}
\sim
N({\bf 0},\Psi_l),
\qquad
\Psi_l
=
\Sigma_l
-
A^{(l)\top}
\Sigma_l
A^{(l)}.
\]
For each simulated data set, we computed the D-trace estimator $\widehat{\Delta}_\Omega(\nu)$ using a tuning parameter satisfying
\(
\nu
\propto
\left(
{\log d}/{n}
\right)^{1/2},
\)
as suggested by Theorem~\ref{thm1}. The Frobenius-norm error
\(
\left\|
\widehat{\Delta}_\Omega(\nu)
-
\Delta_\Omega
\right\|_{\mathrm F}
\)
was then recorded.

\begin{figure}[!t]
    \centering
    \includegraphics[width = 0.45\textwidth]{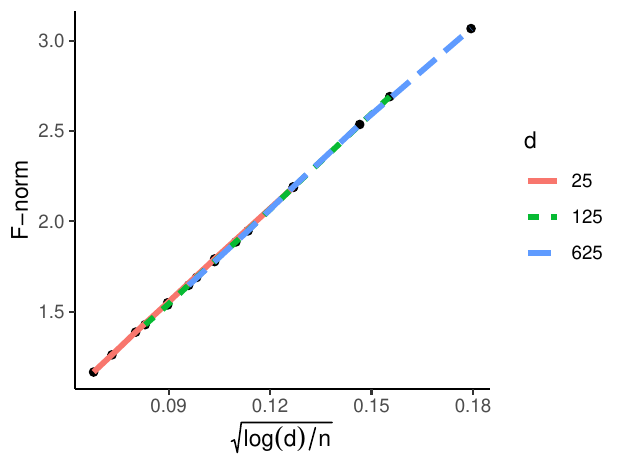}
    \caption{Empirical verification of the convergence rate in Theorem~\ref{thm1}. The average Frobenius-norm error $\|\widehat{\Delta}_\Omega(\nu)-\Delta_\Omega\|_{\mathrm F}$ is plotted against $(\log d/n)^{1/2}$ for $d=25,125,625$ and $n=200,300,\ldots,700$. Each point is averaged over 1000 replications. The approximately linear relationship supports the theoretical convergence rate $O\!\left\{(\log d/n)^{1/2}\right\}$.}
    \label{f:norm}
\end{figure}

Figure~\ref{f:norm} reports the average Frobenius-norm error over 1000 replications for $n=200,300,\ldots,700$ and $d=25,125,625$. Each curve corresponds to a fixed value of $d$ and plots the average estimation error against $(\log d/n)^{1/2}$.
The approximately linear relationship provides empirical support for the convergence rate established in Theorem~\ref{thm1}, indicating that the Frobenius-norm error scales as
\(
O\!\left(\sqrt{{\log d}/{n}}\right).
\)

\subsection{Simulation IV}

 To evaluate the finite-sample performance of the proposed selective inference procedure, we considered a simulation setting closely related to Simulation I. Specifically, we adopted the same transition matrices $A^{(1)}$ and $A^{(2)}$, as well as the same precision matrices $\Sigma_1^{-1}$ and $\Sigma_2^{-1}$, from Simulation I with $d=50$. We set $n_1=n_2=6{,}000$, and time series data were generated according to the same VAR model described in Simulation I.
Following the sample-splitting strategy proposed in Section~\ref{sec:DiffGC:inf}, each time series was divided into three equal segments of length $2{,}000$, corresponding to the estimation sample, the wash-out sample, and the inference sample, respectively. The estimation sample was used to estimate the supports for $\Delta_\Omega$ and $\Delta_A$, while the inference sample was used to perform selective inference conditional on the selected supports.
We applied a threshold of $0.01$ to the estimated DiffNet and DiffGC coefficients when constructing the candidate sets for selective inference.

The choice of $d=50$ and $n_1=n_2=6{,}000$ was motivated by the real-data application in Section~\ref{sec:real}. In particular, the EEG application involves a smaller dimension ($d=34$) and a slightly larger sample size ($n_2\approx7{,}000$). Consequently, this design allows us to assess the robustness of the proposed inferential procedure under conditions that are comparable to those encountered in our real application.

To evaluate both the screening and inferential components of the proposed procedure, we considered five performance metrics.
Let \(S\) denote the true support, \(\widehat{S}\) denote the support selected during the screening stage, and \(R\) denote the set of entries declared significant by the proposed selective inference procedure using a two-sided significance level of \(\alpha=0.05\).
We define the selection recall and selection precision as
\[
\text{Selection Recall}
=
\frac{|S \cap \widehat{S}|}{|S|}
\quad\text{and}\quad
\text{Selection Precision}
=
\frac{|S \cap \widehat{S}|}{|\widehat{S}|},
\]
respectively.
For the inferential stage, we consider the power, conditional power, and false positive rate (FPR), defined by
\[
\text{Power}
=
\frac{|S \cap R|}{|S|},
\qquad
\text{Conditional Power}
=
\frac{|S \cap \widehat{S} \cap R|}
{|S \cap \widehat{S}|},
\]
and
\[
\text{FPR}
=
\frac{|S^c \cap R|}{|S^c|},
\]
respectively. We do not consider the conditional false positive rate
\[
\frac{|S^c \cap \widehat{S} \cap R|}
{|S^c \cap \widehat{S}|},
\]
because the screening procedure achieves high selection precision in all simulation settings considered. Consequently, the quantity \(|S^c \cap \widehat{S}|\) is often very small and occasionally equal to zero, resulting in unstable or undefined estimates of the conditional false positive rate. Since selection precision already quantifies the frequency of screening errors and the FPR summarizes the overall inferential error rate, the conditional false positive rate provides little additional information in the present setting.

Across the 100 simulation replicates, the screening procedure exhibited excellent support recovery performance for the DiffNet matrix $\Delta_\Omega$, achieving an average selection recall of \(0.996\) (SD \(=0.008\)) and an average selection precision of \(0.958\) (SD \(=0.017\)). These results indicate that the candidate set used for selective inference contained nearly all true DiffNet edges while introducing relatively few false positives.
For the inferential stage, we considered a grid of tuning parameters with \(\rho \in \{1/2,1/4,1/8,1/16\}\) and \(\alpha \in \{0.01,0.05,0.10,0.20,0.50\}\), yielding a total of 20 configurations. The resulting inferential performance is summarized in Table~\ref{table:diffnet:inf}. Remarkably, the proposed procedure produced identical results across all 20 tuning-parameter configurations, achieving a power of \(0.956\) and a conditional power of \(0.960\), with a false positive rate of \(0.000\) throughout. The standard deviations were uniformly small, indicating stable performance across simulation replicates. Taken together, these results demonstrate that the proposed selective inference procedure for $\Delta_\Omega$ is insensitive to the choice of tuning parameters and is capable of accurately identifying DiffNet edges.

\begin{table}[!ht]
\centering
\caption{Selective inference performance for the DiffNet matrix $\Delta_\Omega$ under different choices of the tuning parameters $\rho$ and $\alpha$. Entries are reported as mean (SD) across 100 simulation replicates.}
\begin{tabular}{ccccc}
\toprule
$\rho$ & $\alpha$ & Power & Conditional Power & FPR \\
\midrule

\multirow{5}{*}{$1/2$}
& 0.01 & 0.956 (0.009) & 0.960 (0.004) & 0.000 (0.000)\\
& 0.05 & 0.956 (0.009) & 0.960 (0.004) & 0.000 (0.000)\\
& 0.10 & 0.956 (0.009) & 0.960 (0.004) & 0.000 (0.000)\\
& 0.20 & 0.956 (0.009) & 0.960 (0.004) & 0.000 (0.000)\\
& 0.50 & 0.956 (0.009) & 0.960 (0.004) & 0.000 (0.000)\\
\midrule

\multirow{5}{*}{$1/4$}
& 0.01 & 0.956 (0.009) & 0.960 (0.004) & 0.000 (0.000)\\
& 0.05 & 0.956 (0.009) & 0.960 (0.004) & 0.000 (0.000)\\
& 0.10 & 0.956 (0.009) & 0.960 (0.004) & 0.000 (0.000)\\
& 0.20 & 0.956 (0.009) & 0.960 (0.004) & 0.000 (0.000)\\
& 0.50 & 0.956 (0.009) & 0.960 (0.004) & 0.000 (0.000)\\
\midrule

\multirow{5}{*}{$1/8$}
& 0.01 & 0.956 (0.009) & 0.960 (0.004) & 0.000 (0.000)\\
& 0.05 & 0.956 (0.009) & 0.960 (0.004) & 0.000 (0.000)\\
& 0.10 & 0.956 (0.009) & 0.960 (0.004) & 0.000 (0.000)\\
& 0.20 & 0.956 (0.009) & 0.960 (0.004) & 0.000 (0.000)\\
& 0.50 & 0.956 (0.009) & 0.960 (0.004) & 0.000 (0.000)\\
\midrule

\multirow{5}{*}{$1/16$}
& 0.01 & 0.956 (0.009) & 0.960 (0.004) & 0.000 (0.000)\\
& 0.05 & 0.956 (0.009) & 0.960 (0.004) & 0.000 (0.000)\\
& 0.10 & 0.956 (0.009) & 0.960 (0.004) & 0.000 (0.000)\\
& 0.20 & 0.956 (0.009) & 0.960 (0.004) & 0.000 (0.000)\\
& 0.50 & 0.956 (0.009) & 0.960 (0.004) & 0.000 (0.000)\\
\bottomrule
\end{tabular}
\label{table:diffnet:inf}
\end{table}

The support recovery performance for the DiffGC matrix $\Delta_A$ is noticeably weaker than that observed for the DiffNet matrix $\Delta_\Omega$. Across the 100 simulation replicates, the screening procedure achieved an average selection recall of \(0.767\) (SD \(=0.036\)) and an average selection precision of \(0.689\) (SD \(=0.04\)). Nevertheless, once the candidate set was obtained, the subsequent selective inference procedure continued to exhibit stable performance across all tuning-parameter configurations considered.
As shown in Table~\ref{table:DiffGC:inf}, the power ranged from \(0.703\) to \(0.704\), the conditional power ranged from \(0.917\) to \(0.918\), and the false positive rate was uniformly \(0.011\) across all configurations.
Compared with inference for $\Delta_\Omega$, the reduced performance for $\Delta_A$ is expected. Although both analyses employ the same thresholding strategy during the screening stage, estimation and inference for the DiffGC matrix involve several estimated quantities, including the DiffNet matrix $\Delta_\Omega$. Consequently, uncertainty accumulated from multiple estimation steps propagates into the inferential procedure for $\Delta_A$, leading to lower support recovery accuracy and reduced power.
Despite this additional difficulty, the proposed selective inference procedure maintains good error control and remains insensitive to the choice of tuning parameters \(\rho\) and \(\alpha\).

\begin{table}[!ht]
\centering
\caption{Selective inference performance for the DiffGC matrix $\Delta_A$ under different choices of the tuning parameters $\rho$ and $\alpha$. Entries are reported as mean (SD) across 100 simulation replicates.}
\begin{tabular}{ccccc}
\toprule
$\rho$ & $\alpha$ & Power & Conditional Power & FPR \\
\midrule

\multirow{5}{*}{$1/2$}
& 0.01 & 0.704 (0.036) & 0.918 (0.025) & 0.011 (0.002)\\
& 0.05 & 0.703 (0.036) & 0.917 (0.025) & 0.011 (0.002)\\
& 0.10 & 0.703 (0.036) & 0.917 (0.025) & 0.011 (0.002)\\
& 0.20 & 0.703 (0.036) & 0.917 (0.025) & 0.011 (0.002)\\
& 0.50 & 0.703 (0.036) & 0.917 (0.025) & 0.011 (0.002)\\
\midrule

\multirow{5}{*}{$1/4$}
& 0.01 & 0.704 (0.036) & 0.918 (0.025) & 0.011 (0.002)\\
& 0.05 & 0.704 (0.036) & 0.918 (0.025) & 0.011 (0.002)\\
& 0.10 & 0.704 (0.036) & 0.918 (0.025) & 0.011 (0.002)\\
& 0.20 & 0.703 (0.036) & 0.918 (0.025) & 0.011 (0.002)\\
& 0.50 & 0.703 (0.036) & 0.917 (0.025) & 0.011 (0.002)\\
\midrule

\multirow{5}{*}{$1/8$}
& 0.01 & 0.704 (0.036) & 0.918 (0.025) & 0.011 (0.002)\\
& 0.05 & 0.704 (0.036) & 0.918 (0.025) & 0.011 (0.002)\\
& 0.10 & 0.704 (0.036) & 0.918 (0.025) & 0.011 (0.002)\\
& 0.20 & 0.704 (0.036) & 0.918 (0.025) & 0.011 (0.002)\\
& 0.50 & 0.703 (0.036) & 0.918 (0.025) & 0.011 (0.002)\\
\midrule

\multirow{5}{*}{$1/16$}
& 0.01 & 0.704 (0.036) & 0.918 (0.025) & 0.011 (0.002)\\
& 0.05 & 0.704 (0.036) & 0.918 (0.025) & 0.011 (0.002)\\
& 0.10 & 0.704 (0.036) & 0.918 (0.025) & 0.011 (0.002)\\
& 0.20 & 0.704 (0.036) & 0.918 (0.025) & 0.011 (0.002)\\
& 0.50 & 0.704 (0.036) & 0.918 (0.025) & 0.011 (0.002)\\
\bottomrule
\end{tabular}
\label{table:DiffGC:inf}
\end{table}

\section{Real data analysis}\label{sec:real}

We analyzed resting-state electroencephalography (EEG) data from the Stimulus-Selective Response Modulation (SRM) project conducted at the Department of Psychology, University of Oslo, Norway. The SRM resting-state EEG dataset contains recordings from 111 healthy control participants acquired using a BioSemi ActiveTwo system with 64 electrodes arranged according to the extended international 10--20 (10--10) system. Each recording consists of four minutes of uninterrupted eyes-closed resting-state EEG sampled at 1024 Hz. While most participants contributed a single recording session (T1), a subset of participants underwent a second resting-state EEG assessment (T2) at a later date, providing repeated measurements suitable for studying longitudinal changes in brain connectivity. More details about the dataset can be found in \citet{hatlestad2022bids}.

For the present analysis, we focused on Subject 003, who completed two resting-state EEG recordings separated by approximately three months. Our objective was to identify statistically significant changes in both Granger causal relationships (DiffGC) and contemporaneous conditional dependence (DiffNet) between the two recording sessions using the proposed selective inference framework. To reduce computational burden and facilitate high-dimensional inference, the EEG signals were downsampled by a factor of 10 prior to analysis. Channel quality was assessed using the accompanying metadata files, and only channels labeled as ``good'' in both recording sessions were retained. This preprocessing step reduced the analysis from the original 64 channels to 34 common high-quality channels.

To implement the proposed inference procedure, the available observations from each session were divided into three segments.
The first segment was used exclusively for estimation and variable selection.
The second segment served as a washout region and was discarded from subsequent analyses in order to reduce dependence between the estimation and inference stages.
The third segment was reserved exclusively for inference.
Because the asymptotic selective inference procedure requires equal inference-sample sizes across the two time points, the inference samples were matched to the shorter recording (T1). This necessitated reserving a larger portion of the T2 recordings as a washout region.
Following this procedure, the T1 estimation sample comprised approximately 2,870 observations (7 downsampled trials) and the T2 estimation sample comprised approximately 7,380 observations (18 trials). Both inference samples contained approximately 2,870 observations each (7 trials from each session). =

Using the estimation sample, we obtained sparse estimates of both the DiffNet matrix \(\Delta_\Omega\) and the DiffGC matrix \(\Delta_A\), which were subsequently used to define candidate sets for selective inference.
We applied a small threshold of \(0.001\) to the estimated coefficients before constructing the candidate sets. The threshold was used to remove numerically negligible estimates and to limit the size of the candidate sets, thereby reducing the computational burden of the subsequent inference procedure. Importantly, this threshold was used only for screening, while statistical significance was determined separately using the proposed selective inference procedure.
This screening step identified $29$ candidate DiffNet edges and $226$ candidate DiffGC coefficients among all possible contemporaneous and directed connections, respectively.  Subsequent inference was conducted using the held-out inference sample to account for selection effects.

We begin by focusing on the DiffGC matrix \(\Delta_A\), which is the primary inferential target in this application. The selective inference procedure depends on two tuning parameters, \(\rho\) and \(\alpha\). To evaluate the robustness of the resulting inference, similar to Simulation IV, we considered a grid of four values of \(\rho\) at \(1/2, 1/4, 1/8,\) and \(1/16\), and six values of \(\alpha\) at \(0.01, 0.05, 0.10, 0.20,\) \(0.50\) and 0.80, yielding 24 parameter combinations. For each combination, we implemented our DiffGC procedure and recorded both the number of significant entries in \(\Delta_A\) and the corresponding entry-wise $p$-values.

\begin{figure}[!ht]
    \centering
    \includegraphics[width=1\linewidth]{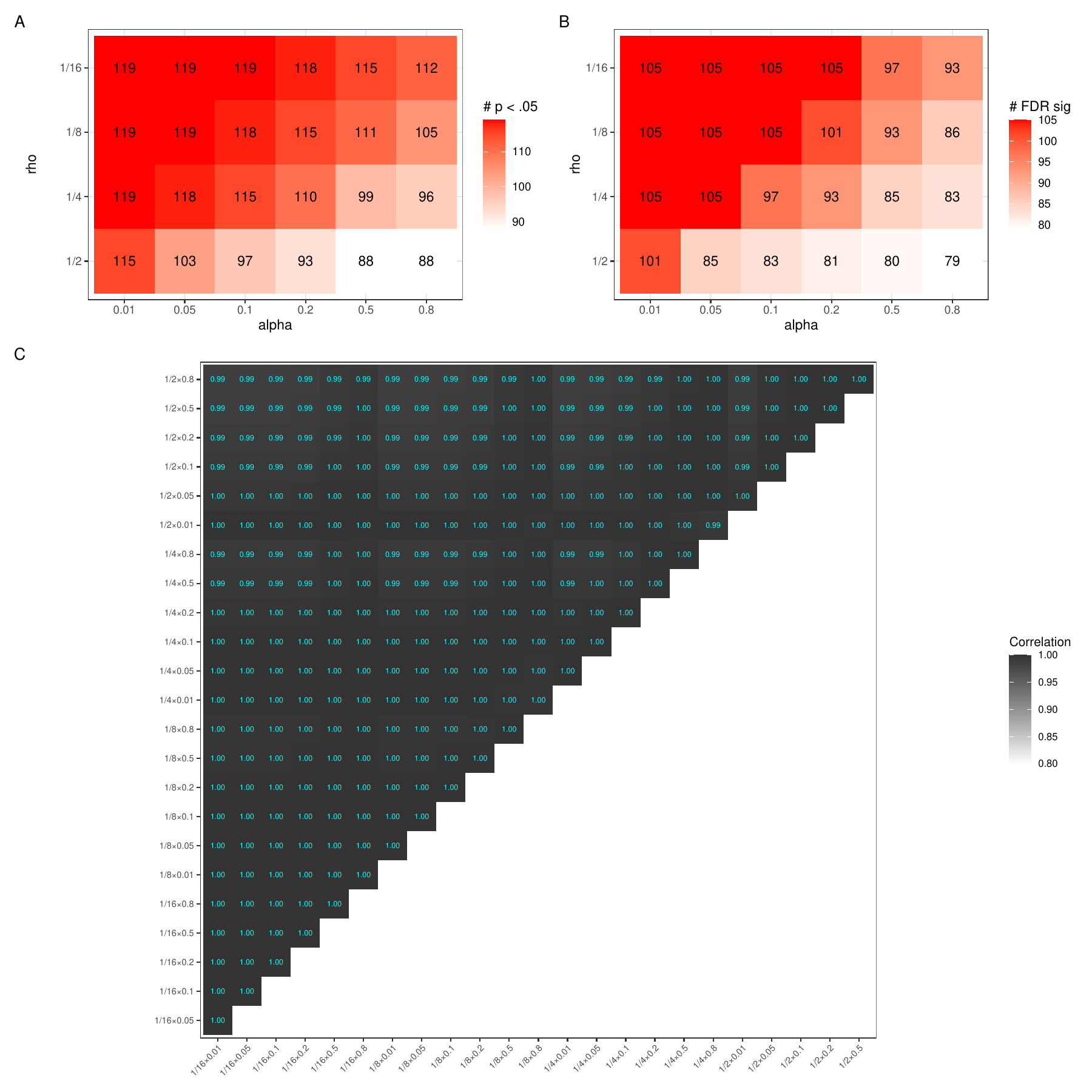}
    \caption{(A) Number of significant DiffGC entries identified using unadjusted \(p < .05\) across a grid of tuning parameter values \(\rho \in \{1/2, 1/4, 1/8, 1/16\}\) and \(\alpha \in \{0.01, 0.05, 0.10, 0.20, 0.50, 0.80\}\).
    (B) Number of significant DiffGC entries after false discovery rate (FDR) correction at \(q < .05\).
    (C) Pairwise correlations of entry-wise \(p\)-value vectors obtained from all 24 parameter combinations.
    Each cell represents the correlation between the DiffGC \(p\)-value profiles from two tuning parameter settings. Although the number of detected edges varies across the parameter grid, the edge-wise inference results exhibit substantial agreement across parameter settings.
    }
    \label{fig:srm003}
\end{figure}

In the simulation studies, the proposed inference procedure exhibited little sensitivity to the choice of tuning parameters. In this real-data application, however, we observe greater variation in the number of detected entries when a hard significance threshold is applied.
The total number of significant DiffGC entries ranges from 88 to 119 using unadjusted $p$-values and from 79 to 105 following FDR correction across the 24 parameter combinations (Figure \ref{fig:srm003}A–B).
Both the unadjusted and FDR-corrected counts exhibit a clear monotonic pattern, with more discoveries in the upper-left region of the parameter grid ($\alpha$ and $\rho$ small) and progressively fewer toward the lower-right region.
This greater sensitivity in the real-data setting, relative to simulations, may reflect model misspecification and weaker signals. Nevertheless, the variation in the number of discoveries appears to arise primarily from systematic shifts in the overall magnitude of the $p$-values: larger values of $\rho$ and $\alpha$ generally produce larger $p$-values and hence fewer discoveries, while leaving their relative ordering nearly unchanged. Indeed, pairwise correlations among the 24 DiffGC $p$-value vectors range from 0.99 to 1.00 (Figure~\ref{fig:srm003}C), indicating that the same connections consistently exhibit the strongest evidence across tuning-parameter combinations. Thus, although the number of significant connections depends on the tuning parameters, the underlying pattern of statistical evidence is highly stable.

To obtain a single unified significance assessment that exploits this high concordance without requiring commitment to any one parameter combination, we aggregated the 24 $p$-value vectors using the Cauchy combination test
\citep{liu2020cauchy}.
The Cauchy combination test is particularly well suited for this purpose because it remains valid under arbitrary dependence structures among the  $p$-values.
Applying the Cauchy combination test yielded a unified significance assessment for the 226 DiffGC coefficients identified during the selection stage.
Figure \ref{fig:srm003:cauchy}A summarizes the relationship between variable selection and subsequent inference.
Of the 226 candidate differential connections selected from the first of three segments, 114 remained significant after aggregation and 94 further survived FDR correction.
Notably, the significant connections were distributed across multiple brain regions rather than being concentrated within a single subnetwork, suggesting widespread differences in transition dynamics between the two recording sessions.
\begin{figure}[!ht]
    \centering
    \includegraphics[width=1\linewidth]{diffvar_results_cauchy_sub003_equal_ninf_Sep26.png}
    \caption{
    (A) Classification of all possible DiffGC connections following variable selection and Cauchy-aggregated inference. White cells denote connections not selected during the screening stage, gray cells denote selected connections that did not achieve statistical significance, orange cells denote marginally significant connections ($p < .05$), and red cells denote connections that survived false discovery rate (FDR) correction ($q < .05$). (B) Volcano plot of the 226 screened DiffGC coefficients. The x-axis shows the estimated \(\widehat{\Delta}_A\) coefficient and the y-axis shows the Cauchy-combined \(-\log_{10}(p)\). Labels identify the five most significant differential connections.
    }
    \label{fig:srm003:cauchy}
\end{figure}

Figure \ref{fig:srm003:cauchy}B displays the estimated DiffGC coefficients alongside their Cauchy-combined $p$-values.
The five most significant DiffGC connections, C3$\rightarrow$C3, Pz$\rightarrow$Pz, FC1$\rightarrow$F3, FC3$\rightarrow$FC3, and Pz$\rightarrow$CP2, include both self-self (autoregressive) entries and directed inter-regional connections.
Across the full set of significant entries, the estimated DiffGC coefficients are predominantly negative, suggesting that temporal predictability, both within and between channels, is generally lower at T1 relative to T2.

\begin{figure}[!ht]
    \centering
    \includegraphics[width=1\linewidth]{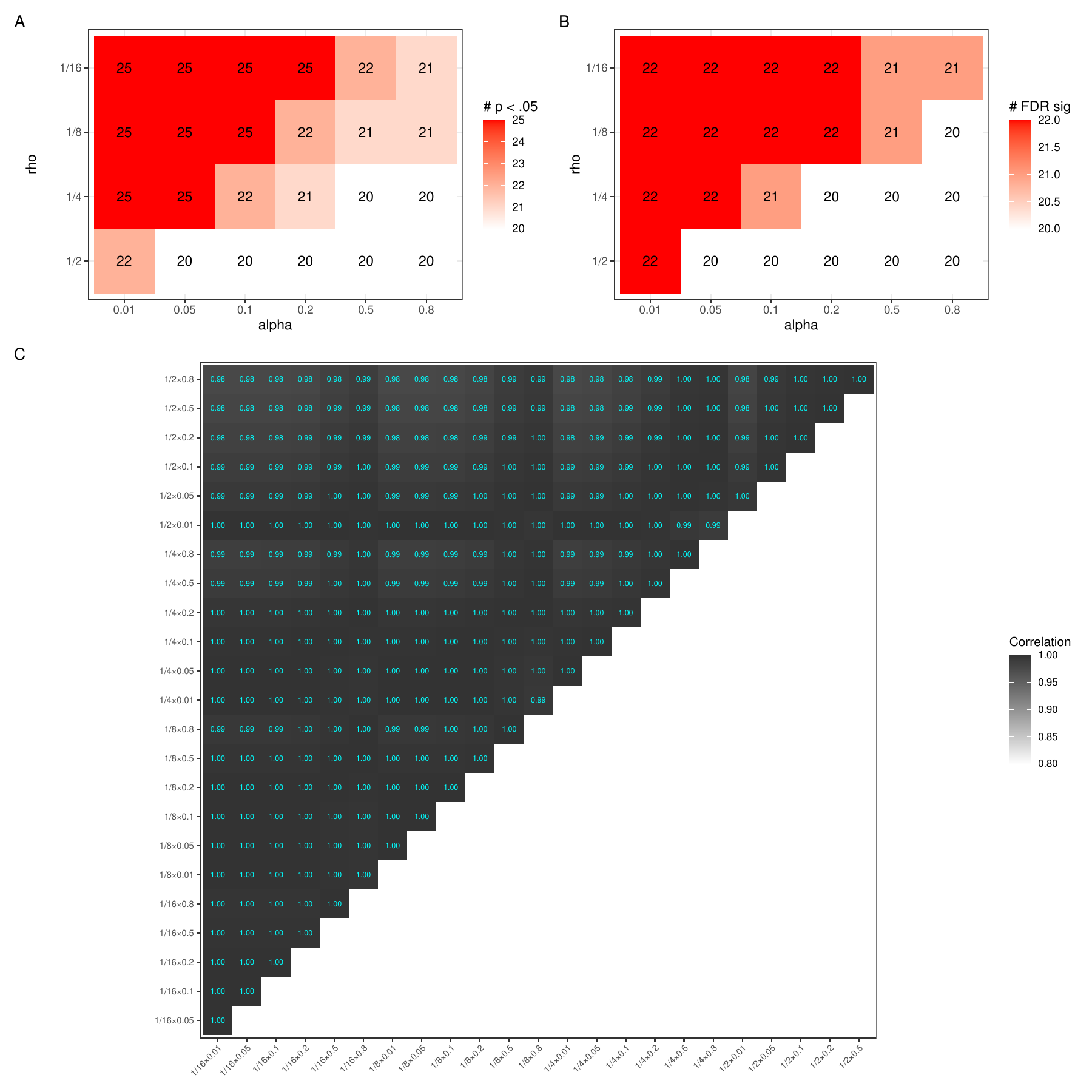}
    \caption{(A) Number of significant DiffNet entries identified using unadjusted \(p < .05\) across a grid of tuning parameter values \(\rho \in \{1/2, 1/4, 1/8, 1/16\}\) and \(\alpha \in \{0.01, 0.05, 0.10, 0.20, 0.50, 0.80\}\).
    (B) Number of significant DiffNet entries after false discovery rate (FDR) correction at \(q < .05\).
    (C) Pairwise correlations of entry-wise \(p\)-value vectors obtained from all 24 parameter combinations.
    Each cell represents the correlation between the DiffNet \(p\)-value profiles from two tuning parameter settings. Although the number of detected edges varies across the parameter grid, the edge-wise inference results exhibit substantial agreement across parameter settings.
    }
    \label{fig:srm003:diffnet}
\end{figure}

We now turn to the DiffNet analysis, which targets changes in the contemporaneous partial correlation structure between channels and provides a complementary, instantaneous perspective on how neural coupling reorganizes between sessions.
Among the 29 selected entries, the total number of significant DiffNet entries varies across the parameter space, ranging from 20 to 25 using unadjusted $p$-values and from 20 to 22 following FDR correction (Figure \ref{fig:srm003:diffnet}A--B), mirroring the monotonic pattern observed for DiffGC.
However, the degree of variation is substantially smaller.
Also, the pairwise correlations among the 24 DiffNet $p$-value vectors are still high, ranging from 0.98 to 1.00 (Figure \ref{fig:srm003:diffnet}C).

Applying the Cauchy combination test to the 29 candidate DiffNet connections yields a unified significance assessment that reflects the underlying stability of the inferential results. Of the 29 candidates identified during the screening stage, 22 remained significant after Cauchy aggregation and 21 remained significant after FDR correction at the 0.05 level. Figure~\ref{fig:srm003:cauchy:diffnet}A displays the final inferential results for the DiffNet matrix. Since $\widehat{\Delta}_\Omega$ is symmetric, only the upper triangular portion of the matrix, including the diagonal, is shown. Figure~\ref{fig:srm003:cauchy:diffnet}B presents the estimated $\widehat{\Delta}_\Omega$ coefficients together with their Cauchy-combined $p$-values. The five most significant DiffNet connections, FC1--FC1, Pz--CPz, C3--C3, CP1--CPz, and CPz--CPz, include both diagonal entries and off-diagonal connections. Similar to the DiffGC analysis, the majority of the estimated DiffNet coefficients are negative. A notable exception is the CPz--CPz diagonal entry, which exhibits a relatively large positive estimate and represents one of the most prominent DiffNet effects identified by the proposed selective inference procedure.

\begin{figure}
    \centering
    \includegraphics[width=1\linewidth]{diffnet_results_cauchy_sub003_equal_ninf_Sep26.png}
    \caption{
(A) DiffNet connections identified by the proposed procedure. Because the DiffNet matrix is symmetric, only the upper triangular portion (including diagonal entries) is shown, with each cell representing a unique network connection.
(B) Volcano plot of the selected DiffNet coefficients corresponding to the upper triangular portion of the DiffNet matrix. The x-axis shows the estimated DiffNet coefficient, and the y-axis shows the aggregated significance level. Labels indicate the five most significant DiffNet connections.
}
     \label{fig:srm003:cauchy:diffnet}
\end{figure}

Taken together, the DiffGC and DiffNet analyses provide complementary evidence for session-to-session reorganization of resting-state EEG dynamics. The strongest DiffGC effects are concentrated at frontocentral and centroparietal electrodes, indicating changes in both channel-specific temporal dynamics and directed interactions between brain regions over the three-month interval. The DiffNet analysis identifies contemporaneous changes involving many of the same electrode locations, revealing altered conditional dependence structure after accounting for the remaining channels. The substantial spatial overlap between the two analyses suggests that the observed differences are not confined to either temporal or contemporaneous connectivity alone, but instead reflect a coordinated reorganization of resting-state brain dynamics.

\section{Discussion}

We studied direct estimation and selective inference for comparing two related high-dimensional VAR models. The proposed methodology avoids separate estimation of the two VAR models and instead directly targets the differential structures of interest. The resulting procedure only requires sparsity of the DiffNet matrix $\Delta_\Omega$ and the DiffGC matrix $\Delta_A$, allowing the individual VAR models to exhibit substantially more complex connectivity patterns.  We established convergence rates and support recovery guarantees, derived asymptotic distributions for the corresponding selective inference targets, and obtained new theoretical results for DiffNet estimation under weaker assumptions than existing approaches.

One limitation of the current theoretical framework is the sparsity assumption on $\Delta_\Omega$. While this assumption has a natural interpretation under VAR(1) models, its interpretation becomes less transparent under VAR($p$) models with $p\ge 2$. To illustrate this point, consider the VAR(2) model. According to the companion representation,
\[
\Delta_\Omega = \left\{\mathrm{cov}\left(\widetilde{\bf x}_t^{(1)}\right)\right\}^{-1} - \left\{\mathrm{cov}\left(\widetilde{\bf x}_t^{(2)}\right)\right\}^{-1},
\]
where $\widetilde{\bf x}_t^{(l)} = ({\bf x}_{t+1}^{(l)\top},{\bf x}_{t}^{(l)\top})^\top$ for $l=1,2$. Letting $\Gamma^{(l)}(h)=\mathrm{cov}({\bf x}_t^{(l)},{\bf x}_{t+h}^{(l)})$ and $\Gamma^{(l)}(-h)=\{\Gamma^{(l)}(h)\}^\top$, we have
\[
\left\{\mathrm{cov}\left(\widetilde{\bf x}_t^{(l)}\right)\right\}^{-1}
=
\begin{pmatrix}
\Xi_1^{(l)}
&
-\Xi_1^{(l)}
\Gamma^{(l)}(-1)
\{\Gamma^{(l)}(0)\}^{-1}
\\
-\{\Gamma^{(l)}(0)\}^{-1}
\Gamma^{(l)}(1)
\Xi_1^{(l)}
&
\Xi_2^{(l)}
\end{pmatrix},
\]
where
\[
\Xi_1^{(l)}
=
\left[
\Gamma^{(l)}(0)
-
\Gamma^{(l)}(-1)
\{\Gamma^{(l)}(0)\}^{-1}
\Gamma^{(l)}(1)
\right]^{-1}
\]
and
\[
\Xi_2^{(l)}
=
\left[
\Gamma^{(l)}(0)
-
\Gamma^{(l)}(1)
\{\Gamma^{(l)}(0)\}^{-1}
\Gamma^{(l)}(-1)
\right]^{-1}.
\]
Consequently, sparsity of $\Delta_\Omega$ implies sparsity of $\Xi_k^{(1)}-\Xi_k^{(2)}$ for $k=1,2$, which is considerably less interpretable than sparsity of the DiffNet matrix for a VAR(1) process. Although Simulation II demonstrates that the proposed method can perform well even when this assumption is violated, developing alternative direct estimation procedures for DiffGC under VAR($p$) models with $p\ge 2$ remains an important direction for future research.

Another promising direction is the development of selective inference procedures beyond the sample-splitting framework considered in this paper. While sample splitting provides a convenient route to valid post-selection inference, it necessarily sacrifices a portion of the available data for estimation. It would be interesting to investigate whether more efficient inferential procedures can be developed while retaining valid uncertainty quantification for DiffNet and DiffGC effects.

\newpage

\acks{Y.W. is supported by NIH grants R35-GM159674 and R03-OD039969.
J.M. is supported by NIH grants R01-GM145772 and R01-GM151301.
A.S. was supported by NIH grant RF1-AG090462.
}

\newpage

\section*{Appendix}
\appendix

This Appendix contains the technical proofs of all theoretical results presented in the main text. We begin with several preliminary results, including a brief review of the unified framework for high-dimensional \(M\)-estimators with decomposable regularizers developed by \citet{negahban2012unified}, which serves as the foundation for our analysis of the proposed penalized estimators.
We then establish the theoretical properties of the DiffNet estimator and the DiffGC estimator, followed by the asymptotic theory for the selective inferential targets introduced in Section~\ref{sec:DiffGC:inf}.

\section{Preliminary results}\label{sec:appenA}
\begin{lemma}\citep{basu2015regularized}\label{s1}
Consider $d$-dimensional random vectors ${\bf z}_1, \ldots, {\bf z}_n$ sampled from a stationary time series $\{ {\bf z}_t\}_{t \in \mathbb{N}}$, where ${\bf z}_t \in \mathbb{R}^d$ and $E({\bf z}_t) = 0$. Suppose that $\{ {\bf z}_t\}_{t \in \mathbb{N}}$ satisfies Assumption 1 with the spectral density function $f(\cdot)$. Let
$\Gamma(h) = E({\bf z}_t{\bf z}_{t+h}^\top)$ for $h \in \mathbb{N}$ and define
 \begin{equation}\label{Upsilon}
 \Upsilon_n =
\begin{bmatrix}
    \Gamma(0) & \Gamma(1) & \Gamma(2) & \dots  & \Gamma(n-1) \\
    \Gamma(-1) & \Gamma(0) & \Gamma(1) & \dots  & \Gamma(n-2) \\
    \vdots & \vdots & \vdots & \ddots & \vdots \\
    \Gamma(1-n) & \Gamma(2-n) & \Gamma(3-n) & \dots  & \Gamma(0)
\end{bmatrix}.
\end{equation}

Then for any $n \geq 1$ and $d \geq 1$, we have
\[
\lambda_{\max}(\Upsilon_n) \leq 2\pi\mathcal{M}(f).
\]
\end{lemma}

\begin{lemma}\citep{negahban2011estimation}\label{s2}
Suppose that $Y \sim N_n(0, Q)$ is an $n$-dimensional Gaussian random vector. For $\eta > 2n^{-1/2}$, we have
\[
\mbox{pr}\left\{\left|\|Y\|_{2}^{2}-E\left(\|Y\|_{2}^{2}\right)\right|>4 n \eta\|Q\|_{2}\right\} \leq 2 \exp \left\{-n\left(\eta-2 n^{-1 / 2}\right)^{2} / 2\right\}+2 \exp (-n / 2).
\]
\end{lemma}

\begin{lemma}\label{lemma1}
Consider $d$-dimensional random vectors ${\bf z}_1, \ldots, {\bf z}_n$ sampled from a stationary time series $\{ {\bf z}_t\}_{t \in \mathbb{N}}$, where ${\bf z}_t \in \mathbb{R}^d$ and $E({\bf z}_t) = 0$. Suppose that $\{ {\bf z}_t\}_{t \in \mathbb{N}}$ satisfies Assumption 1 with the spectral density function $f(\cdot)$. For any non-negative integer $h$, let
$\Gamma(h) = E({\bf z}_t{\bf z}_{t+h}^\top)$, $\widehat{\Gamma}(h) = (n-h)^{-1}\sum_{t=1}^{n-h} {\bf z}_t{\bf z}_{t+h}^\top$, $\Gamma(-h) = \Gamma(h)^\top$ and $\widehat{\Gamma}(-h) = \widehat{\Gamma}(h)^\top$.
If $d > 1$ and $n \geq 24\log d + h$,
then with probability at least $1 - 8d^{-1}$, we have
\begin{align*}
\|\widehat{\Gamma}(h) - \Gamma(h)\|_{\max} &~ \leq 32\pi \left(\frac{6\log d}{n - h}\right)^{1/2}  \mathcal{M}(f), \\
\|\widehat{\Gamma}(-h) - \Gamma(-h)\|_{\max} &~ \leq 32\pi \left(\frac{6\log d}{n - h}\right)^{1/2}  \mathcal{M}(f),
\end{align*}
where
$\mathcal{M}(f)$ is defined in Assumption 1.
\end{lemma}

\begin{proof}
 Since $\Gamma(-h) = \Gamma(h)^\top$,
 we only prove the first inequality.
 For $j,k = 1, \ldots, d$, let $\sigma_{jk,h}$ and $\widehat{\sigma}_{jk,h}$, respectively, denote the $(j,k)$-th entry of $\Gamma(h)$ and $\widehat{\Gamma}(h)$.
 Then, for all $\eta > 0$, we have
 \begin{align*}%
    &~ \mbox{pr}\left(\left|\sigma_{jk,h} - \widehat{\sigma}_{jk,h}\right| > \eta\right) = \mbox{pr}\left(\left|(n-h)^{-1}\sum_{t = 1}^{n-h} {z}_{tj} {z}_{t+h,k} - \sigma_{jk,h}\right| > \eta  \right) \nonumber \\
    = &~ \mbox{pr} \left( \left|\{4(n-h)\}^{-1} \left\{\sum_{t = 1}^{n-h} ({z}_{tj} + {z}_{t+h,k})^2  - \sum_{t = 1}^{n-h} ({z}_{tj} - {z}_{t+h,k})^2 \right\} - \sigma_{jk,h} \right|> \eta  \right) \nonumber \\
    \leq & ~ \mbox{pr}\left( \left|(n-h)^{-1} \sum_{t = 1}^{n-h} ({z}_{tj} + {z}_{t+h,k})^2 - (\sigma_{jj,0} + \sigma_{kk,0} + 2\sigma_{jk,h})\right|> 2\eta \right) \nonumber \\
    + & ~ \mbox{pr}\left( \left|(n-h)^{-1} \sum_{t = 1}^{n-h} ({z}_{tj} - {z}_{t+h,k})^2 - (\sigma_{jj,0} + \sigma_{kk,0} - 2\sigma_{jk,h})\right|> 2\eta \right).
 \end{align*}
 Note that $({z}_{1j} + {z}_{1+h,k}, {z}_{2j} + {z}_{2+h,k}, \ldots, {z}_{n-h,j} + {z}_{n,k})^\top \sim N(0,Q_h)$ %
 where for $r,s = 1, \ldots, n-h$, the $(r,s)$-th entry of $Q_h$ takes the following form:
 \begin{align*}%
 Q_{rs,h} = &~ \mbox{cov}({z}_{rj} + {z}_{r+h,k},{z}_{sj} + {z}_{s+h,k}) \nonumber \\
 = &~ \Gamma(s-r)_{jj} + \Gamma(s-r+h)_{jk} + \Gamma(s-r)_{kk} + \Gamma(s-r-h)_{kj} \nonumber \\
 = &~ {\bf v}_j^\top \Gamma(s-r) {\bf v}_j + {\bf v}_j^\top \Gamma(s-r+h) {\bf v}_k + {\bf v}_k^\top \Gamma(s-r) {\bf v}_k + {\bf v}_k^\top \Gamma(s-r-h) {\bf v}_j %
 \end{align*}
 and where
 ${\bf v}_j \in \mathbb{R}^d$ is a vector with a one in its $j$-th entry and zeros elsewhere.
For any ${\bf w} \in \mathbb{R}^{n-h}$ with $\|{\bf w}\|_2 = 1$, we have
\begin{align}\label{lemma1:pf3}
    {\bf w}^\top Q_h{\bf w} = &~ \sum_{r= 1}^{n-h} \sum_{s = 1}^{n-h} w_rw_s Q_{rs,h} \nonumber \\
    = &~ \sum_{r=1}^{n-h}\sum_{s=1}^{n-h} w_rw_s {\bf v}_j^\top \Gamma(s-r){\bf v}_j  + \sum_{r=1}^{n-h}\sum_{s=1}^{n-h} w_rw_s {\bf v}_k^\top \Gamma(s-r){\bf v}_k + \nonumber \\ &~ \sum_{r=1}^{n-h}\sum_{s=1}^{n-h} w_rw_s {\bf v}_j^\top \Gamma(s-r+h){\bf v}_k +
    \sum_{r=1}^{n-h}\sum_{s=1}^{n-h} w_rw_s {\bf v}_k^\top \Gamma(s-r-h){\bf v}_j.
\end{align}
Letting
\[
\widetilde{\bf w}_1 = \left( {\bf w}^{\top}, \underbrace{0, \ldots, 0}_{h} \right)^\top, ~ \widetilde{\bf w}_2 = \left( \underbrace{0, \ldots, 0}_{h}, {\bf w}^\top \right)^\top,
\]
one can check that
\begin{align*}
 &~ \sum_{r=1}^{n-h}\sum_{s=1}^{n-h} w_rw_s {\bf v}_j^\top \Gamma(s-r){\bf v}_j = \left(\widetilde{\bf w}_1 \otimes {\bf v}_j\right)^\top \Upsilon_n \left(\widetilde{\bf w}_1 \otimes {\bf v}_j\right),   \\
  &~ \sum_{r=1}^{n-h}\sum_{s=1}^{n-h} w_rw_s {\bf v}_k^\top \Gamma(s-r){\bf v}_k = \left(\widetilde{\bf w}_1 \otimes {\bf v}_k\right)^\top \Upsilon_n \left(\widetilde{\bf w}_1 \otimes {\bf v}_k\right),   \\
   &~ \sum_{r=1}^{n-h}\sum_{s=1}^{n-h} w_rw_s {\bf v}_j^\top \Gamma(s-r+h){\bf v}_k = \left(\widetilde{\bf w}_1 \otimes {\bf v}_j\right)^\top \Upsilon_n \left(\widetilde{\bf w}_2 \otimes {\bf v}_k\right),   \\
   &~ \sum_{r=1}^{n-h}\sum_{s=1}^{n-h} w_rw_s {\bf v}_k^\top \Gamma(s-r-h){\bf v}_j = \left(\widetilde{\bf w}_2 \otimes {\bf v}_k\right)^\top \Upsilon_n \left(\widetilde{\bf w}_1 \otimes {\bf v}_j\right).
\end{align*}
Since $\|\widetilde{\bf w}_1 \otimes {\bf v}_j\|_2 = 1, \|\widetilde{\bf w}_1 \otimes {\bf v}_k\|_2 = 1, \|\widetilde{\bf w}_2 \otimes {\bf v}_j\|_2 = 1$, and $\|\widetilde{\bf w}_2 \otimes {\bf v}_k\|_2 = 1$, we have
\[
{\bf w}^\top Q_h {\bf w} \leq 4\lambda_{\max}(\Upsilon_n),
\]
where $\lambda_{\max}(\Upsilon_n)$ denotes the largest eigenvalue of $\Upsilon_n$.
Thus, using Lemma \ref{s1} in Section S.2 of the appendix, we get $\|Q_h\|_2 \leq 4\lambda_{\max}(\Upsilon_n) \leq 8\pi\mathcal{M}(f)$,
where $\mathcal{M}(f)$ is defined in Assumption 1. Then, using Lemma \ref{s2} in Section S.2, for $\eta \geq 32\pi (n-h)^{-1/2}\mathcal{M}(f)$, we get
\begin{align}\label{lemma1:pf5}
  &~\mbox{pr}\left( \left|(n-h)^{-1} \sum_{t = 1}^{n-h} ({z}_{tj} + {z}_{t+h,k})^2 - (\sigma_{jj,h} + \sigma_{kk,h} + 2\sigma_{jk,h})\right|> 2\eta \right) \nonumber \\
  \leq &~ 2\exp\left[ -\frac{n-h}{2}\left\{ \frac{\eta}{16\pi\mathcal{M}(f)} - 2(n-h)^{-1/2}\right\}^2 \right] + 2\exp\left( -\frac{n-h}{2}\right).
\end{align}
Using a similar argument, we have
\begin{align}\label{lemma1:pf6}
  &~\mbox{pr}\left( \left|(n-h)^{-1} \sum_{t = 1}^{n-h} ({z}_{tj} - {z}_{t+h,k})^2 - (\sigma_{jj,h} + \sigma_{kk,h}-2\sigma_{jk,h})\right|> 2\eta \right) \nonumber \\
  \leq &~ 2\exp\left[ -\frac{n-h}{2}\left\{ \frac{\eta}{16\pi\mathcal{M}(f)} - 2(n-h)^{-1/2}\right\}^2 \right] + 2\exp\left( -\frac{n-h}{2}\right).
\end{align}
Combining (\ref{lemma1:pf5}) and (\ref{lemma1:pf6}) and applying the union bound, we have
\begin{align}\label{lemma1:pf7}
    \mbox{pr}\left(\|\Gamma(h) - \widehat{\Gamma}(h)\|_{\max} > \eta\right) \leq &~ 4d^2\exp\left(-\frac{n-h}{2}\right) \nonumber \\
    + &~ 4d^2\exp\left[ -\frac{n-h}{2}\left\{ \frac{\eta}{16\pi\mathcal{M}(f)} - 2(n-h)^{-1/2}\right\}^2 \right].
\end{align}
Taking $\eta = 32\pi\mathcal{M}(f)\{6(n-h)^{-1}\log d\}^{1/2}$, when $d > 1$ and $n \geq 24\log d + h$, it can be checked that $6\log d \geq 4$ and
\[
1 \geq \left|\frac{\eta}{16\pi\mathcal{M}(f)} - 2(n-h)^{-1/2}\right| \geq \left(\frac{6\log d}{n-h}\right)^{1/2}.
\]
Therefore, we have
\begin{align*}%
    \mbox{pr}\left(\|\Gamma(h) - \widehat{\Gamma}(h)\|_{\max} > \eta\right) \leq
    8d^2\exp\left[ -\frac{(n-h)}{2}\left\{ \frac{\eta}{16\pi\mathcal{M}(f)} - 2(n-h)^{-1/2}\right\}^2 \right]
    \leq  8d^{-1}.
\end{align*}
This completes the proof.
\end{proof}

Our theoretical analysis relies on the unified framework for
high-dimensional \(M\)-estimators with decomposable regularizers
developed by \citet{negahban2012unified}.
Consider the generic \(l_1\)-penalized \(M\)-estimator
\begin{equation}
\label{nega:1}
\widehat{\ve\theta}(\lambda)
=
\argmin_{\ve\theta}
\left\{
L_n(\ve\theta)
+
\lambda\|\ve\theta\|_1
\right\},
\end{equation}
where
\(\lambda>0\)
is a tuning parameter and
\(L_n(\ve\theta)\)
is a twice differentiable loss function.
Let
\(\ve\theta^*\)
denote the minimizer of the population risk
\(E\{L_n(\ve\theta)\}\),
and let
\(S^*\)
denote the support of
\(\ve\theta^*\).

The framework of \citet{negahban2012unified}
requires two ingredients:
(i) a decomposable regularizer and
(ii) a restricted strong convexity (RSC) condition on the loss function.
The \(l_1\)-norm is decomposable in the sense of Example 1 of
\citet{negahban2012unified}.
Since all loss functions considered in this paper are twice differentiable,
we adopt the following restricted eigenvalue condition, which is a special
case of the RSC condition.

\begin{assumption}
\label{assump:RE}
There exists a constant
\(\kappa_1>0\)
such that
\[
\ve\theta^\top
\nabla^2 L_n(\ve\theta)
\ve\theta
\ge
\kappa_1
\|\ve\theta\|_2^2,
\qquad
\forall\,
\ve\theta\in\mathcal C(S^*),
\]
where
\[
\mathcal C(S^*)
=
\left\{
\ve\theta:
\|\ve\theta_{S^{*c}}\|_1
\le
3\|\ve\theta_{S^*}\|_1
\right\}.
\]
\end{assumption}

Under Assumption \ref{assump:RE}, the estimation error of
\eqref{nega:1}
can be controlled by the magnitude of the score function evaluated at
\(\ve\theta^*\).
The following result is a direct consequence of Lemma 1 and Theorem 1 of \citet{negahban2012unified}.

\begin{lemma}
\label{lemma4}
Suppose Assumption \ref{assump:RE} holds.
If
\[
\lambda
\ge
2
\|
\nabla L_n(\ve\theta^*)
\|_{\max},
\]
then
\[
\widehat{\ve\theta}(\lambda)
-
\ve\theta^*
\in
\mathcal C(S^*),
\]
and
\[
\|
\widehat{\ve\theta}(\lambda)
-
\ve\theta^*
\|_{2}
\le
\frac{
6\lambda
}{
\kappa_1
}
\Psi(S^*),
\]
where
\[
\Psi(S^*)
=
\sup_{
\substack{
\text{supp}(\ve\theta)\in\mathcal S^*, \ve\theta \neq 0
\\
\|\ve\theta\|_2=1
}
}
\|\ve\theta\|_1
\]
is the subspace compatibility constant.
\end{lemma}

\section{Proofs for DiffNet estimation}\label{sec:appenB}

Recall that
\[
C_M
=
\max
\left\{
\mathcal M(f_1),
\mathcal M(f_2)
\right\},
\qquad
C_\Sigma
=
\max
\left\{
\|\Sigma_1\|_{\max},
\|\Sigma_2\|_{\max}
\right\},
\]
and
\[
\xi(n_1,n_2,d,p)
=
C_M
\left(
\frac{
6\log d
}{
\min(n_1,n_2)-p
}
\right)^{1/2}.
\]

The following lemma shows that
\(\widehat{\Sigma}_1\otimes\widehat{\Sigma}_2\)
satisfies a modified restricted eigenvalue condition in the sense of
\citet{loh2012high}
with high probability.

\begin{lemma}
\label{lemma2}
Suppose that
${\bf x}_t^{(l)}$
satisfies Assumption 1 with spectral density function
$f_l(\cdot)$,
$l=1,2$.
If
\[
\min(n_1,n_2)
\ge
\max
\left\{
6144\pi^2
\max
\left(
\frac{\mathcal M^2(f_1)}
{\|\Sigma_1\|_{\max}^2},
\frac{\mathcal M^2(f_2)}
{\|\Sigma_2\|_{\max}^2}
\right),
1
\right\}
\log d
+
(p-1),
\]
then, with probability at least
\(1-16d^{-1}\),
\[
{\bf m}^\top
\left(
\widehat{\Sigma}_1
\otimes
\widehat{\Sigma}_2
\right)
{\bf m}
\ge
\lambda_{\min}(\Sigma_1)
\lambda_{\min}(\Sigma_2)
\|{\bf m}\|_2^2
-
64\pi
C_\Sigma
\xi(n_1,n_2,d,p-1)
\|{\bf m}\|_1^2
\]
for all
\({\bf m}\in\mathbb R^{d^2}\).
\end{lemma}

\begin{proof}
Under the reformulated VAR($p$) model,
\[
\widetilde{\bf x}_t^{(l)}
=
\widetilde A^{(l)\top}
\widetilde{\bf x}_{t-1}^{(l)}
+
\widetilde{\ve\epsilon}_t^{(l)},
\qquad
l=1,2,
\]
the covariance matrices
\(\Sigma_l\)
and
\(\widehat{\Sigma}_l\)
have the block Toeplitz representations
\[
\Sigma_l
=
\begin{bmatrix}
\Gamma(0) & \Gamma(-1) & \cdots & \Gamma(1-p)
\\
\Gamma(1) & \Gamma(0) & \cdots & \Gamma(2-p)
\\
\vdots & \vdots & \ddots & \vdots
\\
\Gamma(p-1) & \Gamma(p-2) & \cdots & \Gamma(0)
\end{bmatrix}
\]
and
\[
\widehat{\Sigma}_l
=
\begin{bmatrix}
\widehat{\Gamma}(0) & \widehat{\Gamma}(-1) & \cdots & \widehat{\Gamma}(1-p)
\\
\widehat{\Gamma}(1) & \widehat{\Gamma}(0) & \cdots & \widehat{\Gamma}(2-p)
\\
\vdots & \vdots & \ddots & \vdots
\\
\widehat{\Gamma}(p-1) & \widehat{\Gamma}(p-2) & \cdots & \widehat{\Gamma}(0)
\end{bmatrix}.
\]
Consequently,
\[
\left\|\widehat{\Sigma}_l-\Sigma_l\right\|_{\max}
=
\max_{|h|\le p-1}
\left\|\widehat{\Gamma}(h)-\Gamma(h)\right\|_{\max}.
\]

Using the identity
\[
\widehat{\Sigma}_1\otimes\widehat{\Sigma}_2
-
\Sigma_1\otimes\Sigma_2
=
\Sigma_1\otimes\left(\widehat{\Sigma}_2-\Sigma_2\right)
+
\Sigma_2\otimes\left(\widehat{\Sigma}_1-\Sigma_1\right)
+
\left(\widehat{\Sigma}_1-\Sigma_1\right)
\otimes
\left(\widehat{\Sigma}_2-\Sigma_2\right),
\]
we obtain
\begin{align}
\label{lemma2:pf1}
\|
\widehat{\Sigma}_1\otimes\widehat{\Sigma}_2
-
\Sigma_1\otimes\Sigma_2
\|_{\max}
\le
&
\|\Sigma_1\|_{\max}
\|\widehat{\Sigma}_2-\Sigma_2\|_{\max}
\nonumber
\\
&
+
\|\Sigma_2\|_{\max}
\|\widehat{\Sigma}_1-\Sigma_1\|_{\max}
\nonumber
\\
&
+
\|\widehat{\Sigma}_1-\Sigma_1\|_{\max}
\|\widehat{\Sigma}_2-\Sigma_2\|_{\max}.
\end{align}

By Lemma \ref{lemma1}, there exists an event
\(\mathcal A_1\)
with
\[
\Pr(\mathcal A_1)
\ge
1-16d^{-1}
\]
such that
\begin{equation}
\label{lemma2:pf2}
\|\Sigma_1\|_{\max}
\|\widehat{\Sigma}_2-\Sigma_2\|_{\max}
+
\|\Sigma_2\|_{\max}
\|\widehat{\Sigma}_1-\Sigma_1\|_{\max}
\le
64\pi
C_\Sigma
\xi(n_1,n_2,d,p-1).
\end{equation}

Moreover, on
\(\mathcal A_1\),
\begin{equation}
\label{lemma2:pf3}
\|\widehat{\Sigma}_1-\Sigma_1\|_{\max}
\,
\|\widehat{\Sigma}_2-\Sigma_2\|_{\max}
\le
6144\pi^2
\mathcal M(f_1)
\mathcal M(f_2)
\log d
\left\{
(n_1-p+1)(n_2-p+1)
\right\}^{-1/2}.
\end{equation}

Under the stated sample size condition, the right-hand side of
\eqref{lemma2:pf3}
is bounded above by $32\pi
C_\Sigma
\xi(n_1,n_2,d,p-1)$.
Combining
\eqref{lemma2:pf1}--\eqref{lemma2:pf3}
therefore yields
\[
\|
\widehat{\Sigma}_1\otimes\widehat{\Sigma}_2
-
\Sigma_1\otimes\Sigma_2
\|_{\max}
\le
96\pi
C_\Sigma
\xi(n_1,n_2,d,p-1).
\]

Consequently, for any
\({\bf m}\in\mathbb R^{d^2}\),
\begin{align*}
{\bf m}^\top
\left(\widehat{\Sigma}_1\otimes\widehat{\Sigma}_2\right)
{\bf m}
\ge
&
{\bf m}^\top
(\Sigma_1\otimes\Sigma_2)
{\bf m}
-
\|
\widehat{\Sigma}_1\otimes\widehat{\Sigma}_2
-
\Sigma_1\otimes\Sigma_2
\|_{\max}
\|{\bf m}\|_1^2
\\
\ge
&
\lambda_{\min}(\Sigma_1)
\lambda_{\min}(\Sigma_2)
\|{\bf m}\|_2^2
-
64\pi
C_\Sigma
\xi(n_1,n_2,d,p-1)
\|{\bf m}\|_1^2,
\end{align*}
where we use the identity
\[
\lambda_{\min}
(\Sigma_1\otimes\Sigma_2)
=
\lambda_{\min}(\Sigma_1)
\lambda_{\min}(\Sigma_2).
\]
This completes the proof.
\end{proof}

The next lemma establishes a uniform bound on the gradient of the
D-trace loss function evaluated at the true DiffNet
parameter \(\Delta_\Omega\).

\begin{lemma}
\label{lemma3}
Under the conditions of Lemma \ref{lemma2}, with probability at least
\(1-16d^{-1}\),
\[
\left\|
\frac{1}{2}
\left(
\widehat{\Sigma}_1
\Delta_\Omega
\widehat{\Sigma}_2
+
\widehat{\Sigma}_2
\Delta_\Omega
\widehat{\Sigma}_1
\right)
-
\left(
\widehat{\Sigma}_2
-
\widehat{\Sigma}_1
\right)
\right\|_{\max}
\le
96\pi
\xi(n_1,n_2,d,p-1)
\left(
1+C_\Sigma|\Delta_\Omega|_1
\right). %
\]
\end{lemma}

\begin{proof}
Define
\[
H
=
\frac{1}{2}
\left(
\Sigma_2\otimes\Sigma_1
+
\Sigma_1\otimes\Sigma_2
\right)
\]
and
\[
\widehat{H}
=
\frac{1}{2}
\left(
\widehat{\Sigma}_2\otimes\widehat{\Sigma}_1
+
\widehat{\Sigma}_1\otimes\widehat{\Sigma}_2
\right).
\]
Then
\[
\left\|
\frac{1}{2}
\left(
\widehat{\Sigma}_1
\Delta_\Omega
\widehat{\Sigma}_2
+
\widehat{\Sigma}_2
\Delta_\Omega
\widehat{\Sigma}_1
\right)
-
\left(
\widehat{\Sigma}_2
-
\widehat{\Sigma}_1
\right)
\right\|_{\max}
=
\left\|
\widehat{H}
\,\mathrm{vec}(\Delta_\Omega)
-
\left\{
\mathrm{vec}\left(\widehat{\Sigma}_2\right)
-
\mathrm{vec}\left(\widehat{\Sigma}_1\right)
\right\}
\right\|_{\max}. %
\]

By equation \eqref{diffest:add:1} in the main text,
\[
H
\,\mathrm{vec}(\Delta_\Omega)
=
\mathrm{vec}(\Sigma_2)
-
\mathrm{vec}(\Sigma_1). %
\]
Therefore,
\begin{align*}
&
\left\|
\widehat{H}
\,\mathrm{vec}(\Delta_\Omega)
-
\left\{
\mathrm{vec}\left(\widehat{\Sigma}_2\right)
-
\mathrm{vec}\left(\widehat{\Sigma}_1\right)
\right\}
\right\|_{\max} %
\\
\le
&
\left\|
(\widehat{H}-H)
\,\mathrm{vec}(\Delta_\Omega)
\right\|_{\max}
+
\sum_{l=1}^{2}
\|
\widehat{\Sigma}_l-\Sigma_l
\|_{\max}
\\
\le
&
\|
\widehat{H}-H
\|_{\max}
|\Delta_\Omega|_1
+
\sum_{l=1}^{2}
\|
\widehat{\Sigma}_l-\Sigma_l
\|_{\max}.
\end{align*}

By Lemma \ref{lemma1},
\[
\sum_{l=1}^{2}
\|
\widehat{\Sigma}_l-\Sigma_l
\|_{\max}
\le
64\pi
\xi(n_1,n_2,d,p-1)
\]
with probability at least
\(1-16d^{-1}\).

Moreover, arguing exactly as in the proof of Lemma \ref{lemma2},
\[
\|
\widehat{H}-H
\|_{\max}
\le
96\pi
C_\Sigma
\xi(n_1,n_2,d,p-1)
\]
with probability at least
\(1-16d^{-1}\).

Combining the above bounds yields
\[
\left\|
\frac{1}{2}
\left(
\widehat{\Sigma}_1
\Delta_\Omega
\widehat{\Sigma}_2
+
\widehat{\Sigma}_2
\Delta_\Omega
\widehat{\Sigma}_1
\right)
-
\left(
\widehat{\Sigma}_2
-
\widehat{\Sigma}_1
\right)
\right\|_{\max}
\le
96\pi
\xi(n_1,n_2,d,p-1)
\left(
1+C_\Sigma|\Delta_\Omega|_1
\right),
\]
which completes the proof.
\end{proof}

Recall that the Hessian matrix of the loss function
$L_A(\cdot)$
with respect to
$\Delta_A$
is
$\Sigma_1+\Sigma_2$.
The following lemma establishes a restricted eigenvalue condition for its sample counterpart
$\widehat{\Sigma}_1+\widehat{\Sigma}_2$.

\begin{lemma}
\label{lemma6}
Suppose that
${\bf x}_t^{(l)}$
satisfies Assumption 1 with spectral density function
$f_l(\cdot)$,
$l=1,2$.
If
\[
\min(n_1,n_2)
\ge
24\log d + (p-1),
\]
then, with probability at least
\(1-16d^{-1}\),
\[
{\bf m}^\top
\left(
\widehat{\Sigma}_1
+
\widehat{\Sigma}_2
\right)
{\bf m}
\ge
\left\{
\lambda_{\min}(\Sigma_1)
+
\lambda_{\min}(\Sigma_2)
\right\}
\|{\bf m}\|_2^2
-
64\pi
\xi(n_1,n_2,d,p-1)
\|{\bf m}\|_1^2
\]
for all
\({\bf m}\in\mathbb R^d\).
\end{lemma}

\begin{proof}
By Lemma \ref{lemma1}, on the event
\(\mathcal A_1\),
\[
\|\widehat{\Sigma}_1-\Sigma_1\|_{\max}
+
\|\widehat{\Sigma}_2-\Sigma_2\|_{\max}
\le
64\pi
\xi(n_1,n_2,d,p-1).
\]

Therefore, for any
\({\bf m}\in\mathbb R^d\),
\begin{align*}
{\bf m}^\top
\left(
\widehat{\Sigma}_1
+
\widehat{\Sigma}_2
\right)
{\bf m}
=&~
{\bf m}^\top
\left(
\Sigma_1
+
\Sigma_2
\right)
{\bf m}
\\
&
+
{\bf m}^\top
\left[
\left(\widehat{\Sigma}_1-\Sigma_1\right)
+
\left(\widehat{\Sigma}_2-\Sigma_2\right)
\right]
{\bf m}
\\
\ge
&
\left\{
\lambda_{\min}(\Sigma_1)
+
\lambda_{\min}(\Sigma_2)
\right\}
\|{\bf m}\|_2^2
\\
&
-
\left(
\|\widehat{\Sigma}_1-\Sigma_1\|_{\max}
+
\|\widehat{\Sigma}_2-\Sigma_2\|_{\max}
\right)
\|{\bf m}\|_1^2
\\
\ge
&
\left\{
\lambda_{\min}(\Sigma_1)
+
\lambda_{\min}(\Sigma_2)
\right\}
\|{\bf m}\|_2^2
-
64\pi
\xi(n_1,n_2,d,p-1)
\|{\bf m}\|_1^2.
\end{align*}
This completes the proof.
\end{proof}

We next prove Theorem \ref{thm1} based on the above lemmas.

\begin{proof}
As our proof primarily relies on the unified high-dimensional $M$-estimator theory in Lemma \ref{lemma4},
we first verify Assumption \ref{assump:RE}.

Recall that the Hessian matrix of the D-trace loss function is
\[
\frac{1}{2}
\left(
\widehat{\Sigma}_1\otimes\widehat{\Sigma}_2
+
\widehat{\Sigma}_2\otimes\widehat{\Sigma}_1
\right).
\]

Let
\[
\mathcal C(S_\Omega)
=
\left\{
{\bf m}\in\mathbb R^{d^2}
:
\|{\bf m}_{S_\Omega^c}\|_1
\le
3\|{\bf m}_{S_\Omega}\|_1
\right\}.
\]
For any
\({\bf m}\in\mathcal C(S_\Omega)\),
\[
\|{\bf m}\|_1
\le
4\|{\bf m}_{S_\Omega}\|_1
\le
4\sqrt{s_\Omega}
\|{\bf m}\|_2.
\]

By Lemma \ref{lemma2}, on the event
\(\mathcal A_1\),
\begin{align*}
&
{\bf m}^\top
\left\{
\frac{1}{2}
(
\widehat{\Sigma}_1\otimes\widehat{\Sigma}_2
+
\widehat{\Sigma}_2\otimes\widehat{\Sigma}_1
)
\right\}
{\bf m}
\\
\ge
&
\left\{
\lambda_{\min}(\Sigma_1)
\lambda_{\min}(\Sigma_2)
-
1024\pi
s_\Omega
C_\Sigma
\xi(n_1,n_2,d,p-1)
\right\}
\|{\bf m}\|_2^2.
\end{align*}

Define
\begin{equation}
\label{CG}
C_G
=
6
\left\{
3072\pi
C_M
C_\Sigma
\lambda_{\min}^{-1}(\Sigma_1)
\lambda_{\min}^{-1}(\Sigma_2)
\right\}^2.
\end{equation}

If
\(
\min(n_1,n_2)
\ge
C_G
s_\Omega^2
\log d
+
(p-1),
\)
then
\[
1536\pi
s_\Omega
C_\Sigma
\xi(n_1,n_2,d,p-1)
\le
\frac{1}{2}
\lambda_{\min}(\Sigma_1)
\lambda_{\min}(\Sigma_2),
\]
which implies
\[
{\bf m}^\top
\left\{
\frac{1}{2}
(
\widehat{\Sigma}_1\otimes\widehat{\Sigma}_2
+
\widehat{\Sigma}_2\otimes\widehat{\Sigma}_1
)
\right\}
{\bf m}
\ge
\frac{1}{2}
\lambda_{\min}(\Sigma_1)
\lambda_{\min}(\Sigma_2)
\|{\bf m}\|_2^2.
\]

Hence Assumption \ref{assump:RE} holds with
\[
\kappa_1
=
\frac{1}{2}
\lambda_{\min}(\Sigma_1)
\lambda_{\min}(\Sigma_2)
\]
and
\(
S^*
=
S_\Omega.
\)

Furthermore, recall that
\[
\nabla
L_D
(
\Delta_\Omega,
\widehat{\Sigma}_1,
\widehat{\Sigma}_2
)
=
\frac{1}{2}
\left(
\widehat{\Sigma}_1
\Delta_\Omega
\widehat{\Sigma}_2
+
\widehat{\Sigma}_2
\Delta_\Omega
\widehat{\Sigma}_1
\right)
-
(
\widehat{\Sigma}_2
-
\widehat{\Sigma}_1
).
\]

By Lemma \ref{lemma3},
\[
\nu
\ge
2
\left\|
\nabla
L_D
(
\Delta_\Omega,
\widehat{\Sigma}_1,
\widehat{\Sigma}_2
)
\right\|_{\max}.
\]

With the assumptions verified,
we now applying Lemma \ref{lemma4} with
\[
\Psi(S_\Omega)
=
\sqrt{s_\Omega},
\]
we obtain
\[
\|
\widehat{\Delta}_\Omega(\nu)
-
\Delta_\Omega
\|_{\mathrm F}
\le
\frac{
12\nu\sqrt{s_\Omega}
}{
\lambda_{\min}(\Sigma_1)
\lambda_{\min}(\Sigma_2)
}.
\]

This completes the proof.
\end{proof}

Based on Theorem \ref{thm1}, we next prove Theorem \ref{thm2}.

\begin{proof}
By Theorem \ref{thm1}, with probability at least
\(1-16d^{-1}\),
\[
\|\widehat{\Delta}_\Omega(\nu)-\Delta_\Omega\|_{\max}
\le
12s_\Omega^{1/2}
\lambda_{\min}^{-1}(\Sigma_1)
\lambda_{\min}^{-1}(\Sigma_2)
\nu.
\]
Since
\[
\tau_\Omega
\ge
12s_\Omega^{1/2}
\lambda_{\min}^{-1}(\Sigma_1)
\lambda_{\min}^{-1}(\Sigma_2)
\nu,
\]
it follows that, with probability at least
\(1-16d^{-1}\),
\[
\|\widehat{\Delta}_\Omega(\nu)-\Delta_\Omega\|_{\max}
\le
\tau_\Omega.
\]

We now show that
\[
\operatorname{sgn}
\!\left(
\mathrm{HT}_{\tau_\Omega}
\left(
\widehat{\Delta}_\Omega(\nu)
\right)
\right)
=
\operatorname{sgn}
(\Delta_\Omega).
\]

First consider
\((j,k)\)
such that
\[
(\Delta_\Omega)_{jk}=0.
\]
Then
\[
\left|
\widehat{\Delta}_{\Omega,jk}(\nu)
\right|
=
\left|
\widehat{\Delta}_{\Omega,jk}(\nu)
-
(\Delta_\Omega)_{jk}
\right|
\le
\tau_\Omega.
\]
Hence,
\[
\mathrm{HT}_{\tau_\Omega}
\left(
\widehat{\Delta}_{\Omega,jk}(\nu)
\right)
=
0.
\]

Next consider
\((j,k)\)
such that
\[
(\Delta_\Omega)_{jk}\neq 0.
\]
By assumption,
\[
|(\Delta_\Omega)_{jk}|
\ge
\Delta_{\Omega,\min}
>
2\tau_\Omega,
\]
where $\Delta_{\Omega,\min} = \min_{(j,k)\in S_\Omega}|(\Delta_\Omega)_{jk}|$.
Using
\[
\|\widehat{\Delta}_\Omega(\nu)-\Delta_\Omega\|_{\max}
\le
\tau_\Omega,
\]
we obtain
\[
\left|
\widehat{\Delta}_{\Omega,jk}(\nu)
\right|
\ge
|(\Delta_\Omega)_{jk}|
-
\left|
\widehat{\Delta}_{\Omega,jk}(\nu)
-
(\Delta_\Omega)_{jk}
\right|
>
2\tau_\Omega-\tau_\Omega
=
\tau_\Omega.
\]
Therefore,
\[
\mathrm{HT}_{\tau_\Omega}
\left(
\widehat{\Delta}_{\Omega,jk}(\nu)
\right)
\neq 0.
\]

Moreover,
\[
\left|
\widehat{\Delta}_{\Omega,jk}(\nu)
-
(\Delta_\Omega)_{jk}
\right|
<
\frac12
|(\Delta_\Omega)_{jk}|,
\]
which implies
\[
\operatorname{sgn}
\left(
\widehat{\Delta}_{\Omega,jk}(\nu)
\right)
=
\operatorname{sgn}
\left(
(\Delta_\Omega)_{jk}
\right).
\]
Since thresholding does not alter the sign of a nonzero entry whose magnitude exceeds
\(\tau_\Omega\),
we conclude that
\[
\operatorname{sgn}
\left(
\mathrm{HT}_{\tau_\Omega}
\left(
\widehat{\Delta}_\Omega(\nu)
\right)
\right)
=
\operatorname{sgn}
(\Delta_\Omega)
\]
with probability at least
\(1-16d^{-1}\).
This completes the proof.
\end{proof}

\section{Proofs for DiffGC estimation}\label{sec:appenC}
Recall that
\(\ve\beta_j\)
and
\({\bf w}_j\)
respectively denote the \(j\)-th column of
\(\Delta_A\)
and
\(
\Sigma_1\Delta_\Omega\Theta_2
+
\Sigma_2\Delta_\Omega\Theta_1
+
2(\Theta_1-\Theta_2).
\)
The next lemma provides a uniform bound on the gradient of
\(L_A(\cdot)\)
evaluated at the true DiffGC matrix.
To simplify the presentation, we collect the sample size conditions
required in this lemma:
\begin{align}
\label{n:cond}
&~
\min(n_1,n_2)
\ge
C_G s_\Omega^2 \log d + p,
\nonumber
\\
&~
\min(n_1,n_2)
\ge
16s_\Omega C_\Omega^2
\log d
|\Delta_\Omega|_1^{-2}
+
p,
\nonumber
\\
&~
n_2
\ge
6144\pi^2
\mathcal M^2(f_2)
\log d
/
\|\Theta_2\|^2_{\max}
+
p,
\nonumber
\\
&~
n_1
\ge
6144\pi^2
\mathcal M^2(f_1)
\log d
/
\|\Theta_1\|^2_{\max}
+
p,
\nonumber
\\
&~
\min(n_1,n_2)
\ge
24\log d + p,
\nonumber \\
&~ \min(n_1, n_2) \geq \left(\frac{2048 \pi \max_js_{A,j} C_M}{\lambda_{\min}(\Sigma_1) + \lambda_{\min}(\Sigma_2)}\right)^2 \times 6\log d + p %
\end{align}
where $C_M$ is defined in Appendix \ref{sec:appenB},
\begin{align*}
C_G = &~ 6\times \left\{3072\pi C_MC_\Sigma \lambda_{\min}^{-1}(\Sigma_1)\lambda_{\min}^{-1}(\Sigma_2)\right\}^2, \\ C_\Omega = &~ 2304\sqrt{6}\pi C_M (1 + C_\Sigma|\Delta_\Omega|_1) \sqrt{s_\Omega} \lambda_{\min}^{-1}(\Sigma_1)\lambda_{\min}^{-1}(\Sigma_2).
\end{align*}

\begin{lemma}
\label{lemma7}
Suppose that
${\bf x}_t^{(l)}$
satisfies Assumption 1 with spectral density function
$f_l(\cdot)$,
\(l=1,2\).
If
\(n_1\)
and
\(n_2\)
satisfy Condition~\eqref{n:cond},
then, with probability at least
\(1-32d^{-1}\),
\[
\max_{1\le j\le d}
\left\|
\left(\widehat{\Sigma}_1+\widehat{\Sigma}_2\right)\ve\beta_j
-
\widehat{\bf w}_j
\right\|_{\max}
\le
C_1
\left(
\frac{\log d}
{\min(n_1,n_2)-p}
\right)^{1/2},
\]
for some positive constant
\(C_1\) with its explicit form given in the proof.
\end{lemma}

\begin{proof}
Since
\[
(\Sigma_1+\Sigma_2)\ve\beta_j
-
{\bf w}_j
=
0,
\]
we have
\[
\left\|
\left(\widehat{\Sigma}_1+\widehat{\Sigma}_2\right)\ve\beta_j
-
\widehat{\bf w}_j
\right\|_{\max}
\le
\|
\widehat{\bf w}_j-{\bf w}_j
\|_{\max}
+
\|\ve\beta_j\|_1
\sum_{l=1}^{2}
\|
\widehat{\Sigma}_l-\Sigma_l
\|_{\max}.
\]

By Lemma \ref{lemma1}, on the event
\(\mathcal A_1\),
\[
\max_{1\le j\le d}
\|\ve\beta_j\|_1
\sum_{l=1}^{2}
\|
\widehat{\Sigma}_l-\Sigma_l
\|_{\max}
\le
C_{I_0}
\|\Delta_A\|_1
\left(
\frac{\log d}
{\min(n_1,n_2)-p+1}
\right)^{1/2},
\]
where $C_{I_0} = 64\sqrt{6}\pi C_M.$

It therefore remains to bound
\(\|\widehat{\bf w}_j-{\bf w}_j\|_{\max}\).

Under the reformulated VAR($p$) model,
\[
\Theta_l
=
E
\left(
\widetilde{\bf x}_t^{(l)}
\widetilde{\bf x}_{t+1}^{(l)\top}
\right)
=
\begin{bmatrix}
\Gamma(1) & \Gamma(0) & \cdots & \Gamma(2-p)
\\
\Gamma(2) & \Gamma(1) & \cdots & \Gamma(3-p)
\\
\vdots & \vdots & \ddots & \vdots
\\
\Gamma(p) & \Gamma(p-1) & \cdots & \Gamma(1)
\end{bmatrix},
\]
and similarly for
\(\widehat{\Theta}_l\).
Consequently,
\[
\|
\widehat{\Theta}_l-\Theta_l
\|_{\max}
=
\max_{|h|\le p}
\|
\widehat{\Gamma}(h)-\Gamma(h)
\|_{\max}.
\]

By Lemma \ref{lemma1}, there exists an event
\(\mathcal A_2\)
with
\[
\Pr(\mathcal A_2)
\ge
1-16d^{-1},
\]
such that
\[
\max_{l=1,2}
\|
\widehat{\Theta}_l-\Theta_l
\|_{\max}
\le
32 \pi
\xi(n_1,n_2,d,p).
\]

Conditioning on
\(\mathcal A_1\cap\mathcal A_2\),
we write
\[
\|
\widehat{\Sigma}_1
\widehat{\Delta}_\Omega(\nu)
\widehat{\Theta}_2
-
\Sigma_1
\Delta_\Omega
\Theta_2
\|_{\max}
\le
I_1+I_2+I_3,
\]
where
\begin{align*}
I_1
=&~
\|
\left(\widehat{\Sigma}_1-\Sigma_1\right)
\widehat{\Delta}_\Omega(\nu)
\widehat{\Theta}_2
\|_{\max},
\\
I_2
=&~
\|
\Sigma_1
(
\widehat{\Delta}_\Omega(\nu)-\Delta_\Omega
)
\widehat{\Theta}_2
\|_{\max},
\\
I_3
=&~
\|
\Sigma_1
\Delta_\Omega
(
\widehat{\Theta}_2-\Theta_2
)
\|_{\max}.
\end{align*}

For \(I_1\), we have
\begin{align*}
I_1
\le
&~
\|\widehat{\Sigma}_1-\Sigma_1\|_{\max}
|\widehat{\Delta}_\Omega(\nu)|_1
\|\widehat{\Theta}_2\|_{\max}.
\end{align*}
By Theorem \ref{thm1} and Lemma \ref{lemma4},
\[
|\widehat{\Delta}_\Omega(\nu)-\Delta_\Omega|_1
\le
4\sqrt{s_\Omega}
\|
\widehat{\Delta}_\Omega(\nu)-\Delta_\Omega
\|_F.
\]
Hence,
\[
|\widehat{\Delta}_\Omega(\nu)-\Delta_\Omega|_1
\le
48s_\Omega
\lambda_{\min}^{-1}(\Sigma_1)
\lambda_{\min}^{-1}(\Sigma_2)
\nu.
\]
The second condition in \eqref{n:cond} implies
\[
|\widehat{\Delta}_\Omega(\nu)-\Delta_\Omega|_1
\le
|\Delta_\Omega|_1,
\]
and therefore
\[
|\widehat{\Delta}_\Omega(\nu)|_1
\le
2|\Delta_\Omega|_1.
\]

Moreover, by Lemma \ref{lemma1},
\[
\|\widehat{\Theta}_2-\Theta_2\|_{\max}
\le
32\pi
\mathcal M(f_2)
\left(
\frac{6\log d}
{n_2-p}
\right)^{1/2}.
\]
Combining this bound with the third condition in \eqref{n:cond} yields
\[
\|\widehat{\Theta}_2-\Theta_2\|_{\max}
\le
\|\Theta_2\|_{\max},
\]
and hence
\[
\|\widehat{\Theta}_2\|_{\max}
\le
2\|\Theta_2\|_{\max}.
\]

Therefore,
\begin{align*}
I_1
\le
&~
4
\|\widehat{\Sigma}_1-\Sigma_1\|_{\max}
|\Delta_\Omega|_1
\|\Theta_2\|_{\max}
\\
\le
&~
128\sqrt{6}\pi
C_M
|\Delta_\Omega|_1
\|\Theta_2\|_{\max}
\left(
\frac{\log d}
{\min(n_1,n_2)-p}
\right)^{1/2}.
\end{align*}

For \(I_2\), we have
\begin{align*}
I_2
\le
&~
\|\Sigma_1\|_{\max}
|\widehat{\Delta}_\Omega(\nu)-\Delta_\Omega|_1
\|\widehat{\Theta}_2\|_{\max}
\\
\le
&~
2
\|\Sigma_1\|_{\max}
\|\Theta_2\|_{\max}
|\widehat{\Delta}_\Omega(\nu)-\Delta_\Omega|_1.
\end{align*}
Using the bound above together with Theorem \ref{thm1},
\[
I_2
\le
96
\|\Sigma_1\|_{\max}
\|\Theta_2\|_{\max}
s_\Omega
\lambda_{\min}^{-1}(\Sigma_1)
\lambda_{\min}^{-1}(\Sigma_2)
\nu.
\]
Substituting the definition of \(\nu\) gives
\[
I_2
\le
18432\sqrt{6}\pi
C_M
\|\Sigma_1\|_{\max}
\|\Theta_2\|_{\max}
s_\Omega
(1+C_\Sigma|\Delta_\Omega|_1)
\lambda_{\min}^{-1}(\Sigma_1)
\lambda_{\min}^{-1}(\Sigma_2)
\left(
\frac{\log d}
{\min(n_1,n_2)-p}
\right)^{1/2}.
\]

For \(I_3\), we have
\[
I_3
\le
\|\Sigma_1\|_{\max}
|\Delta_\Omega|_1
\|\widehat{\Theta}_2-\Theta_2\|_{\max}.
\]
Applying Lemma \ref{lemma1} yields
\[
I_3
\le
32\sqrt{6}\pi
C_M
\|\Sigma_1\|_{\max}
|\Delta_\Omega|_1
\left(
\frac{\log d}
{\min(n_1,n_2)-p}
\right)^{1/2}.
\]

Combining the above bounds, we obtain
\[
I_1+I_2+I_3
\le
C_I
\left(
\frac{\log d}
{\min(n_1,n_2)-p}
\right)^{1/2},
\]
where
\begin{align*}
C_I
=
&~
128\sqrt{6}\pi
C_M
|\Delta_\Omega|_1
C_\Theta
\\
&
+
18432\sqrt{6}\pi
C_M
C_\Sigma
C_\Theta
s_\Omega
(1+C_\Sigma|\Delta_\Omega|_1)
\lambda_{\min}^{-1}(\Sigma_1)
\lambda_{\min}^{-1}(\Sigma_2)
\\
&
+
32\sqrt{6}\pi
C_M
C_\Sigma
|\Delta_\Omega|_1
\end{align*}
with $C_\Theta = \max\left( \|\Theta_1\|_{\max}, \|\Theta_2\|_{\max} \right)$.

Combining the above bounds yields
\[
I_1+I_2+I_3
\le
C_I
\left(
\frac{\log d}
{\min(n_1,n_2)-p
}
\right)^{1/2}
\]
for some positive constant
\(C_I\).

By symmetry,
\[
\|
\widehat{\Sigma}_2
\widehat{\Delta}_\Omega(\nu)
\widehat{\Theta}_1
-
\Sigma_2
\Delta_\Omega
\Theta_1
\|_{\max}
\le
C_I
\left(
\frac{\log d}
{\min(n_1,n_2)-p}
\right)^{1/2}.
\]

Moreover, by Lemma \ref{lemma1},
\[
2
\|\widehat{\Theta}_1-\Theta_1\|_{\max}
+
2
\|\widehat{\Theta}_2-\Theta_2\|_{\max}
\le
128\sqrt{6}\pi C_M
\left(
\frac{\log d}
{\min(n_1,n_2)-p}
\right)^{1/2}.
\]

Combining the above bounds yields
\[
\max_{1\le j\le d}
\|
\widehat{\bf w}_j
-
{\bf w}_j
\|_{\max}
\le
(2C_I+128\sqrt{6}\pi C_M)
\left(
\frac{\log d}
{\min(n_1,n_2)-p}
\right)^{1/2}.
\]

Therefore,
\begin{align*}
\max_{1\le j\le d}
\left\|
\left(\widehat{\Sigma}_1+\widehat{\Sigma}_2\right)\ve\beta_j
-
\widehat{\bf w}_j
\right\|_{\max}
\le
&
\max_{1\le j\le d}
\|
\widehat{\bf w}_j
-
{\bf w}_j
\|_{\max}
\\
&
+
\max_{1\le j\le d}
\|\ve\beta_j\|_1
\sum_{l=1}^{2}
\|
\widehat{\Sigma}_l-\Sigma_l
\|_{\max}.
\end{align*}

Using the bound established at the beginning of the proof,
\[
\max_{1\le j\le d}
\|\ve\beta_j\|_1
\sum_{l=1}^{2}
\|
\widehat{\Sigma}_l-\Sigma_l
\|_{\max}
\le
C_{I_0}
\|\Delta_A\|_1
\left(
\frac{\log d}
{\min(n_1,n_2)-p+1}
\right)^{1/2},
\]
we conclude that
\[
\max_{1\le j\le d}
\left\|
\left(\widehat{\Sigma}_1+\widehat{\Sigma}_2\right)\ve\beta_j
-
\widehat{\bf w}_j
\right\|_{\max}
\le
C_1
\left(
\frac{\log d}
{\min(n_1,n_2)-p}
\right)^{1/2},
\]
where
\[
C_1
=
2C_I
+
128\sqrt{6}\pi C_M
+
C_{I_0}\|\Delta_A\|_1.
\]

This completes the proof.

\end{proof}

Next, we prove Theorem \ref{thm3}, which is the convergence rate for $\widehat{\Delta}_A$.

\begin{proof}
Like in the proof of Lemma \ref{lemma7}, we condition on the event
\(\mathcal A_1\cap\mathcal A_2\)
throughout the proof.
As we will rely on Lemma \ref{lemma4}, we will first verify Assumption \ref{assump:RE}.
By Lemma \ref{lemma7},
\[
\max_{1\le j\le d}
\|
\nabla L_A(\ve\beta_j)
\|_{\max}
=
\max_{1\le j\le d}
\left\|
\left(\widehat{\Sigma}_1+\widehat{\Sigma}_2\right)\ve\beta_j
-
\widehat{\bf w}_j
\right\|_{\max}
\le
C_1
\left(
\frac{\log d}
{\min(n_1,n_2)-p}
\right)^{1/2}.
\]
Hence, for
\[
\lambda_j
= \bar{\lambda} :=
2C_1
\left(
\frac{\log d}
{\min(n_1,n_2)-p}
\right)^{1/2},
\qquad
j=1,\ldots,d,
\]
we have
\(
\lambda_j
\ge
2
\|
\nabla L_A(\ve\beta_j)
\|_{\max}.
\)
Then, by Lemma \ref{lemma4}, for
\[
\widehat{\ve\beta}_j-\ve\beta_j
\in
\mathcal C(S_{A,j}),
\qquad
j=1,\ldots,d.
\]
we have
\[
\|
\widehat{\ve\beta}_j-\ve\beta_j
\|_1
\le
4
\|
(\widehat{\ve\beta}_j-\ve\beta_j)_{S_{A,j}}
\|_1
\le
4\sqrt{s_{A,j}}
\|
\widehat{\ve\beta}_j-\ve\beta_j
\|_2.
\]

Let
\(
{\bf m}
=
\widehat{\ve\beta}_j-\ve\beta_j.
\)
Notice that
\(
\nabla^2 L_A(\ve\beta_j) = \widehat{\Sigma}_1 + \widehat{\Sigma}_2.
\)
Applying Lemma \ref{lemma6} yields
\[
{\bf m}^\top
\left(\widehat{\Sigma}_1+\widehat{\Sigma}_2\right)
{\bf m}
\ge
\left\{
\lambda_{\min}(\Sigma_1)
+
\lambda_{\min}(\Sigma_2)
-
1024 \pi s_{A,j}
\xi(n_1,n_2,d,p)
\right\}
\|{\bf m}\|_2^2.
\]

Given the last condition in \eqref{n:cond}, i.e.,
\[
\min(n_1,n_2)
\ge
\left(
\frac{
2048 \max_js_{A,j} C_M
}{
\lambda_{\min}(\Sigma_1)
+
\lambda_{\min}(\Sigma_2)
}
\right)^2
6\log d
+
p,
\]
we get
\[
1024 \pi \max_j s_{A,j}
\xi(n_1,n_2,d,p)
\le
\frac12
\left\{
\lambda_{\min}(\Sigma_1)
+
\lambda_{\min}(\Sigma_2)
\right\}.
\]
Hence,
\[
{\bf m}^\top
\left(\widehat{\Sigma}_1+\widehat{\Sigma}_2\right)
{\bf m}
\ge
\frac12
\left\{
\lambda_{\min}(\Sigma_1)
+
\lambda_{\min}(\Sigma_2)
\right\}
\|{\bf m}\|_2^2.
\]
Therefore, Assumption \ref{assump:RE} holds with
\[
\kappa_1
=
\frac12
\left\{
\lambda_{\min}(\Sigma_1)
+
\lambda_{\min}(\Sigma_2)
\right\}.
\]

Next, we apply Lemma \ref{lemma4} and get
\[
\|
\widehat{\ve\beta}_j-\ve\beta_j
\|_2
\le
12 \bar{\lambda}
\left\{
\lambda_{\min}(\Sigma_1)
+
\lambda_{\min}(\Sigma_2)
\right\}^{-1}
\sqrt{s_{A,j}},
\qquad
j=1,\ldots,d.
\]
Summing over
\(j=1,\ldots,d\),
we obtain
\[
\|
\widehat{\Delta}_A(\ve\lambda)-\Delta_A
\|_F
=
\left(
\sum_{j=1}^{d}
\|
\widehat{\ve\beta}_j-\ve\beta_j
\|_2^2
\right)^{1/2}
\le
12 \bar{\lambda}
\left\{
\lambda_{\min}(\Sigma_1)
+
\lambda_{\min}(\Sigma_2)
\right\}^{-1}
\left(
\sum_{l=1}^{d}
s_{A,l}
\right)^{1/2}.
\]

Similarly,
\[
\|
\widehat{\Delta}_A(\ve\lambda)-\Delta_A
\|_{\max}
=
\max_{1\le j\le d}
\|
\widehat{\ve\beta}_j-\ve\beta_j
\|_\infty
\le
\max_{1\le j\le d}
\|
\widehat{\ve\beta}_j-\ve\beta_j
\|_2,
\]
which yields
\[
\|
\widehat{\Delta}_A(\ve\lambda)-\Delta_A
\|_{\max}
\le
12 \bar{\lambda}
\left\{
\lambda_{\min}(\Sigma_1)
+
\lambda_{\min}(\Sigma_2)
\right\}^{-1}
\max_{1\le l\le d}
\sqrt{s_{A,l}}.
\]

This completes the proof.
\end{proof}

Next, we prove Theorem \ref{thm4} based on Theorem \ref{thm3}.

\begin{proof}
By Theorem \ref{thm3}, with probability at least
\(1-32d^{-1}\),
\[
\|
\widehat{\Delta}_A(\ve\lambda)
-
\Delta_A
\|_{\max}
\le
12 \bar{\lambda}
\left\{
\lambda_{\min}(\Sigma_1)
+
\lambda_{\min}(\Sigma_2)
\right\}^{-1}
\max_{1\le l\le d}
\sqrt{s_{A,l}}.
\]
Since
\[
\tau_A
\ge
12 \bar{\lambda}
\left\{
\lambda_{\min}(\Sigma_1)
+
\lambda_{\min}(\Sigma_2)
\right\}^{-1}
\max_{1\le l\le d}
\sqrt{s_{A,l}},
\]
it follows that
\[
\|
\widehat{\Delta}_A(\ve\lambda)
-
\Delta_A
\|_{\max}
\le
\tau_A
\]
with probability at least
\(1-32d^{-1}\).

We now show that
\[
\operatorname{sgn}
\left(
\mathrm{HT}_{\tau_A}
\left(
\widehat{\Delta}_A(\ve\lambda)
\right)
\right)
=
\operatorname{sgn}
(\Delta_A).
\]

If
\[
(\Delta_A)_{jk}=0,
\]
then
\[
\left|
\widehat{\Delta}_{A,jk}(\ve\lambda)
\right|
=
\left|
\widehat{\Delta}_{A,jk}(\ve\lambda)
-
(\Delta_A)_{jk}
\right|
\le
\tau_A,
\]
and therefore
\[
\mathrm{HT}_{\tau_A}
\left(
\widehat{\Delta}_{A,jk}(\ve\lambda)
\right)
=
0.
\]

Now suppose
\[
(\Delta_A)_{jk}\neq 0.
\]
By assumption,
\[
|(\Delta_A)_{jk}|
>
2\tau_A.
\]
Therefore,
\[
\left|
\widehat{\Delta}_{A,jk}(\ve\lambda)
\right|
\ge
|(\Delta_A)_{jk}|
-
\left|
\widehat{\Delta}_{A,jk}(\ve\lambda)
-
(\Delta_A)_{jk}
\right|
>
2\tau_A-\tau_A
=
\tau_A.
\]
Hence
\[
\mathrm{HT}_{\tau_A}
\left(
\widehat{\Delta}_{A,jk}(\ve\lambda)
\right)
\neq 0.
\]

Moreover,
\[
\left|
\widehat{\Delta}_{A,jk}(\ve\lambda)
-
(\Delta_A)_{jk}
\right|
<
\frac12
|(\Delta_A)_{jk}|,
\]
which implies
\[
\operatorname{sgn}
\left(
\widehat{\Delta}_{A,jk}(\ve\lambda)
\right)
=
\operatorname{sgn}
\left(
(\Delta_A)_{jk}
\right).
\]

Therefore,
\[
\operatorname{sgn}
\left(
\mathrm{HT}_{\tau_A}
\left(
\widehat{\Delta}_A(\ve\lambda)
\right)
\right)
=
\operatorname{sgn}
(\Delta_A)
\]
with probability at least
\(1-32d^{-1}\).
This completes the proof.
\end{proof}

\section{Proof for direct inference}

We first justify treating the selection and inference samples as asymptotically independent.
For $l=1,2$, let
$\mathcal{X}_l^{\rm sel}=\{{\bf x}_t^{(l)}:1\leq t\leq n_l^{(1)}\}$ and
$\mathcal{X}_l^{\rm inf}=\{{\bf x}_t^{(l)}:n_l^{(2)}+1\leq t\leq n_l\}$,
and let $g_l=n_l^{(2)}-n_l^{(1)}$ denote the length of the buffer segment.
By the $\beta$-mixing assumption and a standard coupling argument \citep{yu1994rates}, there exists a random block
$\mathcal{X}_l^{\rm inf,*}$ such that
\[
\mathcal{X}_l^{\rm inf,*}\overset{d}{=}\mathcal{X}_l^{\rm inf},
\qquad
\mathcal{X}_l^{\rm inf,*}\perp\mathcal{X}_l^{\rm sel},
\qquad
\Pr\left(
\mathcal{X}_l^{\rm inf,*}\neq\mathcal{X}_l^{\rm inf}
\right)
\leq \beta_l(g_l) = o(1).
\]
In the remainder of the proof, we therefore work with
the coupled inference samples and suppress the superscript $\ast$ for
notational simplicity.

We then prove Lemma \ref{lem:target:equivalence}.

\begin{proof}
Let $V_A^c=\{1,\ldots,d^2\}\setminus V_A$ and
$V_\Omega^c=\{1,\ldots,d^2\}\setminus V_\Omega$.
Restricting the first equation in \eqref{selinf:add:1} to the rows indexed
by $V_\Omega$ gives
\begin{align*}
(\Sigma_1\otimes\Sigma_2)_{(V_\Omega,V_\Omega)}
\{\operatorname{vec}(\Delta_\Omega)\}_{V_\Omega}
&+
(\Sigma_1\otimes\Sigma_2)_{V_\Omega,V_\Omega^c}
\{\operatorname{vec}(\Delta_\Omega)\}_{V_\Omega^c} \\
&=
\{\operatorname{vec}(\Sigma_2-\Sigma_1)\}_{V_\Omega}.
\end{align*}
Since $\mathcal{M}_\Omega\supseteq S_\Omega$,
$\{\operatorname{vec}(\Delta_\Omega)\}_{V_\Omega^c}=0$. Hence,
\begin{align*}
\{\operatorname{vec}(\Delta_\Omega)\}_{V_\Omega}
&=
\left\{
(\Sigma_1\otimes\Sigma_2)_{(V_\Omega,V_\Omega)}
\right\}^{-1}
\{\operatorname{vec}(\Sigma_2-\Sigma_1)\}_{V_\Omega} \\
&=
\psi_\Omega.
\end{align*}

Next, let $H=I_d\otimes(\Sigma_1+\Sigma_2)$ and
$C=(\Theta_1^\top\otimes\Sigma_2)+(\Theta_2^\top\otimes\Sigma_1)$.
Restricting the second equation in \eqref{selinf:add:1} to the rows indexed
by $V_A$ yields
\begin{align*}
H_{V_A,V_A}\{\operatorname{vec}(\Delta_A)\}_{V_A}
&+
H_{V_A,V_A^c}\{\operatorname{vec}(\Delta_A)\}_{V_A^c} \\
&=
C_{V_A,V_\Omega}\{\operatorname{vec}(\Delta_\Omega)\}_{V_\Omega}
+
C_{V_A,V_\Omega^c}\{\operatorname{vec}(\Delta_\Omega)\}_{V_\Omega^c} \\
&\quad+
2\{\operatorname{vec}(\Theta_1-\Theta_2)\}_{V_A}.
\end{align*}
Since $\mathcal{M}_A\supseteq S_A$ and
$\mathcal{M}_\Omega\supseteq S_\Omega$, we have
$\{\operatorname{vec}(\Delta_A)\}_{V_A^c}=0$ and
$\{\operatorname{vec}(\Delta_\Omega)\}_{V_\Omega^c}=0$. Together with
$\{\operatorname{vec}(\Delta_\Omega)\}_{V_\Omega}=\psi_\Omega$, this gives
\[
H_{V_A,V_A}\{\operatorname{vec}(\Delta_A)\}_{V_A}
=
C_{V_A,V_\Omega}\psi_\Omega
+
2\{\operatorname{vec}(\Theta_1-\Theta_2)\}_{V_A}.
\]
By the definition of $\psi_A$, it follows that
$\{\operatorname{vec}(\Delta_A)\}_{V_A}=\psi_A$.
\end{proof}

We then prove Theorem \ref{inference:thm1:101724}.
\begin{proof}
Recall that
\[
\widehat{\psi}_\Omega
=
\widehat{\Phi}^{-1}
\left\{
\operatorname{vec}(\widehat{\Sigma}_2-\widehat{\Sigma}_1)
\right\}_{V_\Omega},
\qquad
\psi_\Omega
=
\Phi^{-1}
\left\{
\operatorname{vec}(\Sigma_2-\Sigma_1)
\right\}_{V_\Omega},
\]
where
\[
\Phi
=
(\Sigma_1\otimes\Sigma_2)_{(V_\Omega,V_\Omega)},
\qquad
\widehat{\Phi}
=
\left(\widehat{\Sigma}_1\otimes\widehat{\Sigma}_2\right)_{(V_\Omega,V_\Omega)}.
\]

Hence,
\begin{align}
\label{020222:4}
\widehat{\psi}_\Omega-\psi_\Omega
=
&
\Phi^{-1}
\left\{
\operatorname{vec}
\left(\widehat{\Sigma}_2-\Sigma_2\right)
-
\operatorname{vec}
\left(\widehat{\Sigma}_1-\Sigma_1\right)
\right\}_{V_\Omega}
\nonumber
\\
&
+
\left(\widehat{\Phi}^{-1}-\Phi^{-1}\right)
\left\{
\operatorname{vec}(\Sigma_2-\Sigma_1)
\right\}_{V_\Omega}
\nonumber
\\
&
+
(\widehat{\Phi}^{-1}-\Phi^{-1})
\left\{
\operatorname{vec}
\left(\widehat{\Sigma}_2-\Sigma_2\right)
-
\operatorname{vec}
\left(\widehat{\Sigma}_1-\Sigma_1\right)
\right\}_{V_\Omega}
\nonumber
\\
\equiv
&~
\mathcal R_1+\mathcal R_2+\mathcal R_3, %
\end{align}
where $\mathcal R_1$ is the desired term in the theorem.

For convenience, let
\[
\Delta_\Phi
=
\widehat{\Phi}-\Phi.
\]
By Lemma \ref{lemma1},
\[
\|\widehat{\Sigma}_l-\Sigma_l\|_{\max}
=
O_p
\left(
\sqrt{
\frac{\log d}
{n^{\rm inf}}
}
\right),
\qquad
l=1,2.
\]
Consequently,
\begin{align}
\label{020222:6.5}
\|\Delta_\Phi\|_{\max}
\le
&
\|\Sigma_1\|_{\max}
\|\widehat{\Sigma}_2-\Sigma_2\|_{\max}
+
\|\Sigma_2\|_{\max}
\|\widehat{\Sigma}_1-\Sigma_1\|_{\max}
\nonumber
\\
&
+
\|\widehat{\Sigma}_1-\Sigma_1\|_{\max}
\|\widehat{\Sigma}_2-\Sigma_2\|_{\max}
\nonumber
\\
=
&
O_p
\left(
\sqrt{
\frac{\log d}
{n^{\rm inf}}
}
\right).
\end{align}
Since
\[
\|\Delta_\Phi\|
\le
m_\Omega
\|\Delta_\Phi\|_{\max},
\]
we obtain
\[
\|\Delta_\Phi\|
=
O_p
\left(
m_\Omega
\sqrt{
\frac{\log d}
{n^{\rm inf}}
}
\right).
\]
Thus, by Assumptions \ref{assump:eigen} and \ref{assump:supp:omega}, we know
\[
\|\Phi^{-1}\Delta_\Phi\|
=
o_p(1).
\]
Therefore, the Neumann expansion applies:
\begin{equation}
\label{020222:5}
\widehat{\Phi}^{-1}
=
(\Phi+\Delta_\Phi)^{-1}
=
\sum_{k=0}^{\infty}
(-1)^k
(\Phi^{-1}\Delta_\Phi)^k
\Phi^{-1}.
\end{equation}

Hence,
\[
\widehat{\Phi}^{-1}-\Phi^{-1}
=
-\Phi^{-1}\Delta_\Phi\Phi^{-1}
+
\sum_{k=2}^{\infty}
(-1)^k
(\Phi^{-1}\Delta_\Phi)^k
\Phi^{-1}
\equiv
\mathcal R_4+\mathcal R_5.
\]

It is easy to see
\[
\|\mathcal R_4\|
=
O_p(\|\Delta_\Phi\|)
=
O_p
\left(
m_\Omega
\sqrt{
\frac{\log d}
{n^{\rm inf}}
}
\right).
\]

For \(\mathcal R_5\), we have
\[
\mathcal R_5
=
\sum_{k=2}^{\infty}
(-1)^k
(\Phi^{-1}\Delta_\Phi)^k
\Phi^{-1}.
\]
Hence,
\begin{align}
\label{020222:6}
\|\mathcal R_5\|
\le
&
\sum_{k=2}^{\infty}
\|\Phi^{-1}\Delta_\Phi\|^k
\|\Phi^{-1}\|.
\end{align}

Since
\[
\|\Phi^{-1}\|
=
O(1),
\]
and by \eqref{020222:6.5},
\[
\|\Delta_\Phi\|
=
O_p
\left(
m_\Omega
\sqrt{
\frac{\log d}
{n^{\rm inf}}
}
\right),
\]
we get
\[
\|\Phi^{-1}\Delta_\Phi\|
=
O_p
\left(
m_\Omega
\sqrt{
\frac{\log d}
{n^{\rm inf}}
}
\right).
\]
Moreover, Assumption \ref{assump:supp:omega} implies
\[
m_\Omega
\sqrt{
\frac{\log d}
{n^{\rm inf}}
}
=
o(1),
\]
and hence
\[
\|\Phi^{-1}\Delta_\Phi\|
=
o_p(1).
\]

Consequently,
\begin{align*}
\|\mathcal R_5\|
\le
&
\|\Phi^{-1}\|
\sum_{k=2}^{\infty}
\|\Phi^{-1}\Delta_\Phi\|^k
\\
=
&
\|\Phi^{-1}\|
\frac{
\|\Phi^{-1}\Delta_\Phi\|^2
}{
1-\|\Phi^{-1}\Delta_\Phi\|
}
\\
=
&
O_p
\left(
\|\Delta_\Phi\|^2
\right)
\\
=
&
O_p
\left(
m_\Omega^2
\frac{\log d}
{n^{\rm inf}}
\right).
\end{align*}

Furthermore,
\[
\frac{\|\mathcal R_5\|}{\|\mathcal R_4\|}
=
O_p(\|\Delta_\Phi\|)
=
o_p(1),
\]
which shows that
\(\mathcal R_5\)
is asymptotically negligible relative to the first-order term
\(\mathcal R_4\).

Next,
\[
\left\|
\operatorname{vec}
\left(\widehat{\Sigma}_2-\Sigma_2\right)
-
\operatorname{vec}
\left(\widehat{\Sigma}_1-\Sigma_1\right)
\right\|
=
O_p
\left(
\sqrt{
\frac{m_\Omega}
{n^{\rm inf}}
}
\right).
\]
Therefore,
\[
\|\mathcal R_3\|
=
O_p
\left(
\frac{\log d}{n^{\rm inf}}
m_\Omega^{3/2}
\right).
\]

Since
\[
\frac{\log d}{n^{\rm inf}}
m_\Omega^{3/2}
=
(n^{\rm inf})^{-1/2}
\left(
\frac{
m_\Omega^3(\log d)^2
}{
n^{\rm inf}
}
\right)^{1/2},
\]
Assumption \ref{assump:supp:omega} yields
\[
\mathcal R_3
=
o_p
\left(
(n^{\rm inf})^{-1/2}
\right).
\]

For \(\mathcal R_2\), recall
\[
\ve\mu
=
\left\{
\operatorname{vec}(\Sigma_2-\Sigma_1)
\right\}_{V_\Omega}.
\]
Using
\[
\operatorname{vec}(ABC)
=
(C^\top\otimes A)
\operatorname{vec}(B),
\]
we obtain
\[
\mathcal R_2
=
-(\ve\mu^\top\otimes I_{m_\Omega})
(\Phi^{-1}\otimes\Phi^{-1})
\operatorname{vec}(\Delta_\Phi)
+
(\ve\mu^\top\otimes I_{m_\Omega})
\operatorname{vec}(\mathcal I_5).
\]

Hence,
\begin{align}
\label{020222:7}
&
\left\|
\mathcal R_2
+
(\ve\mu^\top\otimes I_{m_\Omega})
(\Phi^{-1}\otimes\Phi^{-1})
\operatorname{vec}(\Delta_\Phi)
\right\|
\nonumber
\\
\le
&
\|
(\ve\mu^\top\otimes I_{m_\Omega})
\operatorname{vec}(\mathcal R_5)
\|
\nonumber
\\
=
&
O_p
\left(
\frac{\log d}{n^{\rm inf}}
m_\Omega^2
\|\ve\mu\|
\right).
\end{align}

Since
\[
\|\ve\mu\|
\le
\sqrt{m_\Omega}
\|\Sigma_1-\Sigma_2\|_{\max},
\]
we obtain
\[
\left\|
\mathcal R_2
+
(\ve\mu^\top\otimes I_{m_\Omega})
(\Phi^{-1}\otimes\Phi^{-1})
\operatorname{vec}(\Delta_\Phi)
\right\|
=
O_p
\left(
\frac{\log d}{n^{\rm inf}}
m_\Omega^{5/2}
\right).
\]

Finally,
\begin{align}
\label{022222:7}
&
\left\|
\Delta_\Phi
-
\left\{
\Sigma_1
\otimes
\left(\widehat{\Sigma}_2-\Sigma_2\right)
+
\left(\widehat{\Sigma}_1-\Sigma_1\right)\otimes \Sigma_2
\right\}_{(V_\Omega,V_\Omega)}
\right\|
\nonumber
\\
=
&
\left\|
\left\{
\left(\widehat{\Sigma}_1-\Sigma_1\right)
\otimes
\left(\widehat{\Sigma}_2-\Sigma_2\right)
\right\}_{(V_\Omega,V_\Omega)}
\right\|
\nonumber
\\
=
&
O_p
\left(
\frac{\log d}{n^{\rm inf}}
m_\Omega
\right).
\end{align}

Combining the above display with
\eqref{020222:7}
yields
\begin{align*}
&~ \left\|
\mathcal R_2
+
(\ve\mu^\top\otimes I_{m_\Omega})
(\Phi^{-1}\otimes\Phi^{-1})
\operatorname{vec}
\left[
\left\{
\Sigma_1
\otimes
\left(\widehat{\Sigma}_2-\Sigma_2\right)
+
\left(\widehat{\Sigma}_1-\Sigma_1\right) \otimes \Sigma_2
\right\}_{(V_\Omega,V_\Omega)}
\right]
\right\| \\
= &~
O_p
\left(
\frac{\log d}{n^{\rm inf}}
m_\Omega^{5/2}
\right).
\end{align*}

Since
\[
\frac{\log d}{n^{\rm inf}}
m_\Omega^{5/2}
=
(n^{\rm inf})^{-1/2}
\left(
\frac{
m_\Omega^5(\log d)^2
}{
n^{\rm inf}
}
\right)^{1/2},
\]
Assumption \ref{assump:supp:omega} implies that the above remainder is
\(o_p((n^{\rm inf})^{-1/2})\).

Combining the bounds for
\(\mathcal R_2\)
and
\(\mathcal R_3\)
with
\eqref{020222:4}
establishes the desired linear representation.
\end{proof}

We next use Theorem \ref{inference:thm1:101724} to prove Theorem \ref{asymp:delta:omega:sel:inf}.
The core part of this proof amounts to rewriting the second term in
(\ref{inference:thm1:eq1})
as a linear transformation of
$\mathrm{vec}\left(\widehat{\Sigma}_l-\Sigma_l\right)$.

\begin{proof}
Recall that
\[
V_\Omega
=
\{l_1,\ldots,l_{m_\Omega}\},
\qquad
l_1<\cdots<l_{m_\Omega},
\]
and
\[
\Pi_{V_\Omega,m}
=
\left[
{\bf e}_{l_1,m},
\ldots,
{\bf e}_{l_{m_\Omega},m}
\right].
\]
Then, for any matrix \(A\),
\[
\{A\}_{(V_\Omega,V_\Omega)}
=
\Pi_{V_\Omega,d^2}^{\top}
A
\Pi_{V_\Omega,d^2}.
\]
Hence,
\begin{align}
\label{add:101724:1}
\left\{
\Sigma_1
\otimes
\sqrt{n^{\rm inf}}
\left(\widehat{\Sigma}_2-\Sigma_2\right)
\right\}_{(V_\Omega,V_\Omega)}
= &~
\Pi_{V_\Omega,d^2}^{\top}
\left\{
\Sigma_1
\otimes
\sqrt{n^{\rm inf}}
\left(\widehat{\Sigma}_2-\Sigma_2\right)
\right\}
\Pi_{V_\Omega,d^2} \nonumber \\
\left\{
\sqrt{n^{\rm inf}}
\left(\widehat{\Sigma}_1-\Sigma_1\right) \otimes \Sigma_2
\right\}_{(V_\Omega,V_\Omega)} =  &~
\Pi_{V_\Omega,d^2}^{\top}
\left\{
\sqrt{n^{\rm inf}}
\left(\widehat{\Sigma}_1-\Sigma_1\right) \otimes \Sigma_2
\right\}
\Pi_{V_\Omega,d^2}
\end{align}

Next, define the block diagonal matrix
\[
B_{\Sigma_l}
=
\begin{pmatrix}
\sqrt{n^{\rm inf}}\left(\widehat{\Sigma}_l-\Sigma_l\right)
& 0
& \cdots
& 0
\\
0
& \sqrt{n^{\rm inf}}\left(\widehat{\Sigma}_l-\Sigma_l\right)
& \cdots
& 0
\\
\vdots
& \vdots
& \ddots
& \vdots
\\
0
& 0
& \cdots
& \sqrt{n^{\rm inf}}\left(\widehat{\Sigma}_l-\Sigma_l\right)
\end{pmatrix}.
\]
Then, we have
\begin{align}
\label{add:101724:2}
\Sigma_1
\otimes
\sqrt{n^{\rm inf}}
\left(\widehat{\Sigma}_2-\Sigma_2\right)
= &~
(\Sigma_1\otimes I_d)
B_{\Sigma_2} \nonumber \\
\sqrt{n^{\rm inf}}(\widehat\Sigma_1{-}\Sigma_1)\otimes\Sigma_2 = &~
(I_d\otimes\Sigma_2)K_{d,d}B_{\Sigma_1}K_{d,d},
\end{align}
where $K_{d,d}$ is the $d^2 \times d^2$ commutation matrix.

Combining
\eqref{add:101724:1}
and
\eqref{add:101724:2},
and applying the identity
\(
\mathrm{vec}(ABC)
=
(C^\top\otimes A)
\mathrm{vec}(B),
\)
gives
\begin{align}\label{add:101724:3}
\mathrm{vec}
\left[
\left\{
\Sigma_1
\otimes
\sqrt{n^{\rm inf}}
\left(\widehat{\Sigma}_2-\Sigma_2\right)
\right\}_{(V_\Omega,V_\Omega)}
\right]
\nonumber
=
&~
\left\{
\Pi_{V_\Omega,d^2}^{\top}
\otimes
\left(
\Pi_{V_\Omega,d^2}^{\top}
(\Sigma_1\otimes I_d)
\right)
\right\}
\mathrm{vec}(B_{\Sigma_2}) \nonumber \\
\mathrm{vec}
\left[
\left\{
\sqrt{n^{\rm inf}}
\left(\widehat{\Sigma}_1-\Sigma_1\right) \otimes \Sigma_2
\right\}_{(V_\Omega,V_\Omega)}
\right]
=
&~
\left\{
\left(\Pi_{V_\Omega,d^2}^{\top} K_{d,d} \right)
\otimes
\left(
\Pi_{V_\Omega,d^2}^{\top} K_{d,d}
(\Sigma_2\otimes I_d)
\right)
\right\}
\mathrm{vec}(B_{\Sigma_1}).
\end{align}

Observe that
\(\mathrm{vec}(B_{\Sigma_l})\in\mathbb R^{d^4}\)
contains only
\(d^3\)
nonzero entries for $l=1,2$.
Let
\[
S_B
=
\left\{
(j-1)d^2
+
d\left\lfloor
\frac{j-1}{d}
\right\rfloor
+
k
:
j=1,\ldots,d^2,
\;
k=1,\ldots,d
\right\}
\]
denote its support.
Then
\[
\left(
\mathrm{vec}(B_{\Sigma_l})
\right)_{S_B}
=
\left(
\underbrace{
\mathrm{vec}
\left[
\sqrt{n^{\rm inf}}
\left(\widehat{\Sigma}_l-\Sigma_l\right)
\right]^\top,
\ldots,
\mathrm{vec}
\left[
\sqrt{n^{\rm inf}}
\left(\widehat{\Sigma}_l-\Sigma_l\right)
\right]^\top
}_{d}
\right)^\top.
\]

Rewriting
\eqref{add:101724:3}
according to the support
\(S_B\),
we obtain
\begin{align}
\label{add:101724:4}
\mathrm{vec}
\left[
\left\{
\Sigma_1
\otimes
\sqrt{n^{\rm inf}}
\left(\widehat{\Sigma}_2-\Sigma_2\right)
\right\}_{(V_\Omega,V_\Omega)}
\right]
\nonumber
=
& ~
\left\{
\Pi_{V_\Omega,d^2}^{\top}
\otimes
\left(
\Pi_{V_\Omega,d^2}^{\top}
(\Sigma_1\otimes I_d)
\right)
\right\}_{(*,S_B)}
\left(
\mathrm{vec}(B_{\Sigma_2})
\right)_{S_B}  \nonumber \\
\mathrm{vec}
\left[
\left\{
\sqrt{n^{\rm inf}}
\left(\widehat{\Sigma}_1-\Sigma_1\right) \otimes \Sigma_2
\right\}_{(V_\Omega,V_\Omega)}
\right]
=
&~
\left\{
\left(\Pi_{V_\Omega,d^2}^{\top} K_{d,d} \right)
\otimes
\left(
\Pi_{V_\Omega,d^2}^{\top} K_{d,d}
(\Sigma_2\otimes I_d)
\right)
\right\}_{(*, S_B)}
\left\{\mathrm{vec}(B_{\Sigma_1}) \right\}_{S_B}.
\end{align}

Since for
$j=1,\ldots,d$,
\(\mathcal F_{1j}\) and \(\mathcal F_{2j}\), respectively,
denote the submatrix of
\[
\left\{
\left(\Pi_{V_\Omega,d^2}^{\top} K_{d,d} \right)
\otimes
\left(
\Pi_{V_\Omega,d^2}^{\top} K_{d,d}
(\Sigma_2\otimes I_d)
\right)
\right\}_{(*, S_B)}
\mbox{ and }
\left\{
\Pi_{V_\Omega,d^2}^{\top}
\otimes
\left(
\Pi_{V_\Omega,d^2}^{\top}
(\Sigma_1\otimes I_d)
\right)
\right\}_{(*,S_B)}
\]
with columns indexed by
\[
\left\{
(j-1)d^2+k:
k=1,\ldots,d^2
\right\},
\]
Collecting the repeated copies of
\(\mathrm{vec}\left[\sqrt{n^{\rm inf}}\left(\widehat{\Sigma}_l-\Sigma_l\right)\right]\)
yields
\begin{align}
\label{add:101724:5}
\mathrm{vec}
\left[
\left\{
\Sigma_1
\otimes
\sqrt{n^{\rm inf}}
\left(\widehat{\Sigma}_2-\Sigma_2\right)
\right\}_{(V_\Omega,V_\Omega)}
\right]
= &~
\left(
\sum_{j=1}^{d}
\mathcal F_{2j}
\right)
\mathrm{vec}
\left[
\sqrt{n^{\rm inf}}
\left(\widehat{\Sigma}_2-\Sigma_2\right)
\right] \nonumber \\
\mathrm{vec}
\left[
\left\{
\sqrt{n^{\rm inf}}
\left(\widehat{\Sigma}_1-\Sigma_1\right) \otimes \Sigma_2
\right\}_{(V_\Omega,V_\Omega)}
\right] = &~
\left(
\sum_{j=1}^{d}
\mathcal F_{1j}
\right)
\mathrm{vec}
\left[
\sqrt{n^{\rm inf}}
\left(\widehat{\Sigma}_1-\Sigma_1\right)
\right].
\end{align}

Substituting
\eqref{add:101724:5}
into
\eqref{inference:thm1:eq1}
gives
\begin{align}
\label{add:101724:6}
&
\Bigg\|
\sqrt{n^{\rm inf}}
\left(\widehat{\psi}_\Omega-\psi_\Omega\right)
-
\Phi^{-1}
\Pi_{V_\Omega,d^2}^{\top}
\left\{\mathrm{vec}
\left[
\sqrt{n^{\rm inf}}
\left(\widehat{\Sigma}_2-\Sigma_2\right)
\right]
-
\mbox{vec}\left[
\sqrt{n^{\rm inf}}
\left(\widehat{\Sigma}_1-\Sigma_1\right)
\right]
\right\}
\nonumber
\\
&
\qquad
+
(\ve\mu^\top\otimes I_{m_\Omega})
(\Phi^{-1}\otimes\Phi^{-1})
\sum_{l=1}^{2}
\left(
\sum_{j=1}^{d}
\mathcal F_{ij}
\right)
\mathrm{vec}
\left[
\sqrt{n^{\rm inf}}
\left(\widehat{\Sigma}_l-\Sigma_l\right)
\right]
\Bigg\|
=
o_p(1).
\end{align}

Thus, we can write
\[
\sqrt{n^{\rm inf}}
\left(\widehat{\psi}_\Omega-\psi_\Omega\right)
+
\sum_{l=1}^{2}
G_l
\,
\mathrm{vec}
\left[
\sqrt{n^{\rm inf}}
\left(\widehat\Sigma_l-\Sigma_l\right)
\right]
=
o_p(1),
\]
where
\begin{align*}
G_1
= &~
(\ve\mu^\top\otimes I_{m_\Omega})
(\Phi^{-1}\otimes\Phi^{-1})
\left(
\sum_{j=1}^{d}
\mathcal F_{1j}
\right)
+
\Phi^{-1}
\Pi_{V_\Omega,d^2}^{\top} \\
G_2
= &~
(\ve\mu^\top\otimes I_{m_\Omega})
(\Phi^{-1}\otimes\Phi^{-1})
\left(
\sum_{j=1}^{d}
\mathcal F_{2j}
\right)
-
\Phi^{-1}
\Pi_{V_\Omega,d^2}^{\top}.
\end{align*}

For each
\(l=1,2\),
Theorem 1 of \citet{roy1989asymptotic}
implies that
\[
\mathrm{vec}
\left[
\sqrt{n^{\rm inf}}
\left(\widehat\Sigma_l-\Sigma_l\right)
\right]
\overset d\longrightarrow
N_{d^2}(0,D_l),
\]
where the
\(
((j_1-1)d+k_1,(j_2-1)d+k_2)
\)-th entry of
\(D_l\)
is given by
\eqref{Di:entry}.

Since the two time series are independent,
\[
\mathrm{cov}
\Big(
\mathrm{vec}
[
\sqrt{n^{\rm inf}}
(\widehat\Sigma_1-\Sigma_1)
],
\,
\mathrm{vec}
[
\sqrt{n^{\rm inf}}
(\widehat\Sigma_2-\Sigma_2)
]
\Big)
=
0.
\]
Therefore,
\[
\sum_{l=1}^{2}
G_l
\,
\mathrm{vec}
\left[
\sqrt{n^{\rm inf}}
\left(\widehat\Sigma_l-\Sigma_l\right)
\right]
\overset d\longrightarrow
N_{m_\Omega}
\left(
0,
\sum_{l=1}^{2}
G_lD_lG_l^\top
\right).
\]

Applying Slutsky's theorem to the
\(o_p(1)\)
remainder term yields
\[
\sqrt{n^{\rm inf}}
\left(\widehat{\psi}_\Omega-\psi_\Omega\right)
\overset d\longrightarrow
N_{m_\Omega}
\left(
0,
\sum_{l=1}^{2}
G_lD_lG_l^\top
\right),
\]
which proves the theorem.
\end{proof}

We next prove Theorem \ref{thm:inf:delta_A:101724}.

\begin{proof}
Recall that
\[
\widehat{\Upsilon}
=
\left\{
I_d\otimes\left(\widehat{\Sigma}_1+\widehat{\Sigma}_2\right)
\right\}_{(V_A,V_A)},
\qquad
\Delta_\Upsilon
=
\widehat{\Upsilon}-\Upsilon.
\]

We begin with the decomposition
\begin{align}
\label{022222:13}
\widehat{\psi}_A-\psi_A
=
J_1+J_2+J_3+J_4+J_5,
\end{align}
where
\begin{align*}
J_1
=&~
\Upsilon^{-1}
\Big[
\left(\widehat{\Theta}_1^\top\otimes\widehat{\Sigma}_2
+
\widehat{\Theta}_2^\top\otimes\widehat{\Sigma}_1\right)_{(V_A,V_\Omega)}
-
\left(\Theta_1^\top\otimes\Sigma_2
+
\Theta_2^\top\otimes\Sigma_1\right)_{(V_A,V_\Omega)}
\Big]
\psi_\Omega,
\\
J_2
=&~
\Upsilon^{-1}
(\Theta_1^\top\otimes\Sigma_2
+
\Theta_2^\top\otimes\Sigma_1)_{(V_A,V_\Omega)}
\left(\widehat{\psi}_\Omega-\psi_\Omega\right),
\\
J_3
=&~
\Upsilon^{-1}
\Big[
\left(\widehat{\Theta}_1^\top\otimes\widehat{\Sigma}_2
+
\widehat{\Theta}_2^\top\otimes\widehat{\Sigma}_1\right)_{(V_A,V_\Omega)}
-
\left(\Theta_1^\top\otimes\Sigma_2
+
\Theta_2^\top\otimes\Sigma_1\right)_{(V_A,V_\Omega)}
\Big]
\left(\widehat{\psi}_\Omega-\psi_\Omega\right),
\\
J_4
=&~
2\Upsilon^{-1}
\left\{
\operatorname{vec}
\left(\widehat{\Theta}_1-\Theta_1
-
\widehat{\Theta}_2+\Theta_2\right) %
\right\}_{V_A},
\\
J_5
=&~
\left(\widehat{\Upsilon}^{-1}-\Upsilon^{-1}\right)
\Big[
\left(\widehat{\Theta}_1^\top\otimes\widehat{\Sigma}_2
+
\widehat{\Theta}_2^\top\otimes\widehat{\Sigma}_1\right)_{(V_A,V_\Omega)}
\widehat{\psi}_\Omega
+
2
\left\{
\operatorname{vec}
\left(\widehat{\Theta}_1-\widehat{\Theta}_2\right)
\right\}_{V_A}
\Big].
\end{align*}

We first show that \(J_3\) is asymptotically negligible.
Using the decomposition
\begin{align}
\label{020222:14}
&
\Big\{
(\widehat{\Theta}_1^\top\otimes\widehat{\Sigma}_2
+
\widehat{\Theta}_2^\top\otimes\widehat{\Sigma}_1)
-
(\Theta_1^\top\otimes\Sigma_2
+
\Theta_2^\top\otimes\Sigma_1)
\Big\}_{(V_A,V_\Omega)}
\nonumber
\\
=
&
\underbrace{
\Big\{
\left(\widehat{\Theta}_1-\Theta_1\right)^\top\otimes\Sigma_2
+
\Theta_1^\top\otimes\left(\widehat{\Sigma}_2-\Sigma_2\right)
+
\left(\widehat{\Theta}_2-\Theta_2\right)^\top\otimes\Sigma_1
+
\Theta_2^\top\otimes\left(\widehat{\Sigma}_1-\Sigma_1\right)
\Big\}_{(V_A,V_\Omega)}
}_{J_6}
\nonumber
\\
&
+
\underbrace{
\Big\{
\left(\widehat{\Theta}_1-\Theta_1\right)^\top
\otimes
\left(\widehat{\Sigma}_2-\Sigma_2\right)
+
\left(\widehat{\Theta}_2-\Theta_2\right)^\top
\otimes
\left(\widehat{\Sigma}_1-\Sigma_1\right)
\Big\}_{(V_A,V_\Omega)}
}_{J_7}.
\end{align}
together with Lemma \ref{lemma1}, we obtain
\[
\|J_6\|
=
O_p
\left(
\sqrt{
\frac{\log d}{n^{\rm inf}}
}
\sqrt{m_A m_\Omega}
\right),
\qquad
\|J_7\|
=
O_p
\left(
\frac{\log d}{n^{\rm inf}}
\sqrt{m_A m_\Omega}
\right).
\]
Since
\[
\|\widehat{\psi}_\Omega-\psi_\Omega\|
=
O_p
\left(
m_\Omega^{3/2}
(n^{\rm inf})^{-1/2} (\log d)^{1/2}
\right)
\]
by Theorem \ref{inference:thm1:101724},
Assumptions \ref{assump:supp:omega} and \ref{assump:supp:A} imply
\[
J_3
=
o_p
\left(
(n^{\rm inf})^{-1/2}
\right).
\]

Next, using \eqref{020222:14},
\[
J_1
=
\Upsilon^{-1}
J_6
\psi_\Omega
+
o_p
\left(
(n^{\rm inf})^{-1/2}
\right).
\]
Multiplying by \(\sqrt{n^{\rm inf}}\), we obtain
\begin{align}
\label{JA:first}
\sqrt{n^{\rm inf}}J_1
=
&
\Upsilon^{-1}
\Big[
\sqrt{n^{\rm inf}}\left(\widehat{\Theta}_1-\Theta_1\right)^\top\otimes\Sigma_2
+
\Theta_1^\top\otimes\sqrt{n^{\rm inf}}\left(\widehat{\Sigma}_2-\Sigma_2\right)
\nonumber
\\
&
+
\sqrt{n^{\rm inf}}\left(\widehat{\Theta}_2-\Theta_2\right)^\top\otimes\Sigma_1
+
\Theta_2^\top\otimes\sqrt{n^{\rm inf}}\left(\widehat{\Sigma}_1-\Sigma_1\right)
\Big]_{(V_A,V_\Omega)}
\psi_\Omega
\nonumber
\\
&
+
o_p(1).
\end{align}

For \(J_2\), substituting the linear representation in
Theorem \ref{inference:thm1:101724} yields
\begin{align}
\label{JA:second}
\sqrt{n^{\rm inf}}J_2
=
&~
\Upsilon^{-1}
(\Theta_1^\top\otimes\Sigma_2
+
\Theta_2^\top\otimes\Sigma_1)_{(V_A,V_\Omega)}
\Phi^{-1}
\Pi_{V_\Omega,d^2}^\top
\nonumber
\\
&~ \times \left\{
\operatorname{vec}
\left[
\sqrt{n^{\rm inf}}
\left(\widehat{\Sigma}_2-\Sigma_2\right)
\right]
-
\operatorname{vec}
\left[
\sqrt{n^{\rm inf}}
\left(\widehat{\Sigma}_1-\Sigma_1\right)
\right]
\right\}
-
\Upsilon^{-1}
(\Theta_1^\top\otimes\Sigma_2
+
\Theta_2^\top\otimes\Sigma_1)_{(V_A,V_\Omega)}
\nonumber
\\
\times &~ (\ve\mu^\top\otimes I_{m_\Omega})
(\Phi^{-1}\otimes\Phi^{-1})
\times
\sum_{l=1}^{2}
\left(
\sum_{j=1}^{d}
\mathcal F_{ij}
\right)
\operatorname{vec}
\left[
\sqrt{n^{\rm inf}}
\left(\widehat{\Sigma}_l-\Sigma_l\right)
\right]
+
o_p(1).
\end{align}

For \(J_4\),
\begin{align}
\label{JA:third}
\sqrt{n^{\rm inf}}J_4
=
&
2\Upsilon^{-1}
\operatorname{vec}
\left[
\sqrt{n^{\rm inf}}
\left(\widehat{\Theta}_1-\Theta_1\right)
\right]
\nonumber
\\
&
-
2\Upsilon^{-1}
\operatorname{vec}
\left[
\sqrt{n^{\rm inf}}
\left(\widehat{\Theta}_2-\Theta_2\right)
\right]. %
\end{align}

Finally, applying the same Neumann-series expansion as in the proof of
Theorem \ref{inference:thm1:101724},
\[
\widehat{\Upsilon}^{-1}
-
\Upsilon^{-1}
=
-
\Upsilon^{-1}
\Delta_\Upsilon
\Upsilon^{-1}
+
o_p
\left(
(n^{\rm inf})^{-1/2}
\right).
\]
Let
\[
K_1
=
(\Theta_1^\top\otimes\Sigma_2
+
\Theta_2^\top\otimes\Sigma_1)_{(V_A,V_\Omega)}
\psi_\Omega
+
2
\{
\operatorname{vec}
(\Theta_1-\Theta_2)
\}_{V_A}.
\]
Since
\[
K_1
=
\Upsilon\psi_A,
\]
we obtain
\begin{align}
\label{JA:fourth}
\sqrt{n^{\rm inf}}J_5
=
&
-
\left(
K_1^\top\otimes I_{m_A}
\right)
(\Upsilon^{-1}\otimes\Upsilon^{-1})
\sqrt{n^{\rm inf}}
\operatorname{vec}(\Delta_\Upsilon)
+
o_p(1).
\end{align}

Combining
\eqref{JA:first},
\eqref{JA:second},
\eqref{JA:third},
and
\eqref{JA:fourth},
we next collect the coefficients of
\[
\operatorname{vec}
\left[
\sqrt{n^{\rm inf}}
\left(\widehat{\Sigma}_1-\Sigma_1\right)
\right],
\quad
\operatorname{vec}
\left[
\sqrt{n^{\rm inf}}
\left(\widehat{\Sigma}_2-\Sigma_2\right)
\right],
\quad
\operatorname{vec}
\left[
\sqrt{n^{\rm inf}}
\left(\widehat{\Theta}_1-\Theta_1\right)
\right],
\quad
\operatorname{vec}
\left[
\sqrt{n^{\rm inf}}
\left(\widehat{\Theta}_2-\Theta_2\right)
\right],
\]
respectively.
Specifically, for the coefficient of $\operatorname{vec}
\left[
\sqrt{n^{\rm inf}}
\left(\widehat{\Sigma}_1-\Sigma_1\right)
\right]$, the relevant term from $J_1$ is
\[
\Upsilon^{-1}\left[\Theta_2^{\top}\otimes\sqrt{n^{\rm inf}}\left(\widehat{\Sigma}_1 - \Sigma_1\right)\right]_{(V_A,V_\Omega)}\psi_\Omega = \Upsilon^{-1} \Pi_{V_A}^\top (\Theta_2^\top \otimes I_d) B_{\Sigma_1} \Pi_{V_\Omega} \psi_\Omega
\]
Using the vectorization trick, the above quantity can be further written as
\[
(\psi_\Omega^\top \otimes \Upsilon^{-1}) \left[ \Pi_{V_\Omega}^\top \otimes (\Pi_{V_A}^\top (\Theta_2^\top \otimes I_d) ) \right] \mbox{vec}(B_{\Sigma_1}) = M_1 \left(\sum_{j=1}^d \mathcal{H}_{2j}\right) \mbox{vec}(B_{\Sigma_1}),
\]
where $M_1$ and $\mathcal{H}_{2j}$ are defined in the main text.

The relevant term from $J_2$ is
\begin{align*}
&~ - \Upsilon^{-1}
(\Theta_1^\top\otimes\Sigma_2
+
\Theta_2^\top\otimes\Sigma_1)_{(V_A,V_\Omega)}
\Phi^{-1}
\Pi_{V_\Omega,d^2}^\top \\
- &~ \Upsilon^{-1}
(\Theta_1^\top\otimes\Sigma_2
+
\Theta_2^\top\otimes\Sigma_1)_{(V_A,V_\Omega)}
 (\ve\mu^\top\otimes I_{m_\Omega})
(\Phi^{-1}\otimes\Phi^{-1})
\left(
\sum_{j=1}^{d}
\mathcal F_{1j}
\right) = -M_2 - M_3 \sum_{j=1}^d \mathcal{F}_{1j};
\end{align*}
here, we use the definition of $M_2, M_3$, and $\mathcal{F}_{1j}$.

Since $\Delta_\Upsilon = \{ I_d \otimes (\widehat{\Sigma}_1 - \Sigma_1 + \widehat{\Sigma}_2 - \Sigma_2) \}_{(V_A, V_A)}$, the relevant term from $J_5$ is
\[
-
\left(
K_1^\top\otimes I_{m_A}
\right)
(\Upsilon^{-1}\otimes\Upsilon^{-1}) (\Pi_{V_A}^\top \otimes \Pi_{V_A}^\top) \mbox{vec}\left[ I_d \otimes \sqrt{n^{\text{inf}}} \left(\widehat{\Sigma}_1 - \Sigma_1\right) \right].
\]
Using the definition of $M_4$ and $\mathcal{L}_j$, this can further be written as
\[
-M_4 \left(\sum_{j=1}^d \mathcal{L}_j\right) \mbox{vec} \left[\sqrt{n^{\text{inf}}} \left(\widehat{\Sigma}_1 - \Sigma_1\right)\right].
\]
Combining all the relevant terms for $\operatorname{vec}
\left[
\sqrt{n^{\rm inf}}
\left(\widehat{\Sigma}_1-\Sigma_1\right)
\right]$, we obtain its coefficient
\[
A_{\Sigma, 1} = M_1 \left(\sum_{j=1}^d \mathcal{H}_{2,j}\right) - M_2 - M_3 \left(\sum_{j=1}^d \mathcal{F}_{1j}\right) - M_4 \left(\sum_{j=1}^d \mathcal{L}_j\right),
\]
as defined in the main text.

Similar derivations identify  the matrices
\(A_{\Sigma,2}\),
\(A_{\Theta,1}\),
and
\(A_{\Theta,2}\),
defined in the main text,
as the coefficient of
\[
\operatorname{vec}
\left[
\sqrt{n^{\rm inf}}
\left(\widehat{\Sigma}_2-\Sigma_2\right)
\right],
\quad
\operatorname{vec}
\left[
\sqrt{n^{\rm inf}}
\left(\widehat{\Theta}_1-\Theta_1\right)
\right],
\quad
\operatorname{vec}
\left[
\sqrt{n^{\rm inf}}
\left(\widehat{\Theta}_2-\Theta_2\right)
\right],
\]
respectively.

Therefore,
\[
\sqrt{n^{\rm inf}}
\left(\widehat{\psi}_A-\psi_A\right)
=
A_{\Sigma,1}
\operatorname{vec}
\left[
\sqrt{n^{\rm inf}}
\left(\widehat{\Sigma}_1-\Sigma_1\right)
\right]
+
A_{\Sigma,2}
\operatorname{vec}
\left[
\sqrt{n^{\rm inf}}
\left(\widehat{\Sigma}_2-\Sigma_2\right)
\right]
\]
\[
+
A_{\Theta,1}
\operatorname{vec}
\left[
\sqrt{n^{\rm inf}}
\left(\widehat{\Theta}_1-\Theta_1\right)
\right]
+
A_{\Theta,2}
\operatorname{vec}
\left[
\sqrt{n^{\rm inf}}
\left(\widehat{\Theta}_2-\Theta_2\right)
\right]
+
o_p(1),
\]
which completes the proof.
\end{proof}

Finally, we use Theorem \ref{thm:inf:delta_A:101724} to prove Theorem \ref{asymp:delta:A:sel:inf}.

\begin{proof}
By Theorem \ref{thm:inf:delta_A:101724},
\[
\sqrt{n^{\rm inf}}
\left(
\widehat{\psi}_A-\psi_A
\right)
=
\sum_{l=1}^{2}
A_{\Sigma,l}
\,
\mathrm{vec}
\left[
\sqrt{n^{\rm inf}}
\left(\widehat{\Sigma}_l-\Sigma_l\right)
\right]
+
\sum_{l=1}^{2}
A_{\Theta,l}
\,
\mathrm{vec}
\left[
\sqrt{n^{\rm inf}}
\left(\widehat{\Theta}_l-\Theta_l\right)
\right]
+
o_p(1).
\]

For \(l=1,2\), define
\[
Z_l
=
\begin{pmatrix}
\mathrm{vec}
\left[
\sqrt{n^{\rm inf}}
\left(\widehat{\Sigma}_l-\Sigma_l\right)
\right]
\\
\mathrm{vec}
\left[
\sqrt{n^{\rm inf}}
\left(\widehat{\Theta}_l-\Theta_l\right)
\right]
\end{pmatrix}.
\]

By Theorem 2 of \citet{roy1989asymptotic},
\[
Z_l
\overset{d}{\longrightarrow}
N
\left(
0,
\begin{pmatrix}
D_l & W_l \\
W_l^\top & \Lambda_l
\end{pmatrix}
\right),
\]
where \(D_l\), \(W_l\), and \(\Lambda_l\) are determined by the asymptotic covariance structure of the sample covariance and lag-one covariance estimators.
Specifically, the
\(
((j_1-1)d+k_1,(j_2-1)d+k_2)
\)-th entry of
\(D_l\)
is given by
\eqref{Di:entry},
while the corresponding entries of
\(W_l\)
and
\(\Lambda_l\)
are given by
\eqref{Wi:Lambdai:entry}.

Since the two time series are independent,
\[
\mathrm{cov}(Z_1,Z_2)=0.
\]
Therefore,
\[
\sum_{l=1}^{2}
\begin{bmatrix}
A_{\Sigma,l}
&
A_{\Theta,l}
\end{bmatrix}
Z_l
\overset{d}{\longrightarrow}
N_{m_A}
\left(
0,
\sum_{l=1}^{2}
\begin{bmatrix}
A_{\Sigma,l}
&
A_{\Theta,l}
\end{bmatrix}
\begin{pmatrix}
D_l & W_l \\
W_l^\top & \Lambda_l
\end{pmatrix}
\begin{bmatrix}
A_{\Sigma,l}^\top
\\
A_{\Theta,l}^\top
\end{bmatrix}
\right).
\]

Applying Slutsky's theorem to the
\(o_p(1)\)
remainder term in
Theorem \ref{thm:inf:delta_A:101724}
yields
\[
\sqrt{n^{\rm inf}}
\left(
\widehat{\psi}_A-\psi_A
\right)
\overset{d}{\longrightarrow}
N_{m_A}
\left(
0,
\sum_{l=1}^{2}
\begin{bmatrix}
A_{\Sigma,l}
&
A_{\Theta,l}
\end{bmatrix}
\begin{pmatrix}
D_l & W_l \\
W_l^\top & \Lambda_l
\end{pmatrix}
\begin{bmatrix}
A_{\Sigma,l}^\top
\\
A_{\Theta,l}^\top
\end{bmatrix}
\right),
\]
which completes the proof.
\end{proof}

\bibliography{reference_diffvar, paper-ref, ref_diffvar_inf}

\end{document}